\documentclass[aip, jcp, preprint, showpacs,superscriptaddress,groupedaddress]{revtex4-2}

\usepackage{amsmath}
\usepackage{amssymb}
\usepackage{braket}
\usepackage{bm}

\usepackage{tikz}
\usepackage{subcaption}

\begin{document}

\title{Transition density matrices of Richardson-Gaudin states}
\author{Paul A. Johnson}
 \email{paul.johnson@chm.ulaval.ca}
\author{Hubert Fortin}
\author{Samuel Cloutier}
\author{Charles-Émile Fecteau}
 \affiliation{D\'{e}partement de chimie, Universit\'{e} Laval, Qu\'{e}bec, Qu\'{e}bec, G1V 0A6, Canada}

\date{\today}

\begin{abstract}
Recently, ground state eigenvectors of the reduced Bardeen-Cooper-Schrieffer Hamiltonian, Richardson-Gaudin (RG) states, have been employed as a wavefunction ansatz for strong correlation. This wavefunction physically represents a mean-field of pairs of electrons (geminals) with a constant pairing strength. To move beyond the mean-field, one must develop the wavefunction in the basis of all the RG states. This requires both practical expressions for transition density matrices and an idea of which states are most important in the expansion. In this contribution, we present expressions for the transition density matrix elements and calculate them numerically for half-filled picket fence models. There are no Slater-Condon rules for RG states, though an analogue of the aufbau principle proves to be useful in choosing which states are important.
\end{abstract}

\maketitle

\section{Introduction}
Strongly-correlated electrons remain a problem in quantum chemistry. Active space methods are effective and expedient when the number of important Slater determinants is small, but quickly become infeasible as the number of important Slater determinants gets large. In such cases, it is more productive to start from a different qualitative picture, such as weakly-interacting pairs of electrons (geminals). Geminals are an old idea for quantum chemistry,\cite{hurley:1953,silver:1969,silver:1970} though recent work has been quite promising.\cite{coleman:1997,surjan:1999,kobayashi:2010,surjan:2012,neuscamman:2012,peter:2013,stein:2014,boguslawski:2014a,boguslawski:2014b,boguslawski:2014c,tecmer:2014,henderson:2014a,henderson:2014b,shepherd:2014,bulik:2015,boguslawski:2016,boguslawski:2017,surjan:2019,vu:2020,nowak:2020}

The group of Scuseria are developing a many-body theory for pairs based on the anti-symmetrized geminal power (AGP) as a mean-field. \cite{henderson:2019, khamoshi:2019, henderson:2020a, dutta:2020, harsha:2020, khamoshi:2020a, khamoshi:2020b} Their approach has the advantage that the wavefunction is simple and represents a clear physical picture: condensation of pairs. Their principal application thus far has been the reduced Bardeen-Cooper-Schrieffer (BCS)\cite{bardeen:1957a,bardeen:1957b} pairing model. 

Our approach is to employ the eigenvectors of the reduced BCS Hamiltonian, the Richardson-Gaudin (RG) states,\cite{richardson:1963,richardson:1964,richardson:1965,gaudin:1976} directly as a wavefunction ansatz. Variational mean-field calculations have been performed,\cite{johnson:2020} optimal energy expressions have been derived,\cite{fecteau:2020a} and the connection to single reference coupled-cluster methods has been explored.\cite{fecteau:2020b}  The principal advantage of our approach is clear: a complete set of eigenvectors with which to construct the physical wavefunction. Thus it will be straightforward, though technically difficult, to develop perturbation theories, selective configuration interactions, or coupled-cluster like approaches analogous to those built upon Hartree-Fock (HF). The next clear step in this direction is to obtain the necessary transition density matrix (TDM) elements between RG states. In the HF case, there are Slater-Condon rules which means that only singly- and doubly-excited Slater determinants have any non-zero coupling through the molecular Hamiltonian. This is not the case for RG states and the purpose of this contribution is to both develop expressions for the TDM elements and to evaluate them numerically to demonstrate which RG states have the largest coupling with the ground state through the Hamiltonian. The methods to develop the equations are quite similar, though trickier, than for the reduced density matrix(RDM) elements\cite{fecteau:2020a} so the presentation will be brief.
RG states have been used previously as a wavefunction model in condensed-matter theory\cite{claeys:2017a} and for configuration interaction in nuclear structure theory.\cite{stijn:2017}

A brief summary of the most relevant properties of RG states is presented in section \ref{sec: RG_vec}. Expressions for TDM elements in both the physical basis (PB) and the Gaudin basis (GB) are presented in section \ref{sec:TDM}. TDM elements and transition probabilities were computed numerically and the results are shown in section \ref{sec:results}.

\section{RG Eigenvectors} \label{sec: RG_vec}
The reduced BCS Hamiltonian is built from the set of pair objects in the PB:
\begin{align} \label{eq:su2}
S^+_i = a^{\dagger}_{i\uparrow} a^{\dagger}_{i\downarrow}, \quad S^-_i = a_{i\downarrow} a_{i\uparrow}, \quad S^z_i = \frac{1}{2} \left( a^{\dagger}_{i\uparrow} a_{i\uparrow} + a^{\dagger}_{i\downarrow}a_{i\downarrow} -1 \right) 
\end{align}
where $a^{\dagger}_{i \sigma}$ creates an electron in spatial orbital $i$ with spin $\sigma$ etc. $S^+_i$ creates a pair in spatial orbital $i$, $S^-_i$ removes a pair from spatial orbital $i$, and $S^z_i$ scales a pair-occupied spatial orbital by $\frac{1}{2}$ and an empty spatial orbital by $-\frac{1}{2}$. These objects have the structure of su(2):
\begin{align}
\left[S^z_i, S^{\pm}_j\right] = \pm \delta_{ij}S^{\pm}_i, \quad
\left[S^+_i, S^-_j\right]     = 2 \delta_{ij}S^z_i.
\end{align}
Rather than $S^z_i$, it is useful to employ the number operator 
\begin{align}
\hat{n}_i = 2 S^z_i + 1
\end{align}
which counts the number of electrons in the spatial orbital $i$. The reduced BCS Hamiltonian is written in terms of these objects as
\begin{align} \label{eq:BCSham}
\hat{H}_{BCS} = \frac{1}{2} \sum_i \varepsilon_i \hat{n}_i -\frac{g}{2} \sum_{ij} S^+_iS^-_j
\end{align}
and expresses a competition between filling the lowest energy spatial orbitals with isotropic scattering of pairs between the spatial orbitals. The particular form of the interaction, $-\frac{g}{2}$, is a consequence of the choice made for the definition of $S^z(u)$ in \eqref{eq:galg}.\cite{sklyanin:1989}

The eigenvectors of the reduced BCS Hamiltonian are built from a set of pair objects in the GB:
\begin{align} \label{eq:galg}
S^+(u) = \sum_i \frac{S^+_i}{u-\varepsilon_i}, \quad S^-(u) = \sum_i \frac{S^-_i}{u-\varepsilon_i}, \quad S^z(u) = \frac{1}{g} - \sum_i \frac{S^z_i}{u-\varepsilon_i}
\end{align}
where $u$ is an arbitrary complex number. For distinct $u_1, u_2$ these objects have commutation relations
\begin{align}
[S^z(u_1),S^{\pm}(u_2) ] = \pm \frac{S^{\pm}(u_1) - S^{\pm}(u_2)}{u_1-u_2}, \quad
[S^+(u_1),S^-(u_2) ] = 2 \frac{S^z(u_1) - S^z(u_2)}{u_1-u_2},
\end{align}
and for $u_1 =u_2 = u$ these may be analytically continued to give\cite{ortiz:2005}:
\begin{align}
[S^z(u),S^{\pm}(u) ] = \pm \frac{\partial S^{\pm}(u)}{\partial u}, \quad
[S^+(u), S^-(u)] = 2 \frac{\partial S^z(u)}{\partial u}.
\end{align}

The vacuum is denoted $\ket{\theta}$ and is usually the state with no particles. It may also contain unpaired electrons that do not participate in the pairing dynamics of the reduced BCS model. The vacuum is destroyed by the lowering operators:
\begin{subequations}
\begin{align}
S^-_i  \ket{\theta} &= 0 \\
S^-(u) \ket{\theta} &= 0
\end{align}
\end{subequations}
and is an eigenvector of the spin measurement operators
\begin{subequations}
\begin{align}
S^z_i  \ket{\theta} &= -\frac{1}{2} \ket{\theta} \\
S^z(u) \ket{\theta} &= \alpha(u) \ket{\theta}.
\end{align}
\end{subequations}
The eigenvalue $\alpha(u)$ is important, so we highlight its value
\begin{align} \label{eq:vac_eig}
\alpha(u) = \frac{1}{g} + \frac{1}{2} \sum_i \frac{1}{u-\varepsilon_i}.
\end{align}

Linearly independent vectors are produced by acting with the raising operators on the vacuum. States of the type
\begin{align}
\ket{\Phi} = S^+_1 S^+_2 \dots S^+_M \ket{\theta}
\end{align}
are Slater determinants, specifically closed shell (restricted Hartree-Fock) though the pairing scheme may be engineered to allow for more general cases. Acting twice with one $S^+_i$ results in zero. 

Eigenvectors of the reduced BCS Hamiltonian have the form
\begin{align} \label{eq:RGvec}
\ket{\{v\}} = S^+(v_1) S^+(v_2) \dots S^+(v_M) \ket{\theta}
\end{align}
where the set of \emph{rapidities} $\{v\}$ are a solution of Richardson's equations
\begin{align} \label{eq:Rich}
\frac{2}{g} + \sum_i \frac{1}{v_a - \varepsilon_i} + \sum_{b\neq a} \frac{2}{v_b - v_a} = 0.
\end{align}
Here we fix our notation: $\{u\}$ will always denote an arbitrary set of complex numbers, $\{v\}$ denotes \emph{a} solution of Richardson's equations and $\{w\}$ denotes the \emph{ground state solution} of Richardson's equations. The eigenvalues of \eqref{eq:RGvec} are
\begin{align}
\hat{H}_{BCS} \ket{\{v\}} = \sum_a v_a \ket{\{v\}}
\end{align}
and thus just the sum of the rapidities defining the state.

Our long term goal is develop the electronic wavefunction in the basis of RG eigenvectors in the hope of having a short expansion when the correct mean-field picture is of weakly-interacting pairs of electrons. The RG mean-field has already been reported,\cite{johnson:2020} along with optimal expressions for the density matrix elements required to compute it.\cite{fecteau:2020a} To calculate pertubative corrections, we will need to evaluate coupling of distinct RG eigenvectors through the Coulomb Hamiltonian
\begin{align}
\hat{H}_C = \sum_{ij} h_{ij} \sum_{\sigma} a^{\dagger}_{i\sigma}a_{j\sigma} + \frac{1}{2} \sum_{ijkl} V_{ijkl} \sum_{\sigma \tau} a^{\dagger}_{i\sigma}a^{\dagger}_{j\tau} a_{l\tau}a_{k \sigma}.
\end{align}
where $\sigma$ and $\tau$ are spin labels. The molecular integrals have been calculated in an orthonormal spatial orbital basis $\{\phi\}$, and the two-body integrals are in \emph{physicists'} notation:
\begin{align}
h_{ij} &= \int d\mathbf{r} \phi^*_i (\mathbf{r}) \left( - \frac{1}{2} \nabla^2 - \sum_I \frac{Z_I}{| \mathbf{r} - \mathbf{R}_I |} \right) \phi_j (\mathbf{r}) \\
V_{ijkl} &= \int d\mathbf{r}_1 d\mathbf{r}_2 \frac{\phi^*_i(\mathbf{r}_1)  \phi^*_j(\mathbf{r}_2)  \phi_k(\mathbf{r}_1)  \phi_l(\mathbf{r}_2)  }{| \mathbf{r}_1 - \mathbf{r}_2|}.
\end{align}
RG states have a definite number of unpaired electrons, or \emph{seniority}. Most often, they have no unpaired electrons, though non-interacting unpaired electrons may be added. We will henceforth assume that all RG states have seniority zero. No generality is lost as adding unpaired electrons affects all the RG states in precisely the same manner. Thus, the only non-zero contributions arise from the seniority-conserving or seniority-zero piece of the Coulomb Hamiltonian:
\begin{align} \label{eq:sen0_ham}
\hat{H}_{SZ} = \sum_i h_{ii} \hat{n}_i + \frac{1}{4} \sum_{i\neq j} \left( 2V_{ijij} - V_{ijji} \right) \hat{n}_i \hat{n}_j +\sum_{ij} V_{iijj} S^+_i S^-_j.
\end{align}
In the physical basis, we will therefore require the (unnormalized) TDM elements:
\begin{align}
D^{vw}_i &= \frac{1}{2} \braket{ \{v\} | \hat{n}_i | \{w\}}  \\
D^{vw}_{ij}  &= \frac{1}{4} \braket{ \{v\} | \hat{n}_i \hat{n}_j | \{w\}}  \\
P^{vw}_{ij}  &=  \braket{ \{v\} | S^+_i S^-_j | \{w\}} 
\end{align}

We will also evaluate TDM elements in the Gaudin basis. The Gaudin pair creators are a rectangular linear transformation of the local pair operators:
\begin{align}
\begin{pmatrix}
S^+(u_1) \\
S^+(u_2) \\
\vdots \\
S^+(u_M)
\end{pmatrix} =
\begin{pmatrix}
\frac{1}{u_1 - \varepsilon_1} & \frac{1}{u_1 - \varepsilon_2} & \dots & \frac{1}{u_1 - \varepsilon_N} \\
\frac{1}{u_2 - \varepsilon_1} & \frac{1}{u_2 - \varepsilon_2} & \dots & \frac{1}{u_2 - \varepsilon_N} \\
\vdots & \vdots & \ddots & \vdots \\
\frac{1}{u_M - \varepsilon_1} & \frac{1}{u_M - \varepsilon_2} & \dots & \frac{1}{u_M - \varepsilon_N} \\
\end{pmatrix}
\begin{pmatrix}
S^+_1 \\
S^+_2 \\
\vdots \\
S^+_N
\end{pmatrix}
\end{align}
or more succinctly
\begin{align} \label{eq:rect}
S^+(\mathbf{u}) = C \mathbf{S}^+.
\end{align}
The Cauchy matrix $C$ has a known\cite{Schechter:1959} explicit right-inverse $C^R$
\begin{align}
[C^R]_{ia} = (u_a - \varepsilon_i) \prod_{k \neq i} \left( \frac{u_a - \varepsilon_k}{\varepsilon_i - \varepsilon_k} \right) \prod_{b \neq a} \left( \frac{\varepsilon_i - u_b}{u_a - u_b} \right)
\end{align}
such that $CC^R = I_M$. The system of equations \eqref{eq:rect} has more unknowns than equations, thus has infinitely-many solutions (provided it has any). A particular solution of \eqref{eq:rect} is
\begin{align}
\mathbf{S}^+ = C^R S^+(\mathbf{u})
\end{align}
as is easily verified by substitution. By the exact same argument, for the Gaudin pair removal operator we arrive at
\begin{align}
\mathbf{S}^- = C^R S^-(\mathbf{u}).
\end{align}
It is helpful to rewrite $S^z(u)$ as 
\begin{align}
S^z(u) = \alpha(u) - \frac{1}{2} \sum_i \frac{\hat{n}_i}{u - \varepsilon_i},
\end{align}
with $\alpha(u)$ the vacuum eigenvalue \eqref{eq:vac_eig} to get
\begin{align}
\hat{\mathbf{n}} = -2 C^R \left( S^z(\mathbf{u}) - \alpha(\mathbf{u}) \right).
\end{align}
The seniority-zero Hamiltonian can then be written in the GB as:
\begin{align}
\hat{H}_{SZ} (\{u\}) = E_0 + \sum_a \tilde{h}_{a} S^z(u_a) + \sum_{ab} \tilde{V}^z_{ab} S^z(u_a) S^z(u_b) + \sum_{ab} \tilde{V}^p_{ab} S^+(u_a) S^-(u_b)
\end{align}
with the transformed integrals
\begin{align}
E_0 &= 2 \sum_{ia} h_{ii} C^R_{ia} \alpha(u_a) + \sum_{i\neq j} \sum_{ab}(2V_{ijij} - V_{ijji}) C^R_{ia} C^R_{jb} \alpha(u_a) \alpha(u_b) \\
\tilde{h}_{a} &= -2  \sum_i h_{ii} C^R_{ia} -2 \sum_{i\neq j} \sum_b (2V_{ijij} - V_{ijji}) C^R_{ia}C^R_{jb} \alpha(u_b) \\
\tilde{V}^z_{ab} &= \sum_{i\neq j} (2V_{ijij} - V_{ijji}) C^R_{ia} C^R_{jb} \\
\tilde{V}^p_{ab} &= \sum_{ij} V_{iijj} C^R_{ia} C^R_{jb}.
\end{align}

The Hamiltonian may be transformed to any set of rapidities $\{u\}$, which may satisfy Richardson's equations or not. The transformation depends only upon the Cauchy structure. We will choose the ground state rapidities $\{w\}$. Thus, in the Gaudin basis, we will evaluate the unnormalized TDM elements:
\begin{align}
Z^{vw} (w_a)      &= \braket{ \{v\} | S^z(w_a) | \{w\}} \\
Z^{vw} (w_a, w_b) &= \braket{ \{v\} | S^z(w_a) S^z(w_b) | \{w\}} \\
P^{vw} (w_a, w_b) &= \braket{ \{v\} | S^+(w_a) S^-(w_b) | \{w\}}
\end{align}
The notation employed here is different from that we employed for the RDM elements in the GB. As we must choose a set of rapidities in which to work, the present notation is meant to emphasize this choice.

\subsection{Nature of RG excited states}
Slater determinants are uniquely defined by which orbitals are occupied, with the lowest in energy generally being that with the lowest energy orbitals filled. This is not the case for RG states, but there is a useful analogue of the aufbau principle, and one of the aims of the present contribution is to reinforce it. Numerical algorithms to solve Richardson's equations\cite{rombouts:2004,faribault:2011,elaraby:2012,guan:2012,pogosov:2012,stijn:2012,claeys:2015} begin with a coupling strength $g$ of zero, for which the reduced BCS Hamiltonian is diagonal in a basis of Slater determinants, before turning on the interaction adiabatically to evolve to the interacting RG solutions. Thus, we can identify an RG state with its Slater determinant initial guess as the evolution is smooth. The ground state is associated with the lowest energy Slater determinant, defined by filling the $M$ lowest sites. It is natural to refer to this RG state as a string of $M$ 1's and $N-M$ 0's. The next lowest energy RG states evolve from Slater determinants which are single pair excitations on top of the lowest energy Slater determinant. We will refer to these RG states as \emph{single excitations} or \emph{singles}, and label them individually with strings of 1's and 0's based on the occupations of the pairs at zero coupling. This nomenclature easily extends to doubles, triples, etc. Strings of 1's and 0's uniquely label each RG state: Richardson established that the $\binom{N}{M}$ distinct solutions of Richardson's equations correspond to the $\binom{N}{M}$ distinct eigenvectors of the reduced BCS Hamiltonian, so provided that each site can only hold one pair this labelling scheme is a one-to-one correspondence. In the case of degeneracies, the situation is a little more complicated, but has been solved as well.\cite{elaraby:2012}

In an analogous Bethe ansatz for electrons,\cite{laurie} Richardson's equations reduce to $N$ independent conditions. In that case, each state is uniquely labelled in the usual occupied/virtual sense familiar to HF. However, \emph{each RG state is defined by a different set of rapidities}, which are coupled by Richardson's equations. As a result it is not possible to write RG excited states as the action of a simple second-quantized operator acting on the RG ground state. 

\section{Transition density matrices} \label{sec:TDM}
\subsection{Scalar products}
Slavnov's theorem\cite{Slavnov:1989,Belliard:2019} expresses the scalar product between two RG states with at least one of the sets (in this case $\{v\}$) a solution of Richardson's equations, while the set $\{u\}$ is arbitrary:
\begin{align} \label{eq:slavnov}
\braket{\{v\} | \{u\}} = K(\{v\},\{u\}) \det J(\{v\},\{u\})
\end{align}
where $K(\{v\},\{u\})$ is 1 over the determinant of a Cauchy matrix
\begin{align}
K(\{v\},\{u\}) = \frac{\prod_{ab} (v_a-u_b)}{\prod_{a<b} (u_a-u_b)(v_b-v_a)}
\end{align}
and the elements of the matrix $J$ are
\begin{align}
J_{ab} = \frac{1}{(v_a -u_b)^2} \left(\frac{2}{g} + \sum_i \frac{1}{(u_b-\varepsilon_i)} - \sum_{c\neq a} \frac{2}{(u_b - v_c)}  \right).
\end{align}
$K$ and $J$ are understood to be functions of two sets of rapidities. As this dependence will always be clear, it will be suppressed. Slavnov's result original result concerned the more general case of the six-vertex model, while the present expression was first written by Zhou et al.\cite{Zhou:2002}

There are two particular limits of \eqref{eq:slavnov} of immediate relevance. The first occurs when $\{u\} \rightarrow \{v\}$ in which case we obtain the square of the norm. The result is the determinant of the Gaudin matrix $G^v$, which is the Jacobian of Richardson's equations:
\begin{align} \label{eq:gmat}
\braket{\{v\} | \{v\} } &= \det G^v \\
G^v_{ab} &= 
\begin{cases}
\sum_i \frac{1}{(v_a -\varepsilon_i)^2} - \sum_{c\neq a} \frac{2}{(v_a - v_c)^2}, \quad &a=b \\
\frac{2}{(v_a-v_b)^2}, \quad &a\neq b.
\end{cases}
\end{align}
The norm of the state $\{w\}$ is similarly obtained, substituting $\{w\}$ for $\{v\}$.

The second case occurs when $\{u\} \rightarrow \{w\}$. In this case, we obtain the scalar product between distinct (non-degenerate) eigenvectors of a Hermitian operator, which must therefore be zero. Using Richardson's equations, one can see that the columns of $J$ become linearly dependent:
\begin{align}
c_1 J_1 + c_2 J_2 + \dots + c_N J_N = 0
\end{align}
with coefficients
\begin{align}
c_k = \prod_a (w_k - v_a)^2
\end{align}
so the determinant of $J$, and hence the overlap $\braket{\{v\}|\{w\}}$, is zero.

\subsection{Physical basis}
We follow the same approach as for the RDM case,\cite{GB:2011,fecteau:2020a} so we pass directly to the results. By moving the PB operators to the right past the Gaudin pair creators until they act upon the vacuum, the TDM elements are:
\begin{subequations}\label{eq:PB_TDM}
\begin{align}
D^{vw}_i &= \sum_a \frac{F^i_a}{w_a - \varepsilon_i} \\
D^{vw}_{ij}  &=  \sum_{a\neq b} \frac{F^{ij}_{ab}}{(w_a-\varepsilon_i)(w_b-\varepsilon_j)} \\
P^{vw}_{ij}  &=  \sum_a \frac{F^i_a}{w_a-\varepsilon_i} - \sum_{a \neq b} \frac{F^{ij}_{ab}}{(w_a-\varepsilon_j)(w_b-\varepsilon_j)}
\end{align}
\end{subequations}
where the common elements are the form factors:
\begin{align}
F^i_a &= \braket{\{v\} | S^+_i | \{w\}_a} \\
F^{ij}_{ab} &= \braket{\{v\} | S^+_i S^+_j | \{w\}_{a,b}}.
\end{align}
The notation $\{w\}_a$ denotes the set $\{w\}$ without the element $w_a$ etc. As in the RDM case, the form factors are evaluated as limits of Slavnov's theorem. First, notice that the PB operators are the residues of the Gaudin pair creators
\begin{align}
S^+_i = \lim_{u_a \rightarrow \varepsilon_i} (u_a - \varepsilon_i) S^+(u).
\end{align}
Now, keep the set $\{u\}$ as arbitrary, and evaluate the corresponding residues of the scalar products
\begin{align}
\braket{\{v\} | S^+_i | \{u\}_a} = \lim_{u_a \rightarrow \varepsilon_i} (u_a - \varepsilon_i) \braket{\{v\}|\{u\}}
\end{align}
and take the limit $\{u\} \rightarrow \{w\}$ to get:
\begin{align} \label{eq:Fai}
F^i_a = \frac{\prod_c (v_c-\varepsilon_i)}{\prod_c (v_c-w_a)} \frac{\prod_{b\neq a} (w_a-w_b)}{\prod_{b \neq a} (\varepsilon_i - w_b)} K \det J^i_a
\end{align}
Here, $J^i_a$ is the matrix $J$ with the $a$th column replaced by the column:
\begin{align} \label{eq:local_RHS}
\mathbf{q}_i = \begin{pmatrix}
\frac{1}{(v_1 - \varepsilon_i)^2} \\
\frac{1}{(v_2 - \varepsilon_i)^2} \\
\vdots \\
\frac{1}{(v_{M} - \varepsilon_i)^2}
\end{pmatrix}.
\end{align}

Similarly,
\begin{align}
\braket{ \{v\}| S^+_i S^+_j | \{u\}_{a,b}} = 
\lim_{u_a \rightarrow \varepsilon_i} \lim_{u_b \rightarrow \varepsilon_j} (u_a-\varepsilon_i)(u_b-\varepsilon_j) \braket{\{v\}|\{u\}}
\end{align}
and taking the limit $\{u\} \rightarrow \{w\}$ gives
\begin{align} \label{eq:Fabij}
F^{ij}_{ab} = \frac{(w_a-w_b)}{(\varepsilon_i-\varepsilon_j)} \frac{\prod_c (v_d-\varepsilon_i)(v_d-\varepsilon_j)}{\prod_c (v_d-w_a)(v_d-w_b)}
\frac{\prod_{d\neq a,b} (w_a-w_d)(w_b-w_d)}{\prod_{d \neq a,b} (\varepsilon_i - w_d)(\varepsilon_j - w_d)} K \det J^{ij}_{ab},
\end{align}
where $J^{ij}_{ab}$ is the matrix $J$ with the $a$th column replaced by the $i$th version of \eqref{eq:local_RHS} and the $b$th column replaced by the $j$th version of \eqref{eq:local_RHS}.

Thus, to compute the TDM elements in the physical basis, we compute the form factors \eqref{eq:Fai} and \eqref{eq:Fabij} then compute the sums \eqref{eq:PB_TDM}. In the RDM case, substantial further simplification was possible due to Cramer's rule and a theorem of Jacobi\cite{vein_book} for scaled minors: dividing by the norm of the state yielded ratios of determinants differing by one or two columns. In this case, the form factors differ substantially from the Gaudin matrices. For $\{u\}$ arbitrary, there is a relevant Jacobi identity
\begin{align}
\frac{\det J^{ij}_{ab}}{\det J} = \frac{\det J^i_a}{\det J}\frac{\det J^j_b}{\det J} - \frac{\det J^i_b}{\det J}\frac{\det J^j_a}{\det J}
\end{align}
but when $\{u\} \rightarrow \{w\}$, $J$ becomes singular and this identity no longer holds. Therefore, while it may be possible to simplify the TDM elements, for this contribution we content ourselves with \eqref{eq:PB_TDM} to evaluate numerically. 

The result for $D^{vw}_{ij}$ is valid provided $i\neq j$. The diagonal element corresponds to $D^{vw}_i$ (and to $P^{vw}_{ii}$). It is not difficult to see that $D^{vw}_i$ and $D^{vw}_{ij}$ sum to zero.
\begin{align}
\sum_i D^{vw}_i &= 0 \\
\sum_{ij} D^{vw}_{ij} &= 0
\end{align}

\subsection{Gaudin basis}

TDM elements in the GB are obtained in a similar manner to that for the RDM elements: first use the Gaudin algebra to write the TDM elements as a sum of form factors, then evaluate the form factors. As the Gaudin algebra actions are exactly the same as in the RDM case, they are listed in appendix \ref{sec:GAA}. The form factors required are more tricky than for the RDM case, but the same procedure is to be followed. Expressions for the form factors are listed in appendix \ref{sec:GFF}.

The final expression for the $S^z(w_a)$ TDM elements in the GB is:
\begin{align} \label{eq:GBZ}
Z^{vw} (w_a) &= K \sum_c \lambda^a_c  \det J^a_c
\end{align}
with
\begin{align}
\lambda^a_c =  \frac{\prod_k (v_k - w_a)  }{\prod_k (v_k - w_c)} \frac{\prod_{l \neq c} (w_c - w_l)}{\prod_{l \neq a} (w_a - w_l)},
\end{align}
and the matrix $J^a_c$ is the matrix $J$ with the $c$th column replaced with the derivative of the $a$th column with respect to $w_a$: $\frac{\partial J_a}{\partial w_a}$. The derivatives are:
\begin{align}
\left(\frac{\partial J_a}{\partial w_a} \right)_b &= \frac{2}{(v_b-w_a)^3} \left( \frac{2}{g} + \sum_i \frac{1}{(w_a-\varepsilon_i)} - \sum_{c\neq b} \frac{2}{(w_a-v_c)} \right) \nonumber \\
&- \frac{1}{(v_b-w_a)^2} \left( \sum_i \frac{1}{(w_a -\varepsilon_i)^2} - \sum_{c\neq b} \frac{2}{(w_a-v_c)^2} \right) 
\end{align}

Clearly, $\lambda^a_a =1$. Further, with
\begin{align}
k(w_a) = \sum_c \frac{1}{w_a - v_c} - \sum_{c \neq a} \frac{1}{w_a - w_c}
\end{align}
and the elements of the second derivatives of $J$
\begin{align}
\frac{1}{2} \left(\frac{\partial^2 J_a}{\partial w_a^2} \right)_b &=
 \frac{3}{(v_b-w_a)^4} \left( \frac{2}{g} + \sum_i \frac{1}{(w_a-\varepsilon_i)} - \sum_{c\neq b} \frac{2}{(w_a-v_c)} \right) \nonumber \\
&- \frac{2}{(v_b-w_a)^3} \left( \sum_i \frac{1}{(w_a -\varepsilon_i)^2} - \sum_{c\neq b} \frac{2}{(w_a-v_c)^2} \right) \nonumber \\
&+ \frac{1}{(v_b-w_a)^2} \left( \sum_i \frac{1}{(w_a-\varepsilon_i)^3} - \sum_{c\neq b} \frac{2}{(w_a-v_c)^3} \right)
\end{align}
the expressions for the other transition density matrix elements are:

\begin{align}
\frac{1}{K} Z^{vw} (w_a, w_a) &= \det J^{\bar{a}}_a + k(w_a) \det J^a_a \nonumber \\
&+ \sum_{c\neq a} \lambda^a_c \left( 3 \det J^{\bar{a}}_c + \left( 3 k(w_a) + \frac{4}{(w_a-w_c)} \right) \det J^a_c \right) \nonumber \\
&+ \sum_{c,d\neq a} \lambda^a_c \lambda^a_d \frac{(w_a-w_c)(w_a-w_d)}{(w_d-w_c)} \det J^{\bar{a}a}_{cd}
\end{align}

\begin{align}
\frac{1}{K} Z^{vw} (w_a, w_b) &= \det J^{ab}_{ab} + k(w_a) \det J^b_b + k(w_b) \det J^a_a + \frac{\left( \lambda^a_b \det J^a_b - \lambda^b_a \det J^b_a \right)}{(w_a-w_b)} \nonumber \\
&+ \sum_{c \neq a,b} \lambda^b_c \left( \det J^{ab}_{ac} + \left( k(w_a) + \frac{3(w_c-w_b)}{(w_a-w_c)(w_a-w_b)} \right) \det J^b_c \right) \nonumber \\
&+ \sum_{c \neq a,b} \lambda^a_c \left( \det J^{ba}_{bc} + \left( k(w_b) + \frac{3(w_c-w_a)}{(w_b-w_c)(w_b-w_a)} \right) \det J^a_c \right) \nonumber \\
&+ \frac{1}{2} \sum_{c,d \neq a,b} \lambda^a_c \lambda^b_d \frac{(w_a-w_d)(w_b-w_c)+(w_a-w_c)(w_b-w_d)}{(w_a-w_b)(w_d-w_c)}  \det J^{ab}_{cd}
\end{align}

\begin{align}
\frac{1}{K} P^{vw} (w_a, w_a) &= -2 \sum_{c \neq a} \lambda^a_c 
\left(
\det J^{\bar{a}}_c + \left(k(w_a) - \frac{2}{(w_c-w_a)} \right) \det J^a_c \right) \nonumber \\
&- \sum_{c,d \neq	a} \lambda^a_c \lambda^a_d \frac{(w_a-w_c)(w_a-w_d)}{(w_d-w_c)} \det J^{\bar{a}a}_{cd}
\end{align}

\begin{align}
\frac{1}{K} P^{vw} (w_a, w_b) &= 
\left( \sum_i \frac{1}{(w_b-\varepsilon_i)^2} - \sum_{c\neq b} \frac{2}{(w_c-w_b)^2} \right) \lambda^a_b (w_a-w_b) \det J^a_b 
+ \frac{2}{w_a-w_b} \det J^b_b \nonumber \\
&- 2 \sum_{c\neq a, b} \frac{(w_c-w_a)}{(w_c-w_b)}\lambda^a_c 
\left(
\det J^{ba}_{bc} + \left( k(w_b) - \frac{(w_c-w_a)+(w_b-w_a)}{(w_c-w_b)(w_b-w_a)}\right)\det J^a_c 
\right) \nonumber \\
&-2 \sum_{c \neq a,b} \frac{\lambda^b_c \det J^b_c}{w_b-w_a}
+ \sum_{c,d\neq b} \lambda^a_c \lambda^b_d \frac{(w_a-w_c)(w_a-w_d)}{(w_a-w_b)(w_c-w_d)} \det J^{ab}_{cd}
\end{align}

$J^{ab}_{cd}$ is the matrix $J$ with the $c$th column replaced with $\frac{\partial J_a}{\partial w_a}$ and the $d$th column replaced with $\frac{\partial J_b}{\partial w_b}$. Similarly, $J^{\bar{a}}_c$ is $J$ with the $c$th column replaced by $\frac{1}{2}\frac{\partial^2 J_a}{\partial w_a^2}$, etc. Again, the expressions for the TDM elements may not be optimal as further simplification from Cramer's rule and Jacobi's identity are not possible as in the RDM case.

\subsection{Consistency checks}
The TDM elements in the GB have been verified numerically against direct transformations of the corresponding elements in the PB. Specifically,

\begin{align}
P^{vw} (w_a, w_b)  = \sum_{ij} \frac{P^{vw}_{ij}}{(w_a-\varepsilon_i)(w_b-\varepsilon_j)}
\end{align}

\begin{align}
Z^{vw} (w_a) = - \sum_i \frac{D^{vw}_i}{w_a-\varepsilon_i}
\end{align}
and
\begin{align}
Z^{vw} (w_a, w_b) = \sum_{ij} \frac{D^{vw}_{ij} - \frac{1}{2} D^{vw}_i - \frac{1}{2}D^{vw}_j}{(w_a-\varepsilon_i)(w_b-\varepsilon_j)} - \frac{1}{g}\sum_i \frac{D^{vw}_i}{w_a-\varepsilon_i} - \frac{1}{g}\sum_j \frac{D^{vw}_j}{w_b-\varepsilon_j}.
\end{align}

The TDM elements in the GB satisfy sum rules, which also serve as a consistency check. With the eigenvalue based variables
\begin{align}
W_i = \sum_a \frac{1}{\varepsilon_i - w_a},
\end{align}
the sum of the $Z^{vw} (w_a)$ elements is
\begin{align}
\sum_a Z^{vw} (w_a) = \sum_i W_i D^{vw}_i
\end{align}
the sum of the $P^{vw} (w_a, w_b)$ elements gives,
\begin{align}
\sum_{ab}  P^{vw} (w_a, w_b) = \sum_{ij} W_iW_j P^{vw}_{ij}
\end{align}
and the sum of the $Z^{vw} (w_a, w_b)$ elements is
\begin{align}
\sum_{ab} Z^{vw} (w_a, w_b) = \frac{2M}{g}\sum_i W_iD^{vw}_i + \sum_{ij} W_i W_j \left( D^{vw}_{ij} - \frac{1}{2}D^{vw}_i - \frac{1}{2}D^{vw}_j \right).
\end{align}

Defining
\begin{align}
\Delta^{vw} (w_a) = \braket{ \{v\} | \frac{\partial S^z (w_a)}{\partial w_a} | \{w\}} = K \sum_c \lambda^a_c \left( \det J^{\bar{a}}_c + k(w_a)\det J^a_c \right),
\end{align}
another consistency check can be performed since $\ket{\{v\}}$ and $\ket{\{w\}}$ are both eigenvectors of $S^2(w_a)$, so that 
\begin{align}
\braket{\{v\} | S^2(w_a) | \{w\}} = 0.
\end{align}
This is easily verified, as
\begin{align}
S^2(w_a) &= S^z(w_a) S^z(w_a) + S^+(w_a)S^-(w_a) - \frac{\partial S^z(w_a)}{\partial w_a}
\end{align}
and therefore,
\begin{align}
\braket{\{v\} | S^2(w_a) | \{w\}} = Z^{vw} (w_a,w_a) + P^{vw} (w_a,w_a) - \Delta^{vw} (w_a) = 0
\end{align}
by a direct computation.

\section{Numerical Results} \label{sec:results}
To understand which RG excited states would be most important in a perturbation theory, we have numerically computed the \emph{normalized} TDM elements in both the PB and the GB for a picket-fence model, i.e. the reduced BCS model with constant energy spacing between the levels. The dynamics of this model are a function of the ratio of the pairing strength $g$ to the spacing $\Delta \varepsilon$. In our chosen convention \eqref{eq:BCSham}, the pairing interaction is $-\frac{g}{2}$ and thus a negative pairing interaction corresponds to a repulsive interaction. 

Rather than plot individual TDM elements, we present sums of the corresponding transition probabilities, i.e. the sum of the absolute squares of the TDM elements. Transition probabilities are real and positive definite, and are thus useful to measure relative importance of RG excited states. Further, in a Rayleigh-Schr\"{o}dinger pertubation theory (RSPT) expansion, contributions from excited states are weighted with the difference in zeroth-order energy differences between the two states. Using the reduced BCS model as a starting point, the energy denominators are the sums of differences of rapidities. Therefore, we also plot transition probabilities with energy denominators to judge which RG excited states would be important for RSPT.

All curves are plotted for small systems, half-filled 4- and 6-site models, while excitations are grouped together for the half-filled 10-site model. The size of the system is not a problem, but the number of states grows combinatorially. Additional results for half-filled 8-site model are included in the supplementary material. 

\subsection{Physical Basis}
We compute the TDM elements in the PB from the results of section \ref{sec:TDM} before taking their absolute squares to plot transition probabilities. In particular, we plot $\sum_i |D^{vw}_i |^2 $, $\sum_{ij} |D^{vw}_{ij} |^2 $ and $\sum_{ij} |P^{vw}_{ij} |^2 $, along with their weighting by energy denominators: $\frac{ \sum_i |D^{vw}_i |^2 }{\sum_a w_a - v_a}$, $\frac{ \sum_{ij} |D^{vw}_{ij} |^2 }{\sum_a w_a - v_a}$ and $\frac{ \sum_{ij} |P^{vw}_{ij} |^2 }{\sum_a w_a - v_a}$. 

\begin{figure} 
	\begin{subfigure}{\textwidth}
		\includegraphics[width=0.325\textwidth]{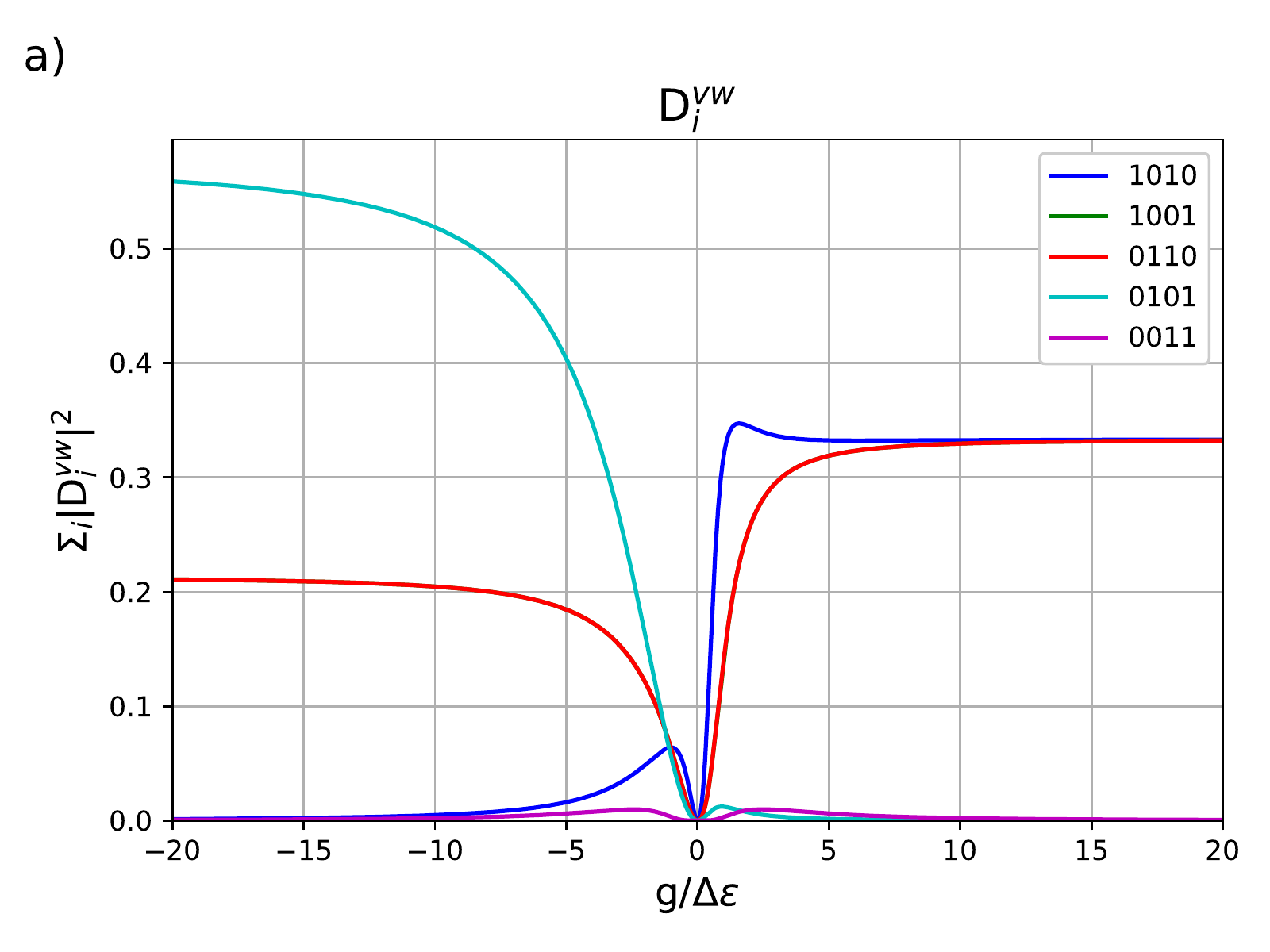}  \hfill
		\includegraphics[width=0.325\textwidth]{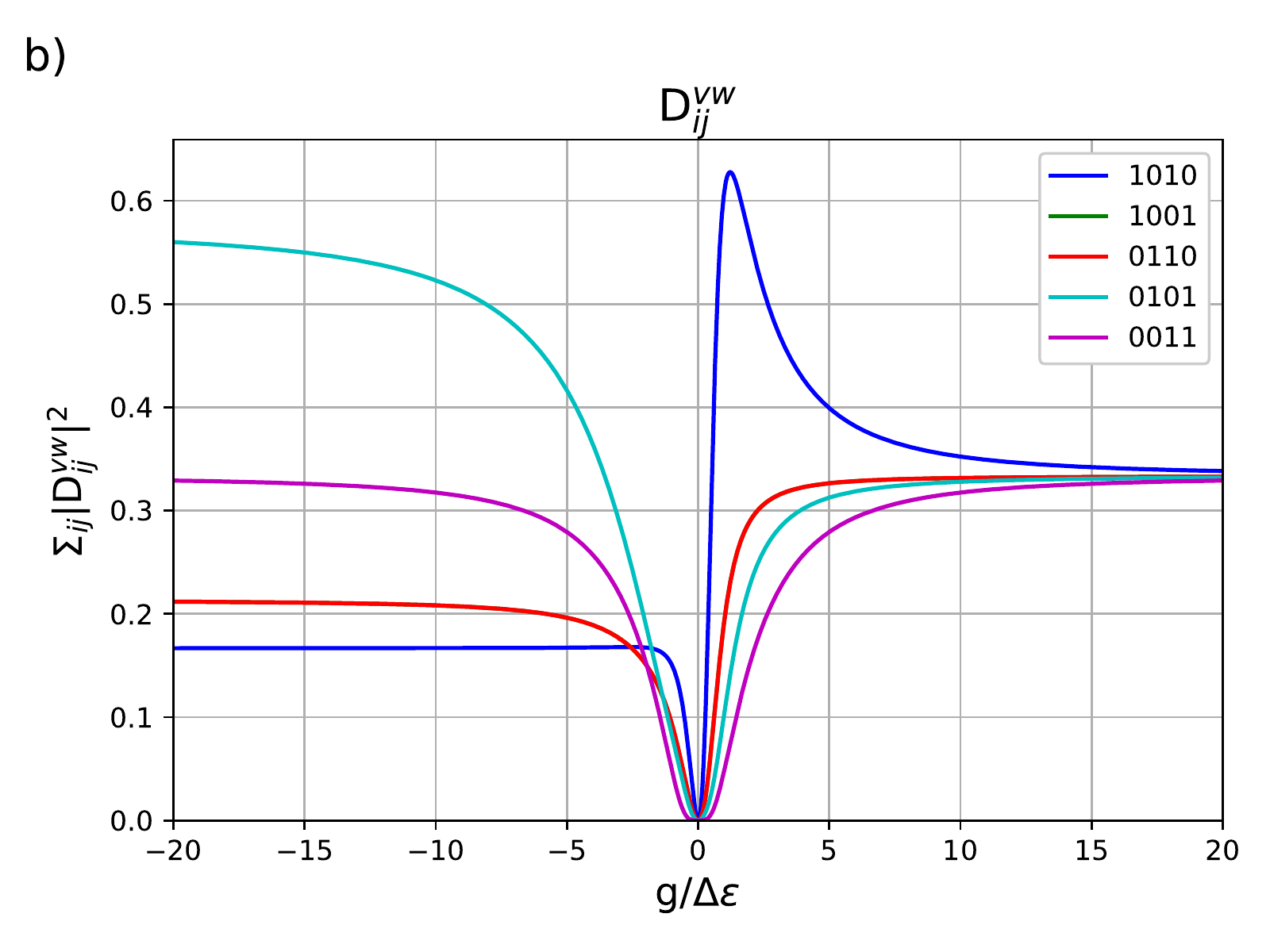} \hfill
		\includegraphics[width=0.325\textwidth]{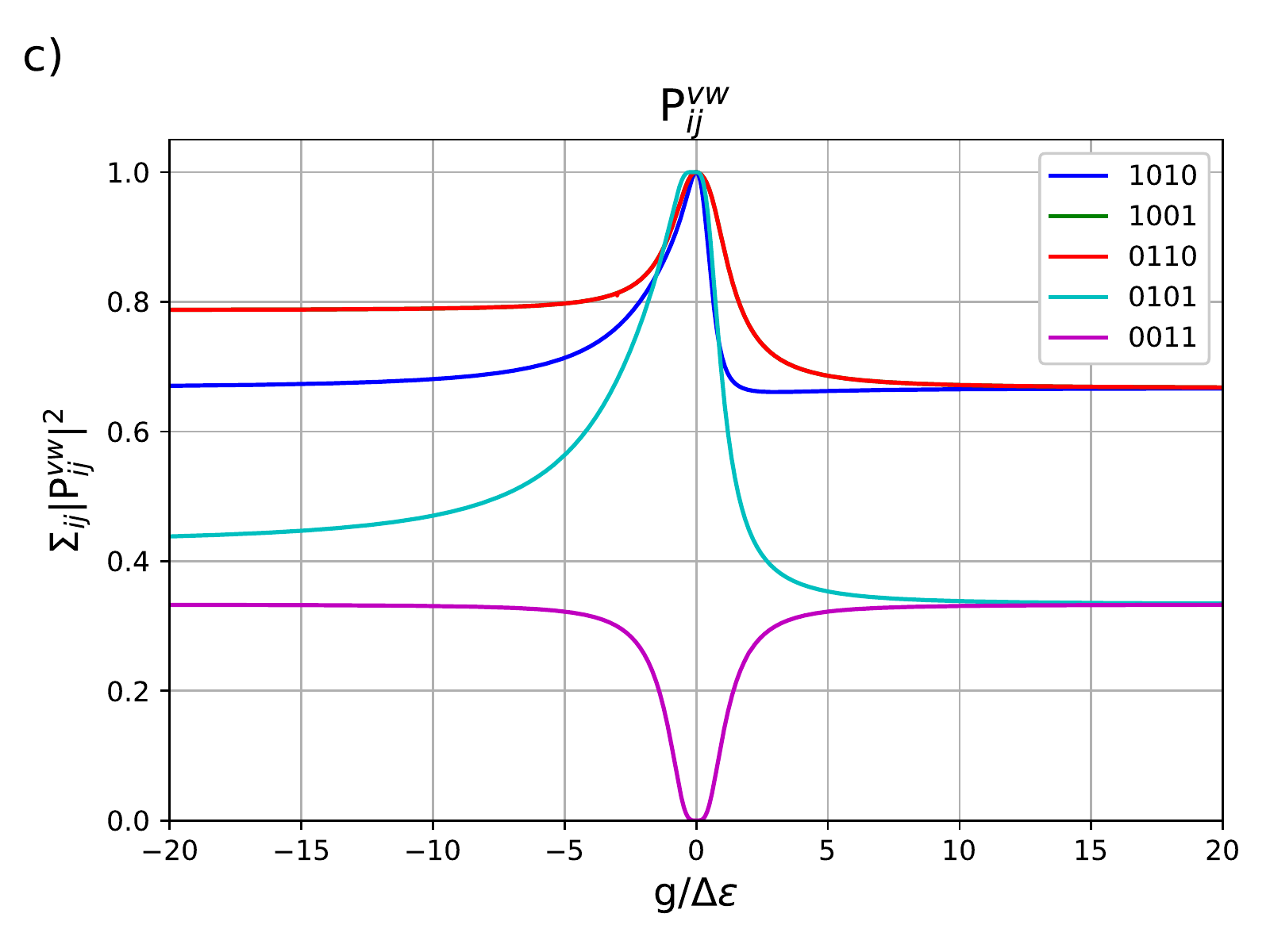}				
	\end{subfigure}
	\begin{subfigure}{\textwidth}
		\includegraphics[width=0.325\textwidth]{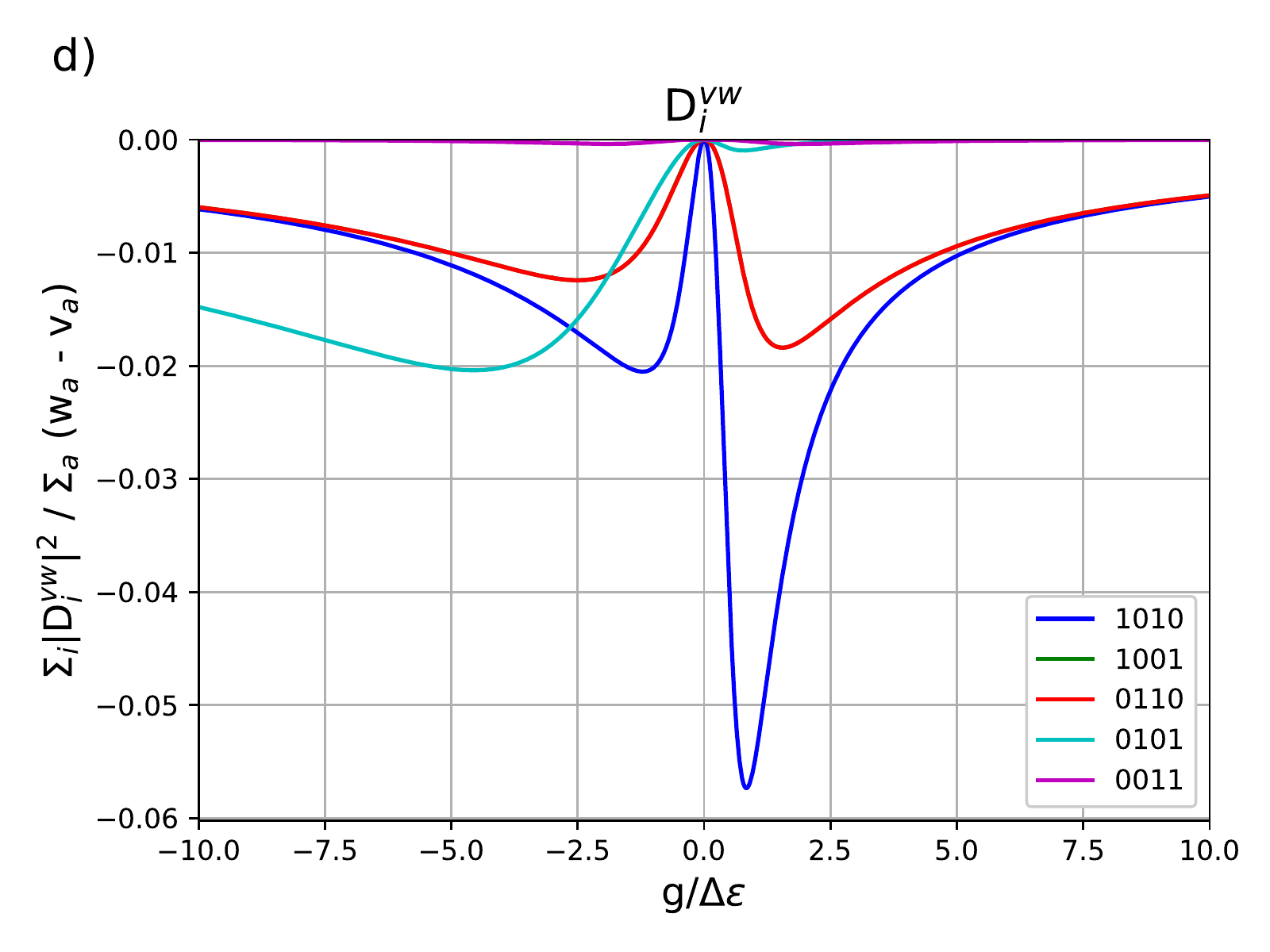}  \hfill
		\includegraphics[width=0.325\textwidth]{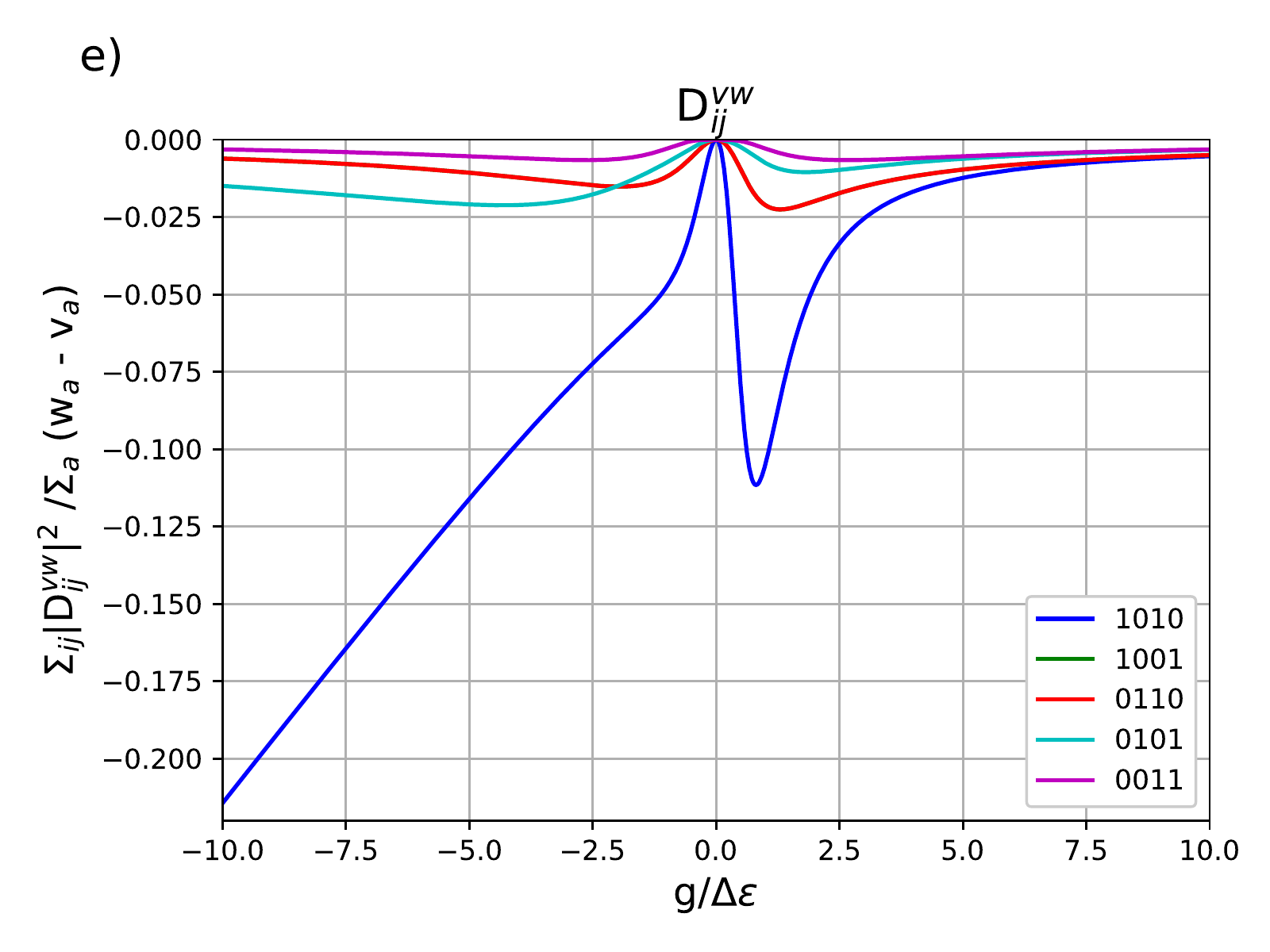} \hfill
		\includegraphics[width=0.325\textwidth]{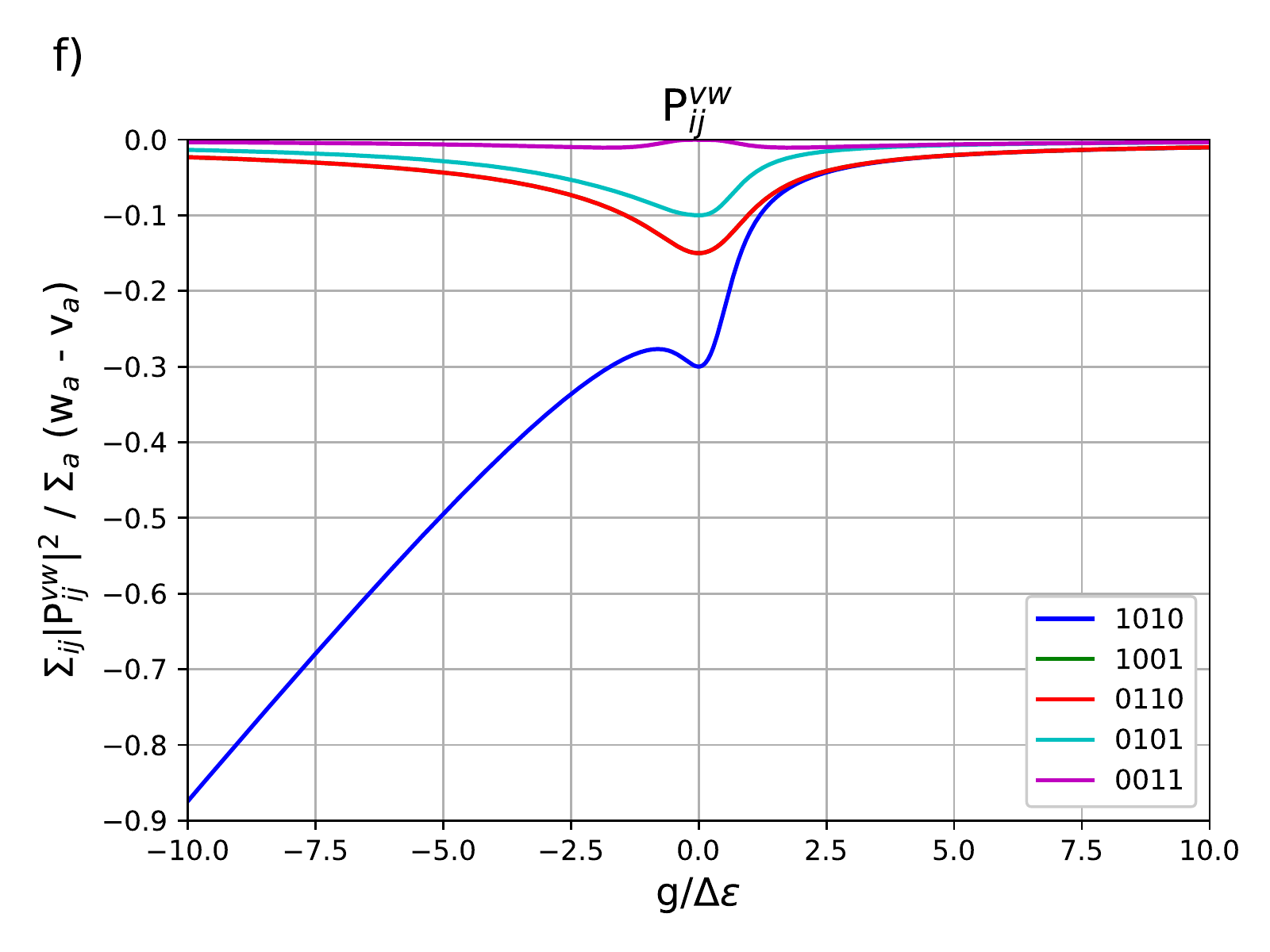}		
	\end{subfigure}
	\caption{a-c) Sums of transition probabilities for half-filled 4-site picket fence model, with energy denominators included d-f).}	
	\label{fig:4_2_curves}
\end{figure}
Transition probabilities for the half-filled 4-site picket fence model are shown in figure \ref{fig:4_2_curves}. In each plot the curves for the states 1001 and 0110 are indistinguishable. The TDM elements and transition probabilities for these states are \emph{not} identical, but their sums are. For the $D^{vw}_i$ elements, the important contributions come from the single excitations, while the double excitation gives almost no contribution. In the attractive regime the most important elements come from the singles close to the Fermi level. We will refer to these excitations as Fermi states. In the repulsive regime the most important contribution is from 0101, which puts the lowest energy pair in the highest level at zero coupling. We will refer to these types of states as Rydberg states. For $D^{vw}_{ij}$ at weak positive coupling the lowest Fermi state is the most important, but contributions from all excited states become the same at large positive coupling. In the repulsive regime, the highest Rydberg single is the most important, followed by the double excitation. For $P^{vw}_{ij}$, the singles all pass through 1 at zero coupling, and are more important than the double at all couplings. 

Adding the energy denominator to $D^{vw}_i$, the state 1010 is the most important at weak couplings but becomes asymptotically the same as 1001 and 0110 in both attractive and repulsive limits. The Rydberg state 0101 becomes the most important in the repulsive regime. The double excitation gives almost no contribution. All transition probabilities are finite at all couplings. Including the energy denominator for $D^{vw}_{ij}$ makes 1010 the only important state in the repulsive regime: the numerator remains constant but the denominator goes to zero linearly. This is because the rapidities for the ground state and for the state 1010 asymptotically approach each other at large negative couplings. In the attractive regime everything remains finite as the rapidities for each state remain distinct (and in fact become large). The same is true for $P^{vw}_{ij}$. 

\begin{figure} 
	\begin{subfigure}{\textwidth}
		\includegraphics[width=0.325\textwidth]{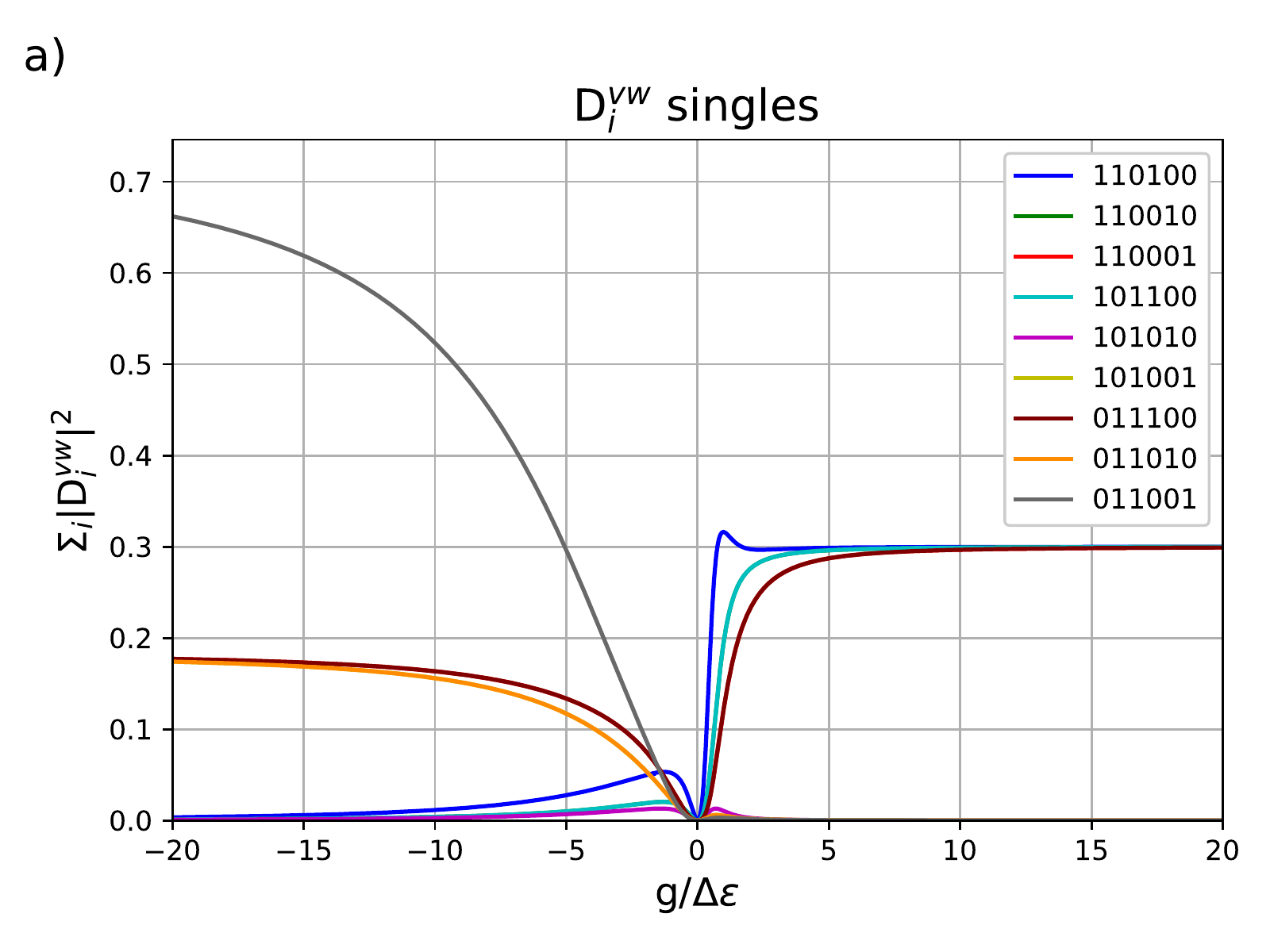}  \hfill
		\includegraphics[width=0.325\textwidth]{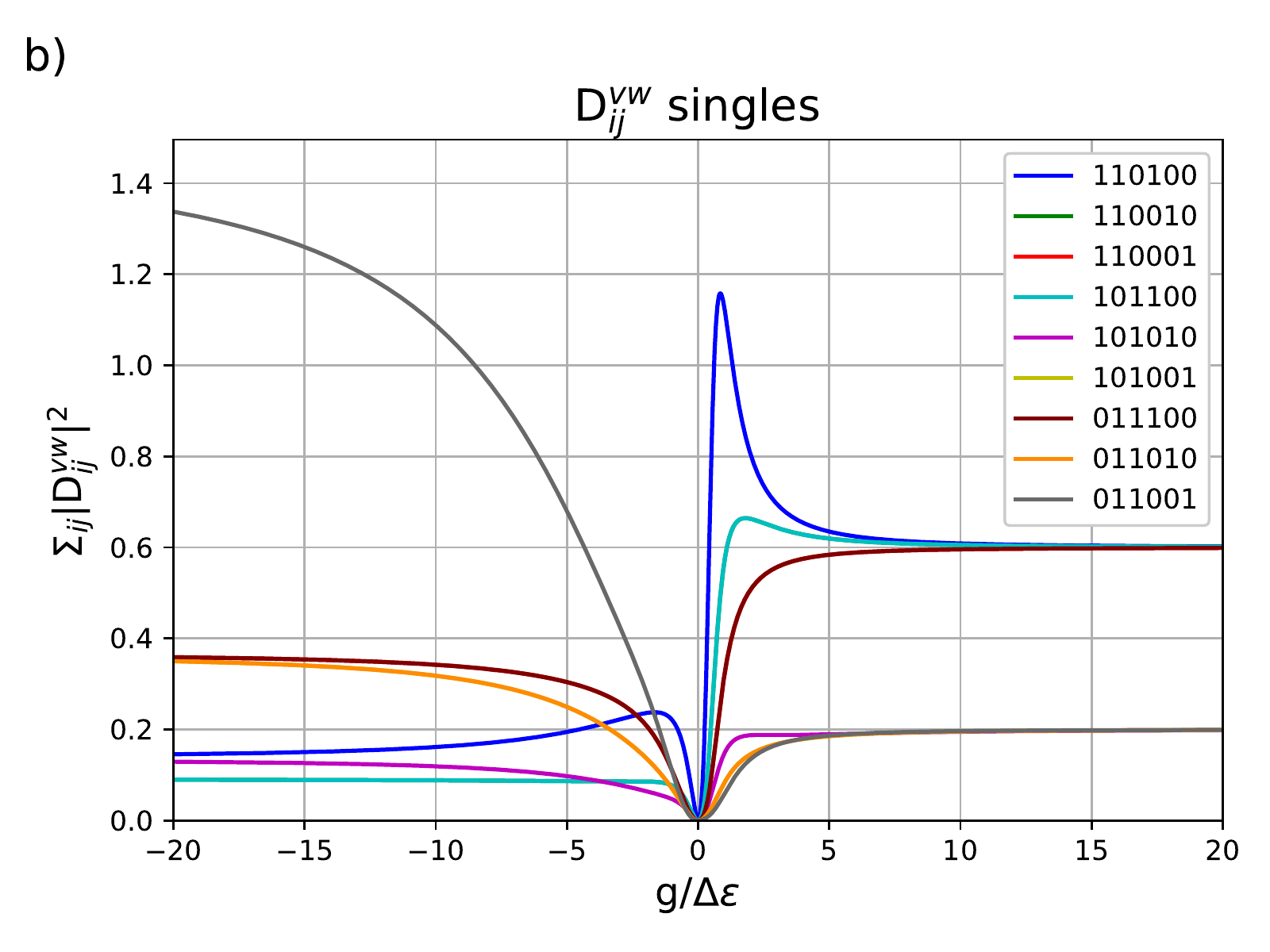} \hfill
		\includegraphics[width=0.325\textwidth]{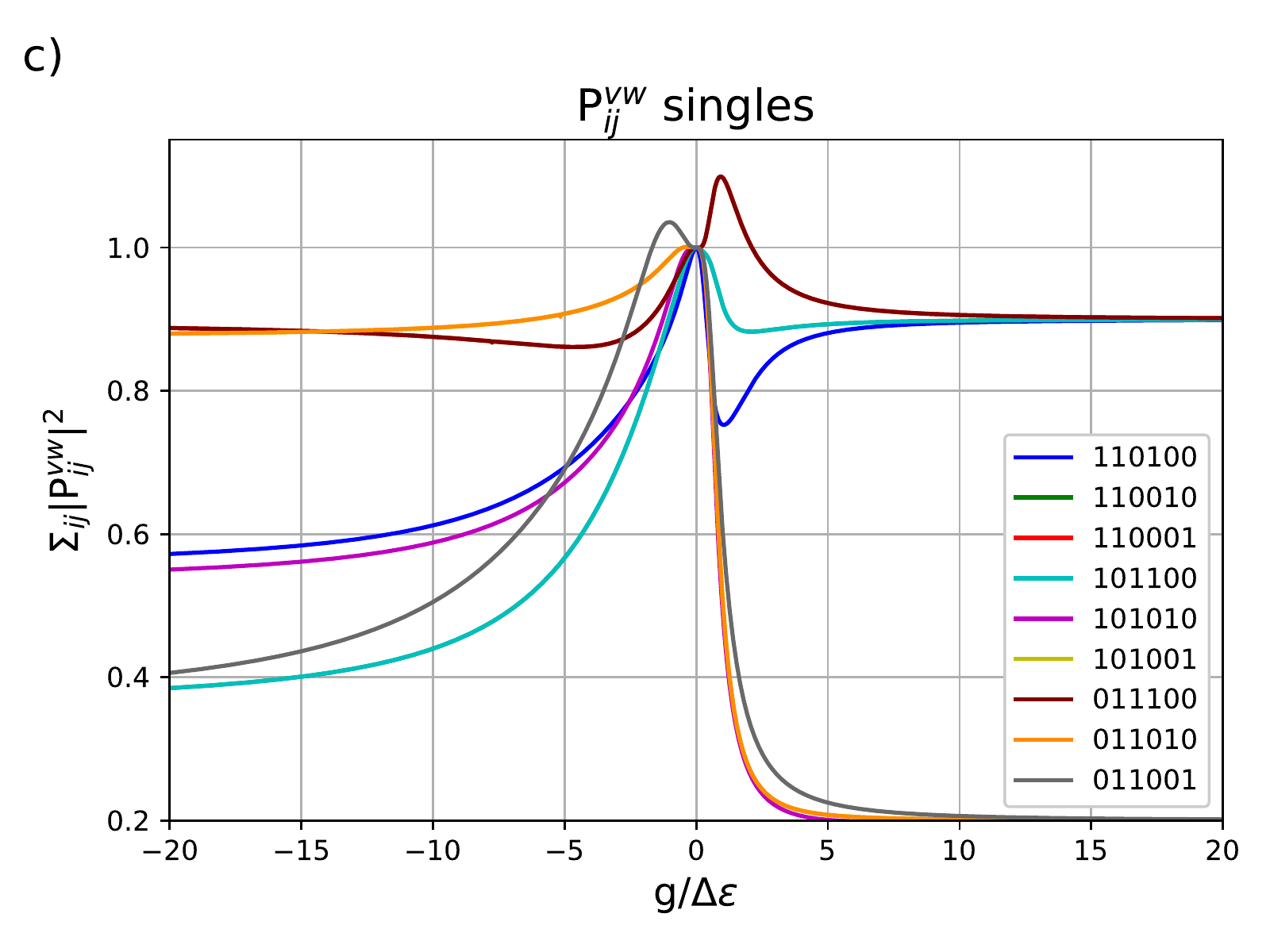}		
	\end{subfigure}
	\begin{subfigure}{\textwidth}
		\includegraphics[width=0.325\textwidth]{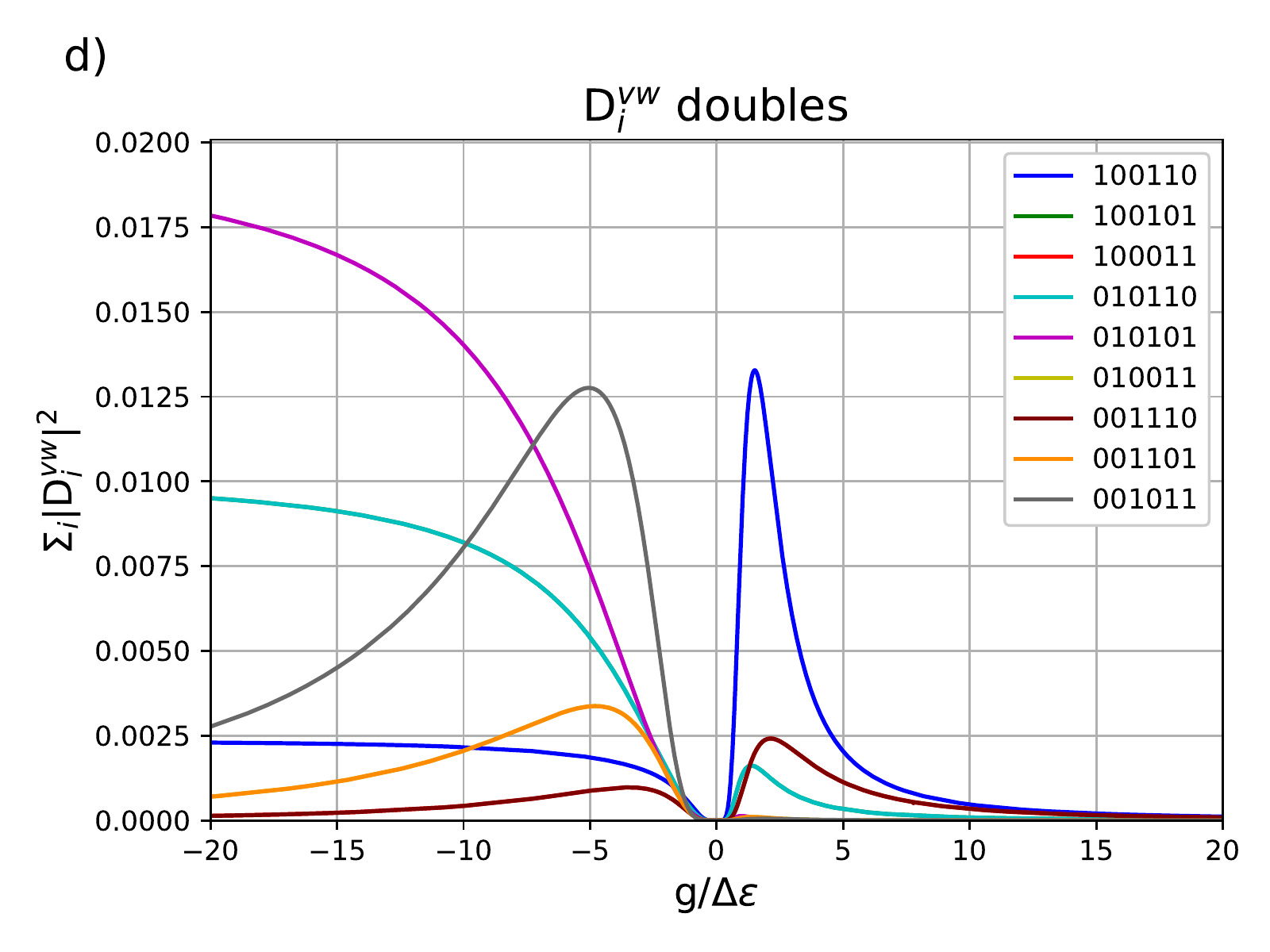}  \hfill
		\includegraphics[width=0.325\textwidth]{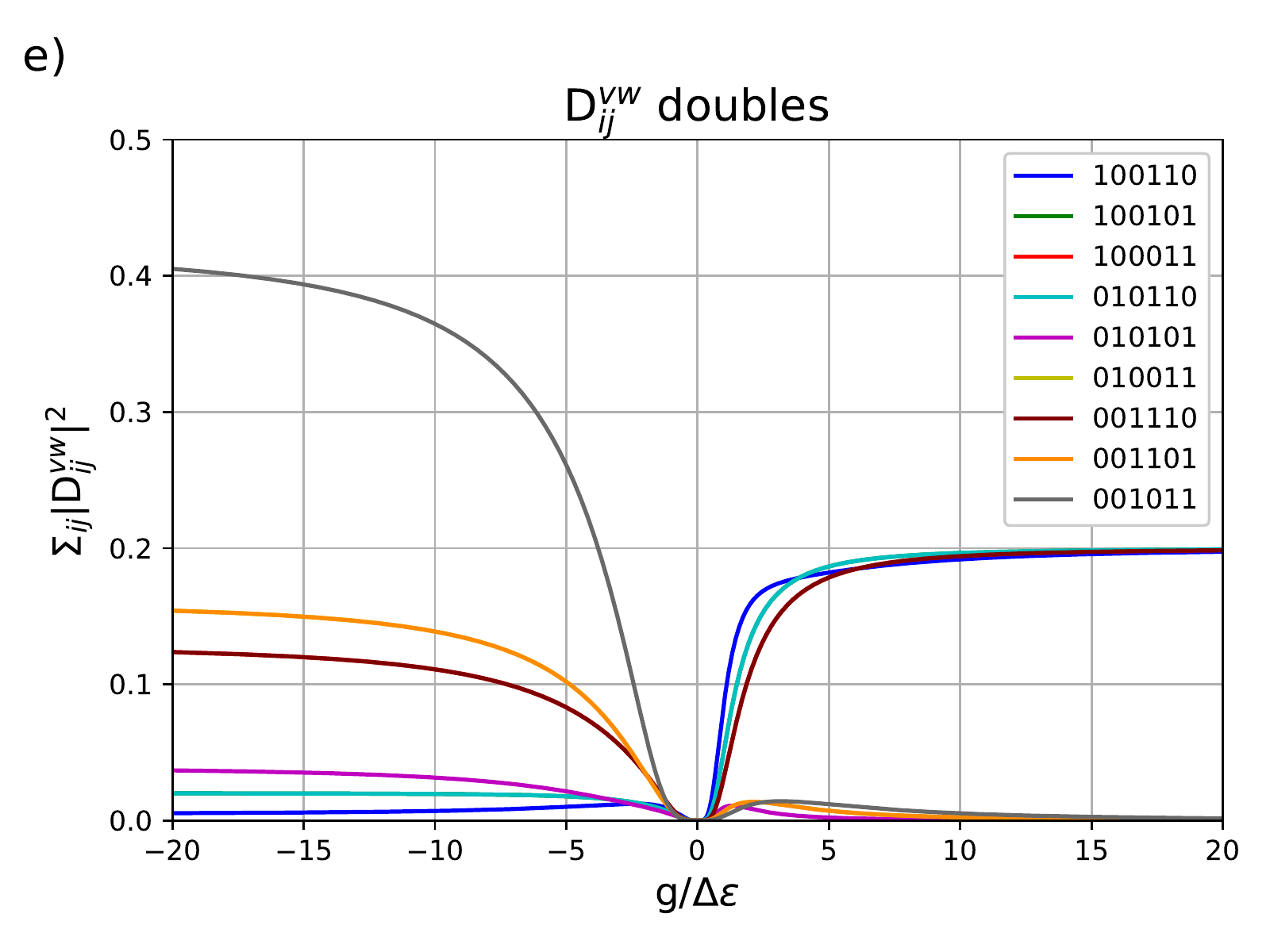} \hfill
		\includegraphics[width=0.325\textwidth]{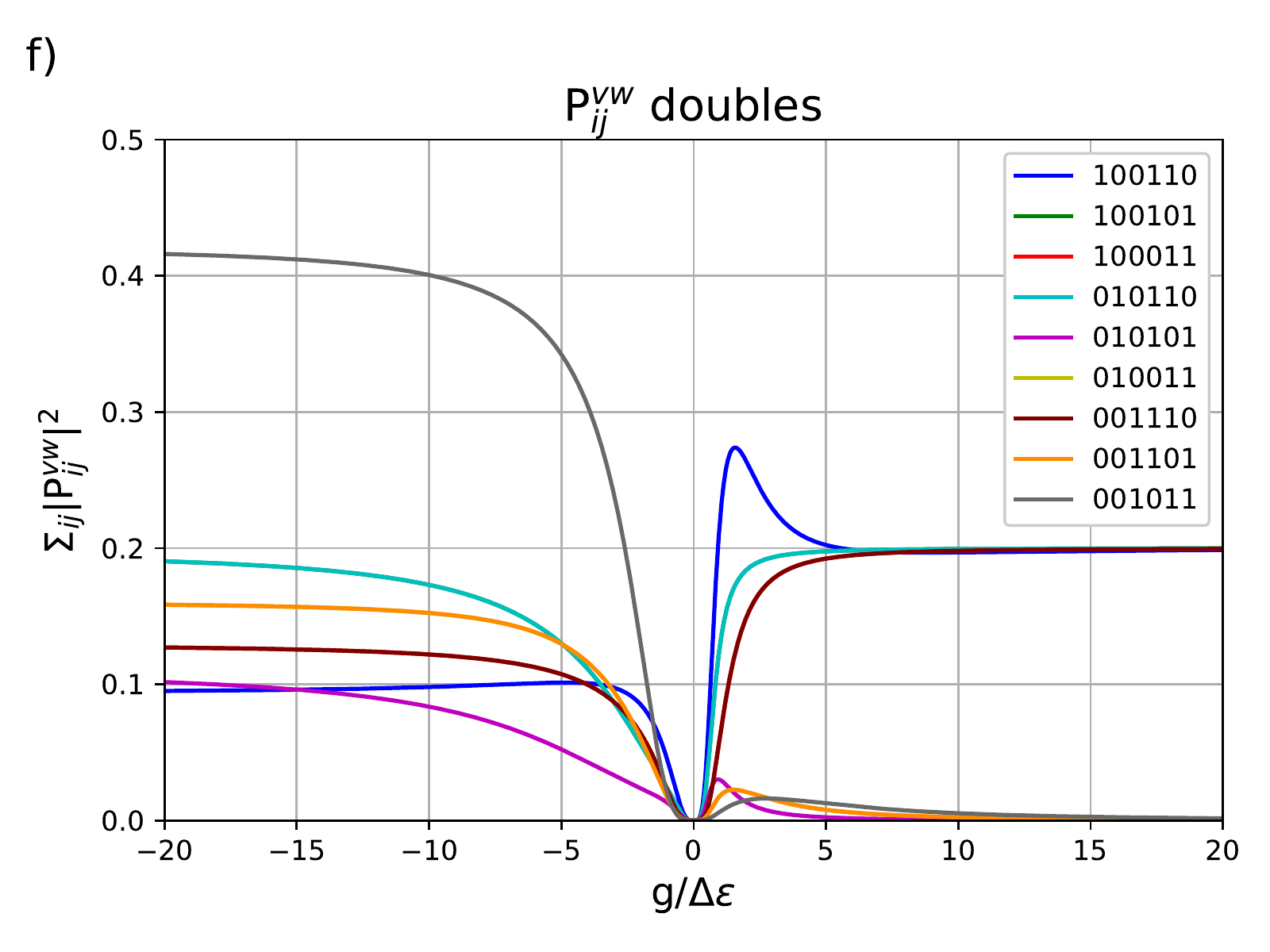}		
	\end{subfigure}
	\begin{subfigure}{\textwidth}
		\includegraphics[width=0.325\textwidth]{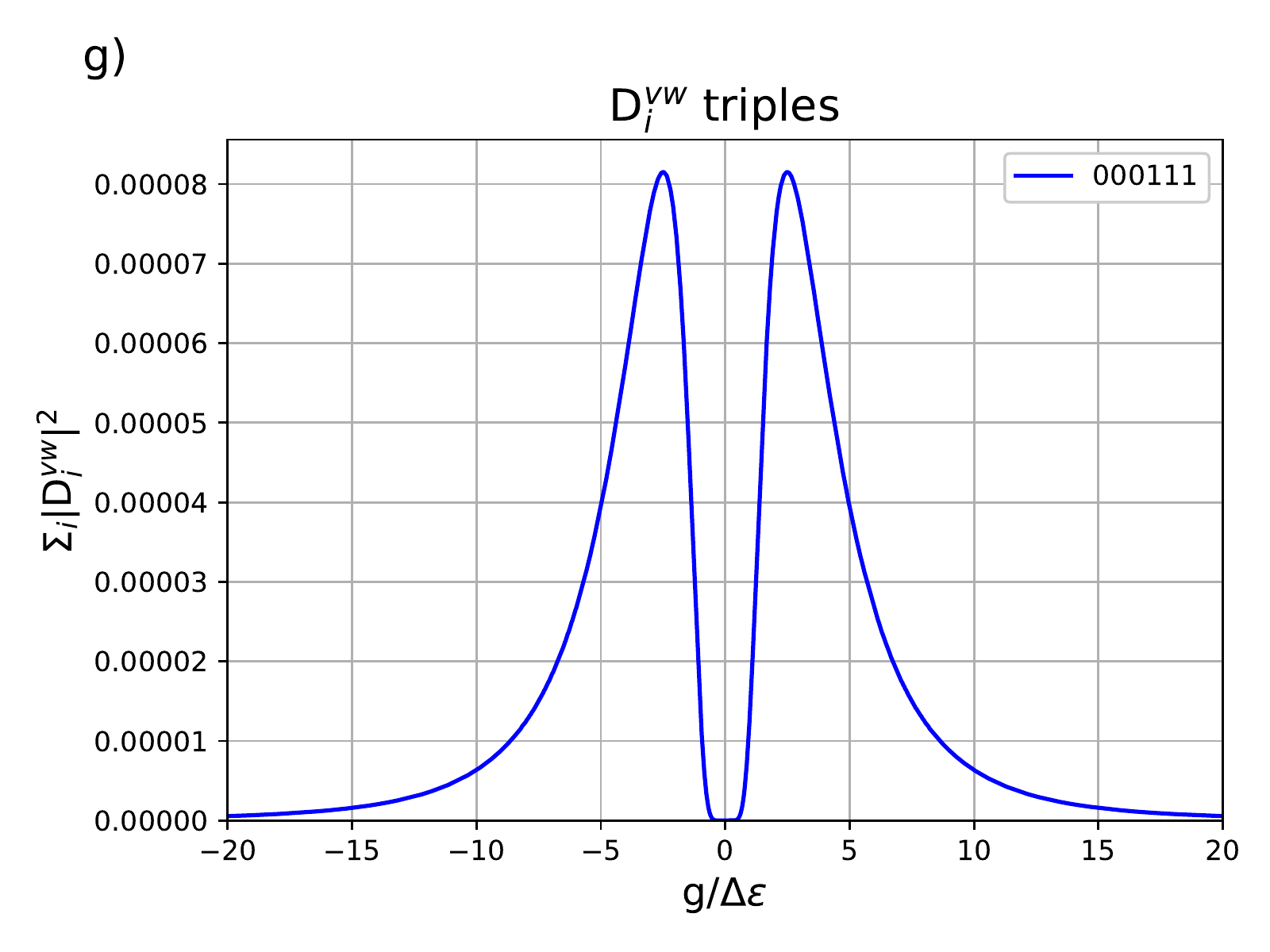}  \hfill
		\includegraphics[width=0.325\textwidth]{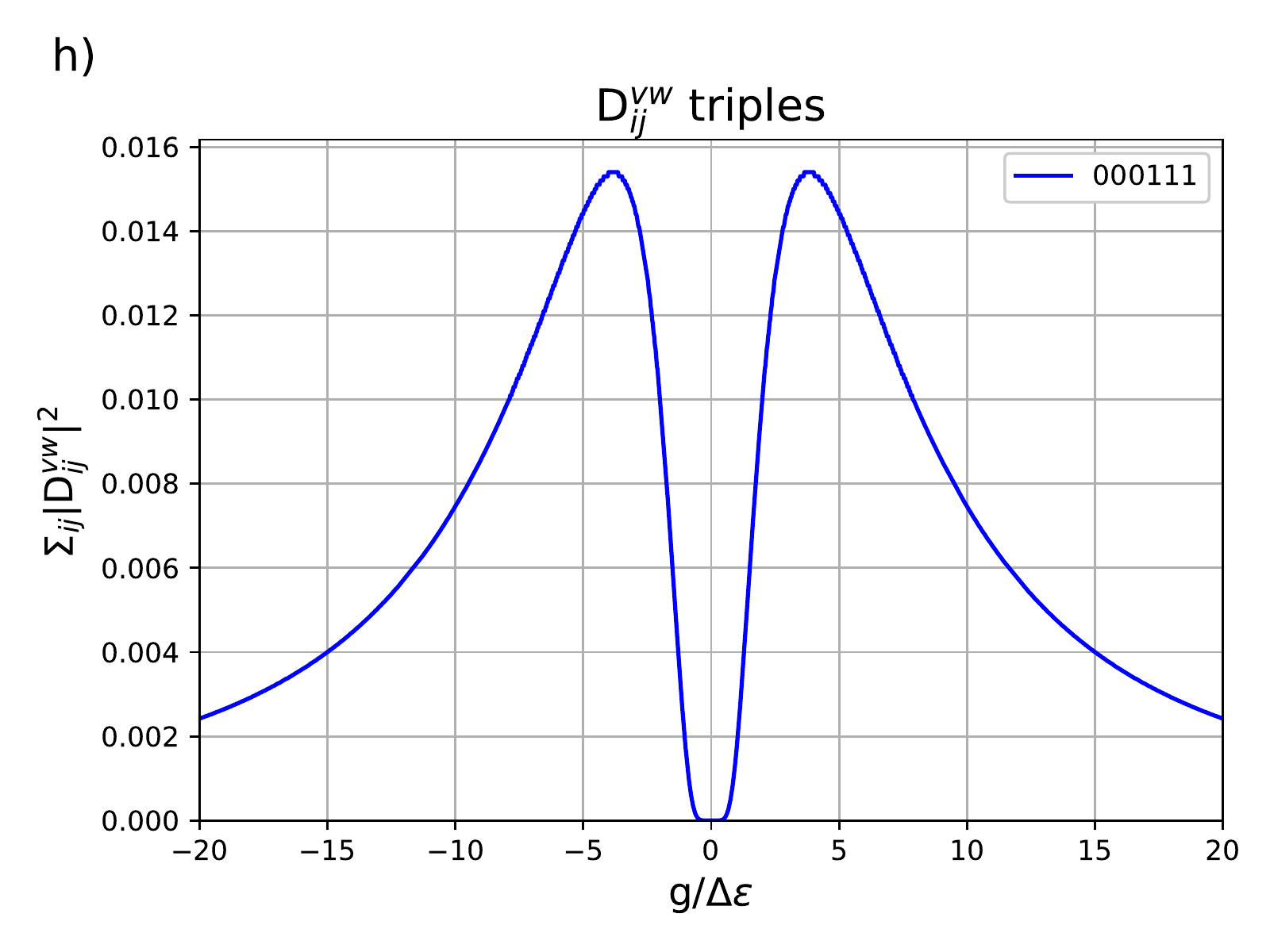} \hfill
		\includegraphics[width=0.325\textwidth]{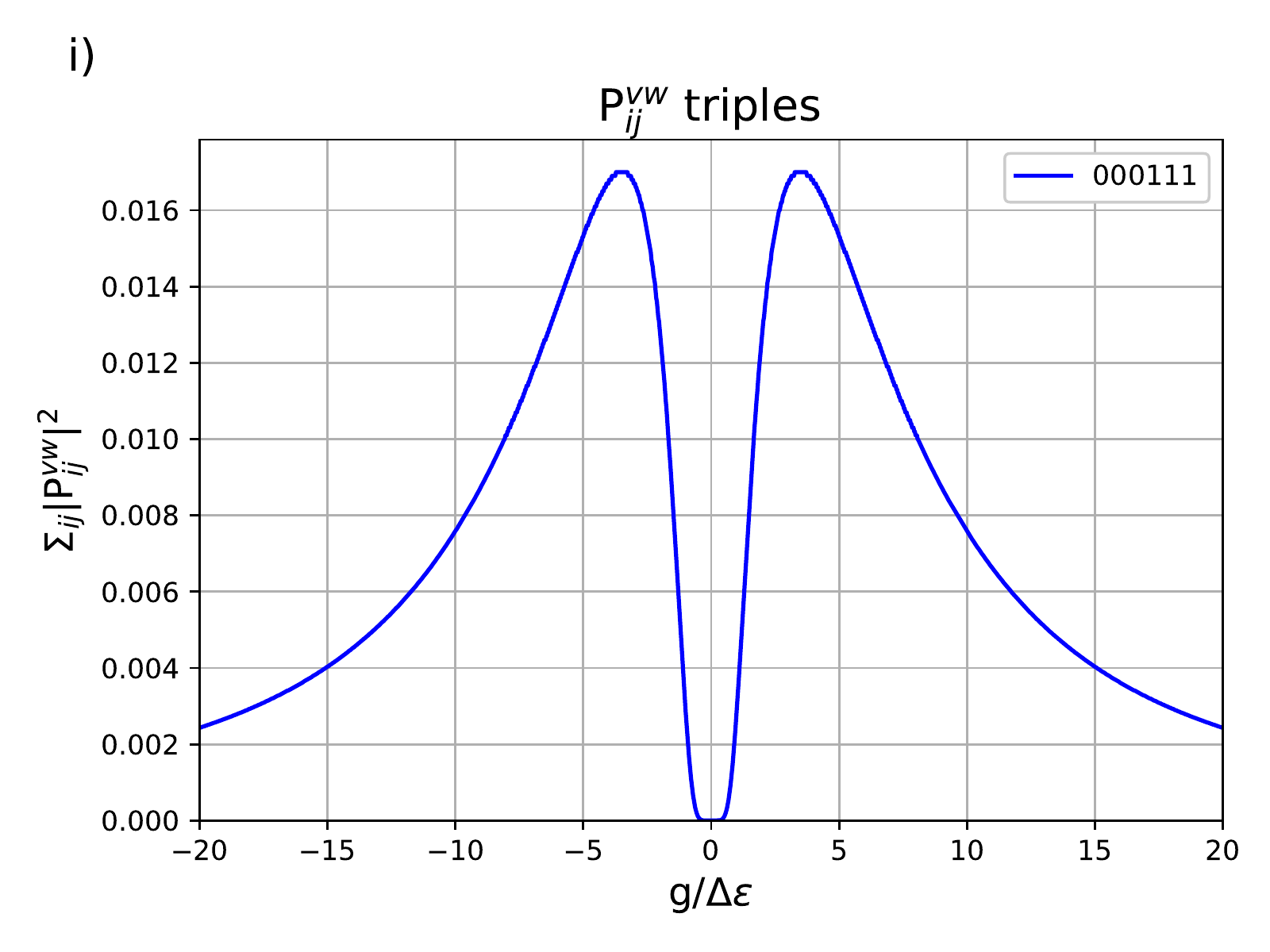}				
	\end{subfigure}
	\caption{Sums of transition probabilities of half-filled 6-site picket fence model for a-c) single, d-f) double and g-i) triple excitations.}
	\label{fig:6_3_curves}
\end{figure}

Transition probabilities for the half-filled 6-site model are shown in figure \ref{fig:6_3_curves}: the first row shows the single excitations, the second row the doubles, and the third row the lone triple excitation. As for the half-filled 4 site model, there are overlapping curves. In particular, each state is degenerate with the state obtained by reading the 1's and 0's backwards then exchanging 1's for 0's. For example, 110010 gives the same contribution as 101100, 110001 gives the same contribution as 011100 etc. Concretely, the green curves are hidden beneath the cyan curves, the red curves are hidden behing the maroon curves, and the yellow curves are hidden behind the orange curves. For $D^{vw}_i$, only the singles matter, with the Fermi states being the most important in the attractive limit and the Rydberg states the most important in the repulsive regime. The same states are the most important for $D^{vw}_{ij}$ but now the highest Rydberg double excitation is the second most important contribution in the repulsive limit. The Fermi doubles are important in the attractive limit. For $P^{vw}_{ij}$ all the singles are important, along with all the doubles in the repulsive limit and the Fermi doubles in the attractive limit. The triple excitation gives no important contributions.

\begin{figure} 
	\begin{subfigure}{\textwidth}
		\includegraphics[width=0.325\textwidth]{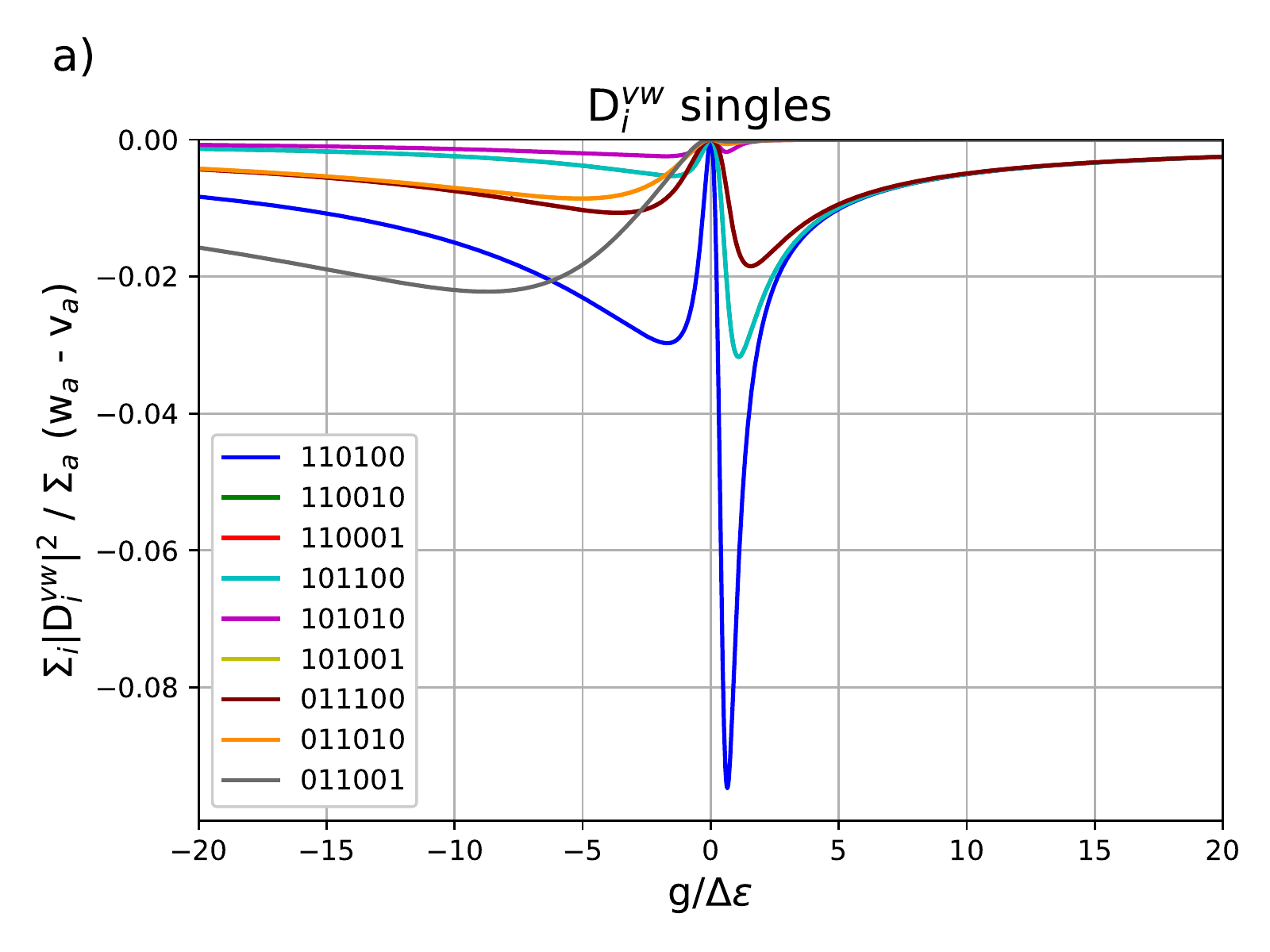}  \hfill
		\includegraphics[width=0.325\textwidth]{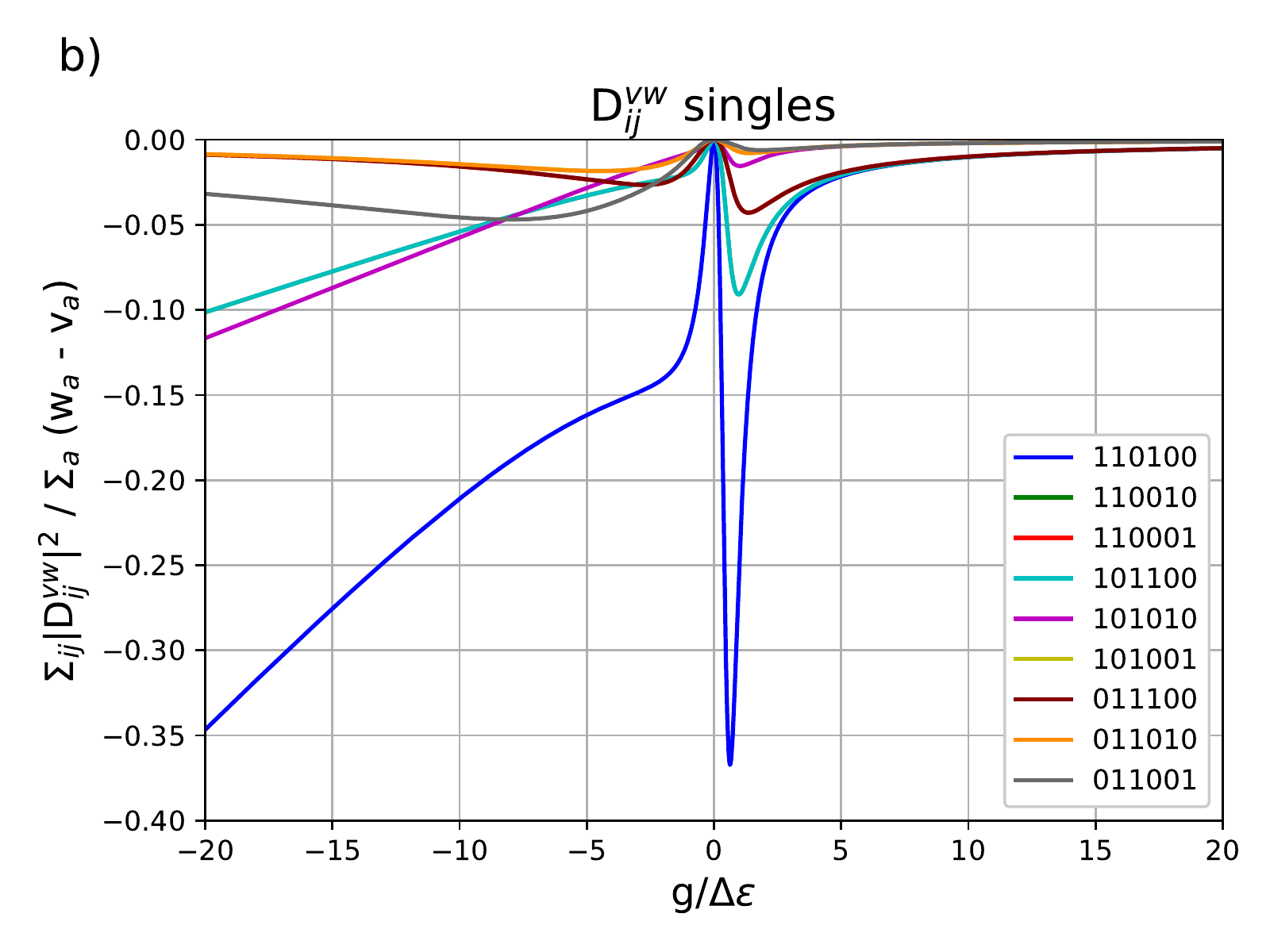} \hfill
		\includegraphics[width=0.325\textwidth]{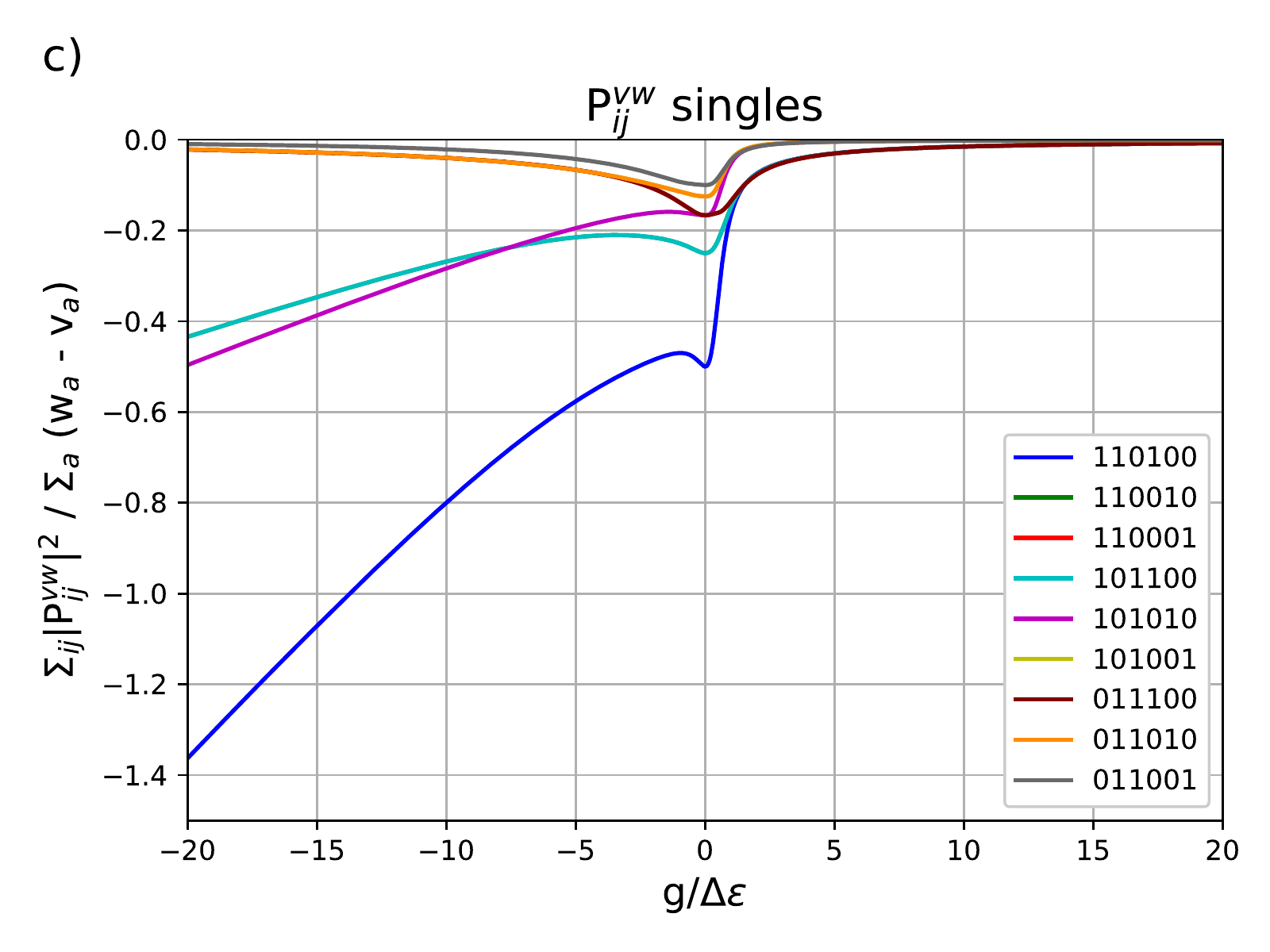}		
	\end{subfigure}
	\begin{subfigure}{\textwidth}
		\includegraphics[width=0.325\textwidth]{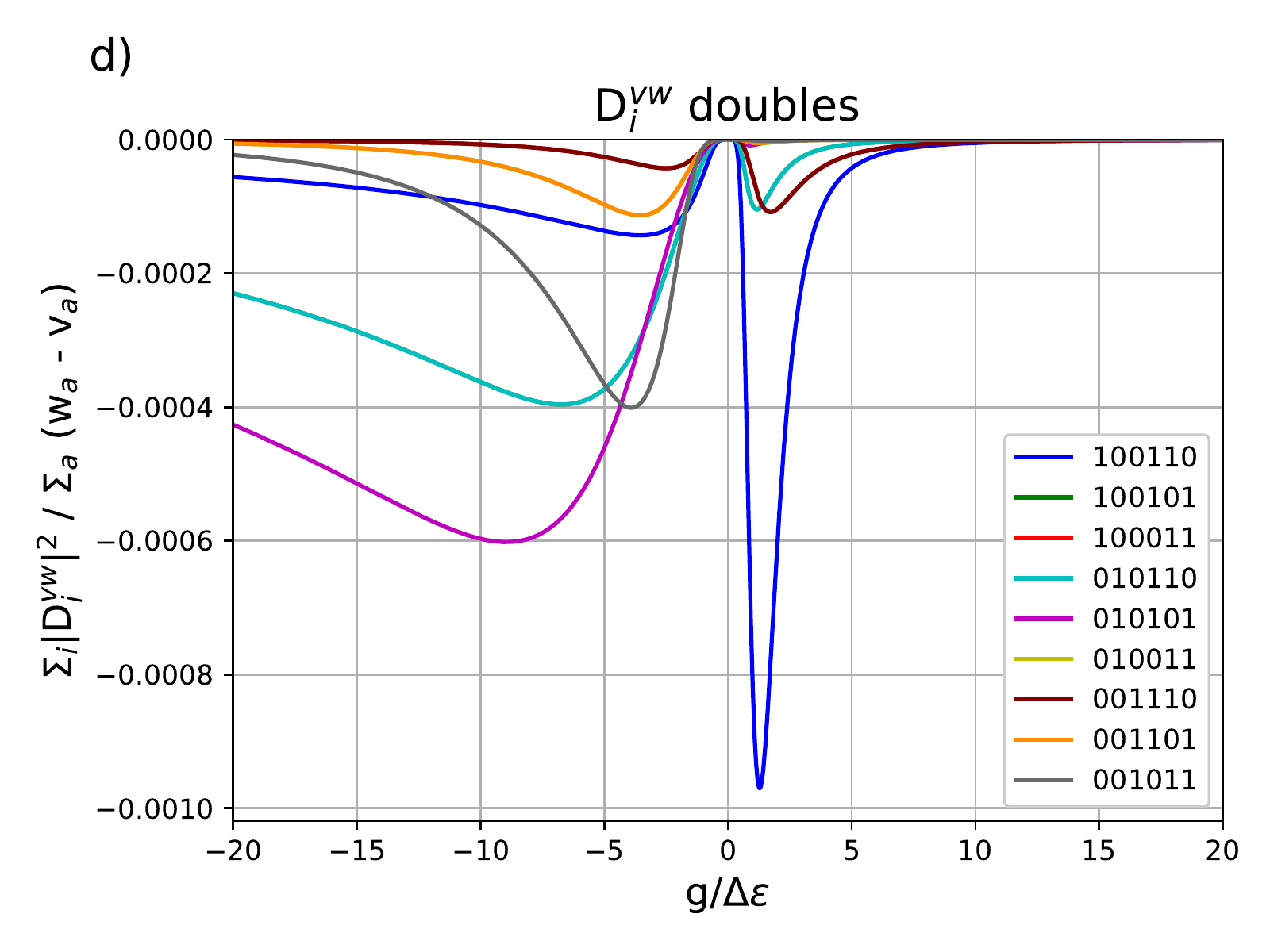}  \hfill
		\includegraphics[width=0.325\textwidth]{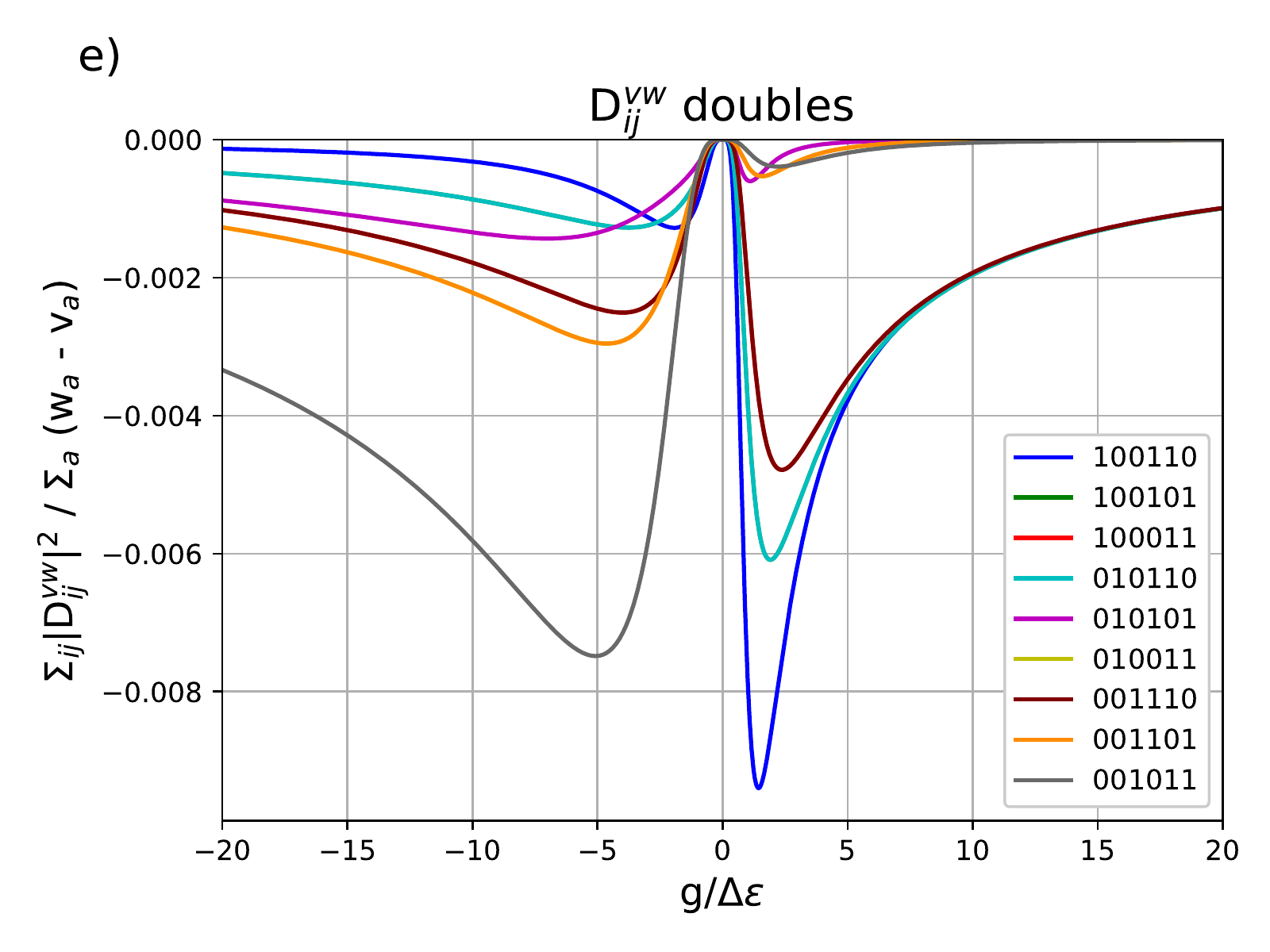} \hfill
		\includegraphics[width=0.325\textwidth]{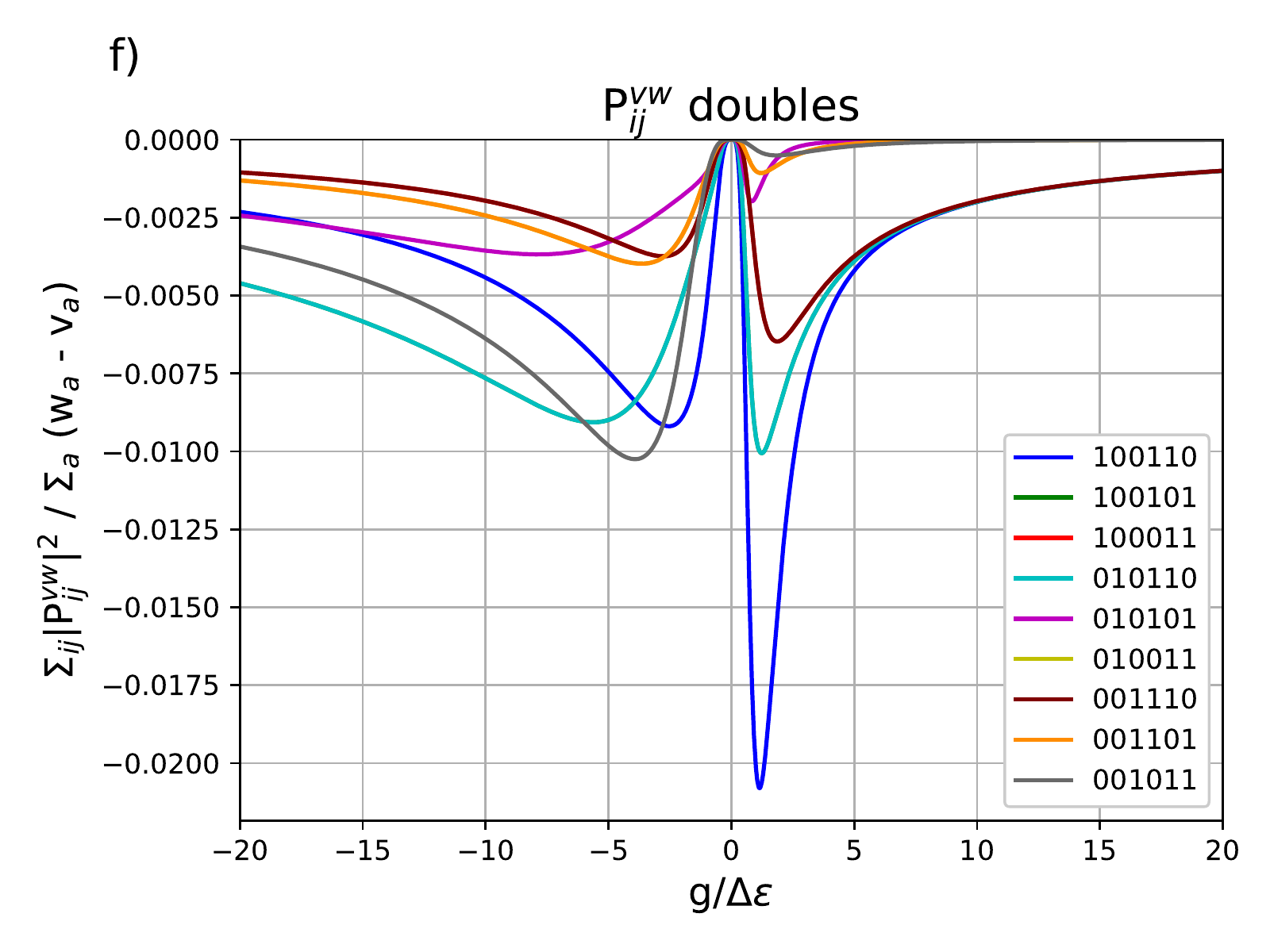}
	\end{subfigure}
	\begin{subfigure}{\textwidth}
		\includegraphics[width=0.325\textwidth]{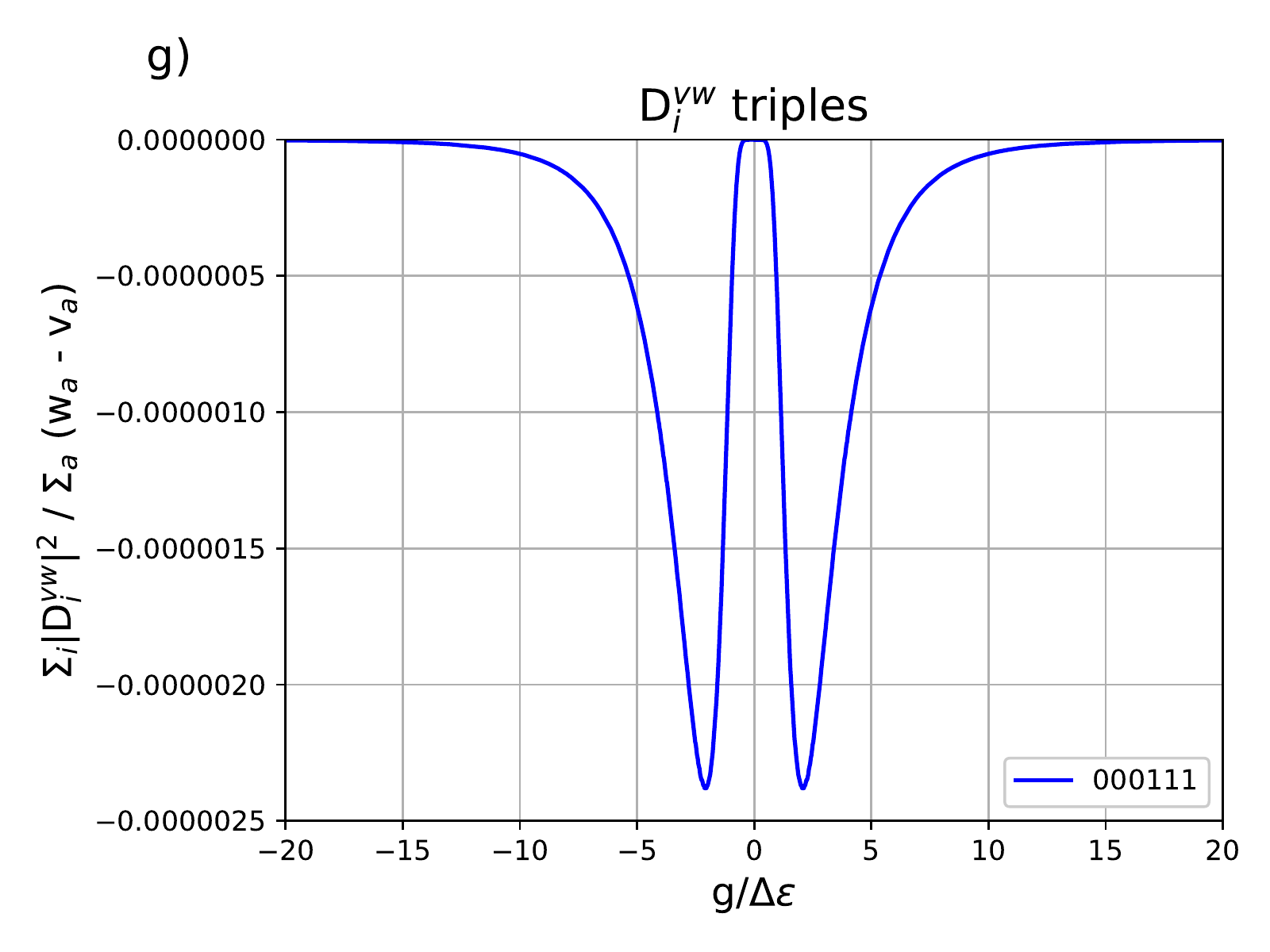}  \hfill
		\includegraphics[width=0.325\textwidth]{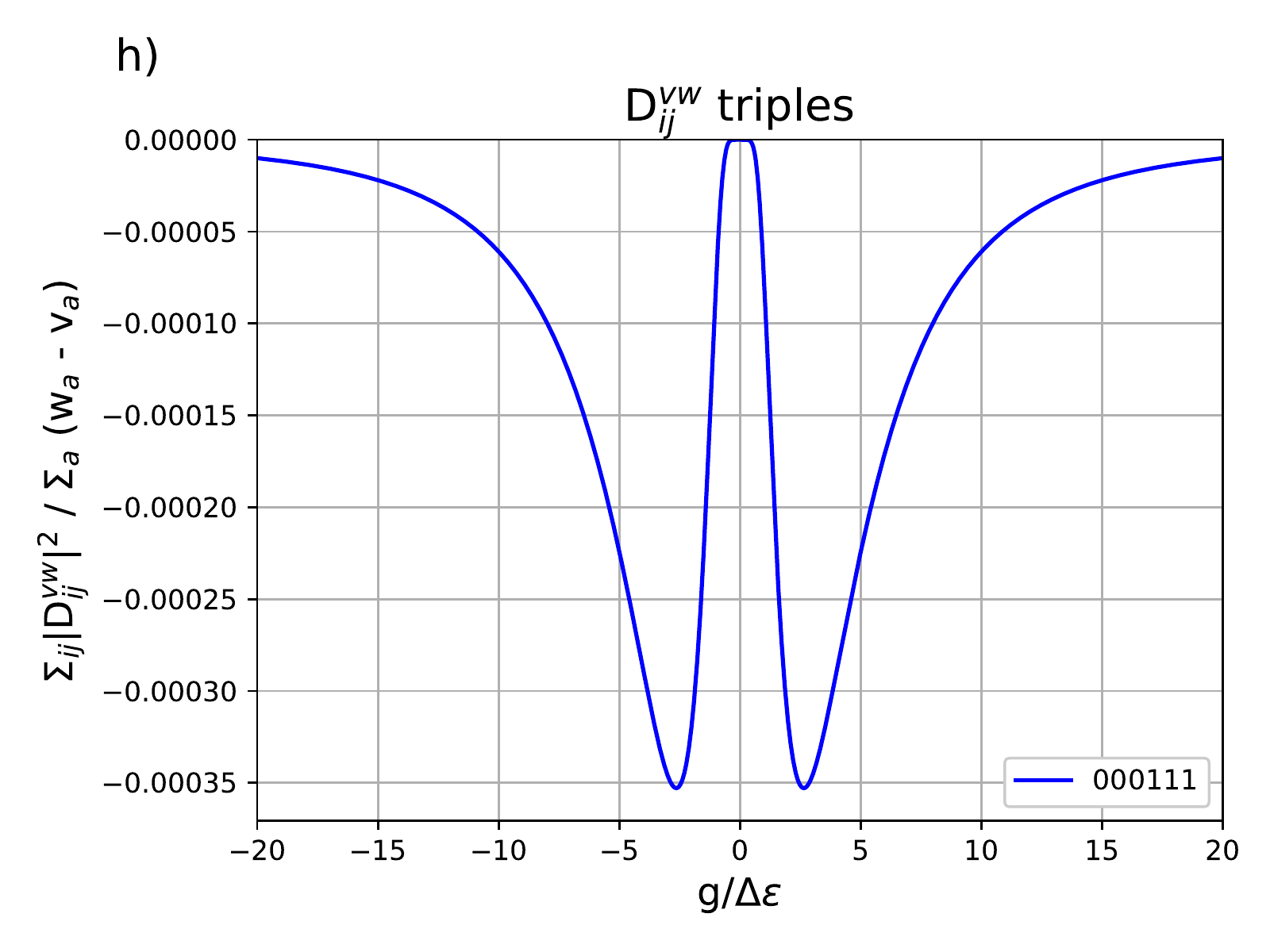} \hfill
		\includegraphics[width=0.325\textwidth]{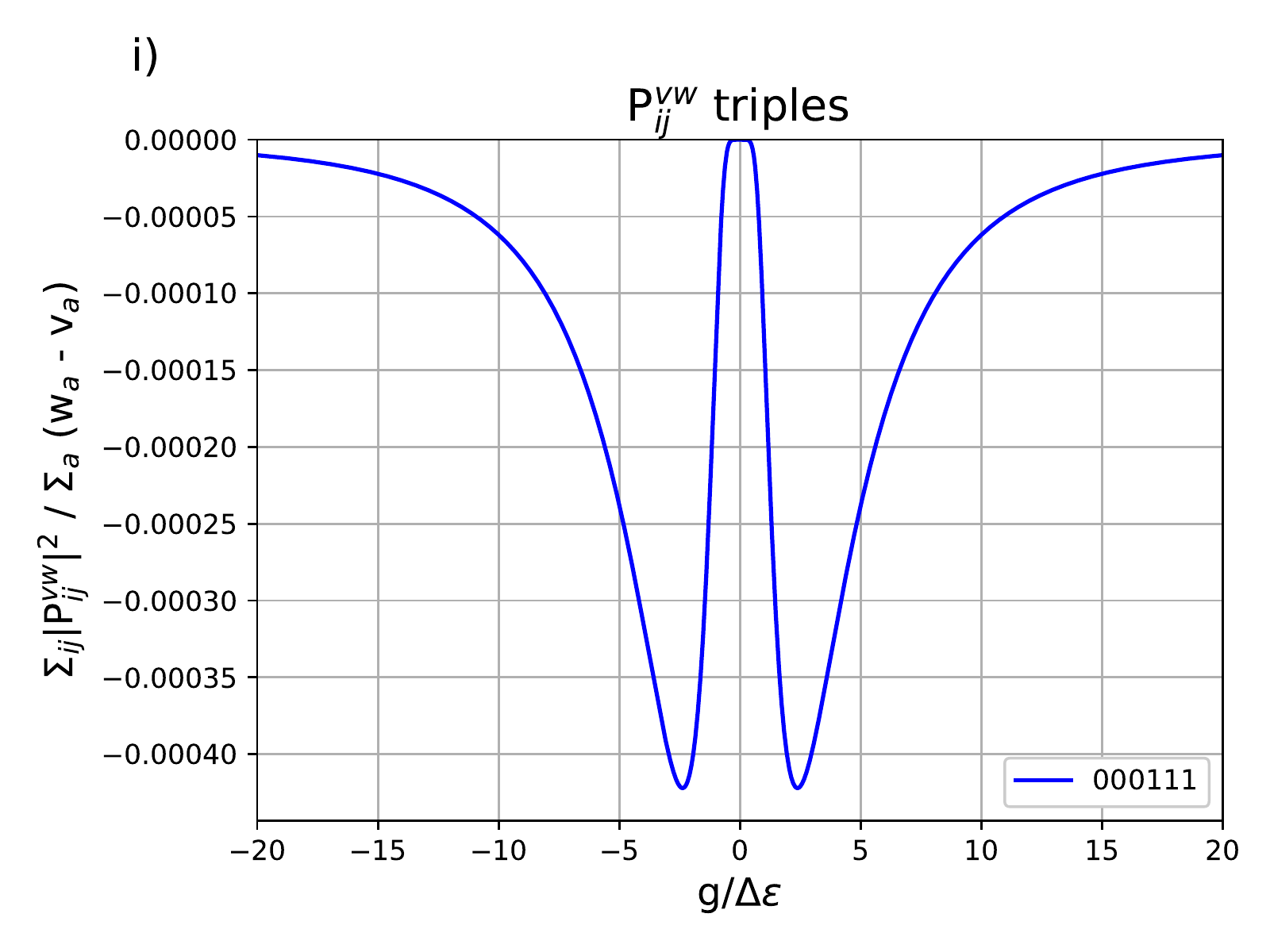}
	\end{subfigure}
	\caption{Sums of transition probabilities with energy denominators of half-filled 6-site picket fence model for a-c) single, d-f) double and g-i) triple excitations.}
	\label{fig:6_3_curves_ed}
\end{figure}

With energy denominators, shown in figure \ref{fig:6_3_curves_ed}, $D^{vw}_i$ is dominated by the singles and is finite at all couplings. $D^{vw}_{ij}$ and $P^{vw}_{ij}$ are dominated by the singles, with a few going linearly to minus infinity.

\begin{figure} 
	\begin{subfigure}{\textwidth}
		\includegraphics[width=0.325\textwidth]{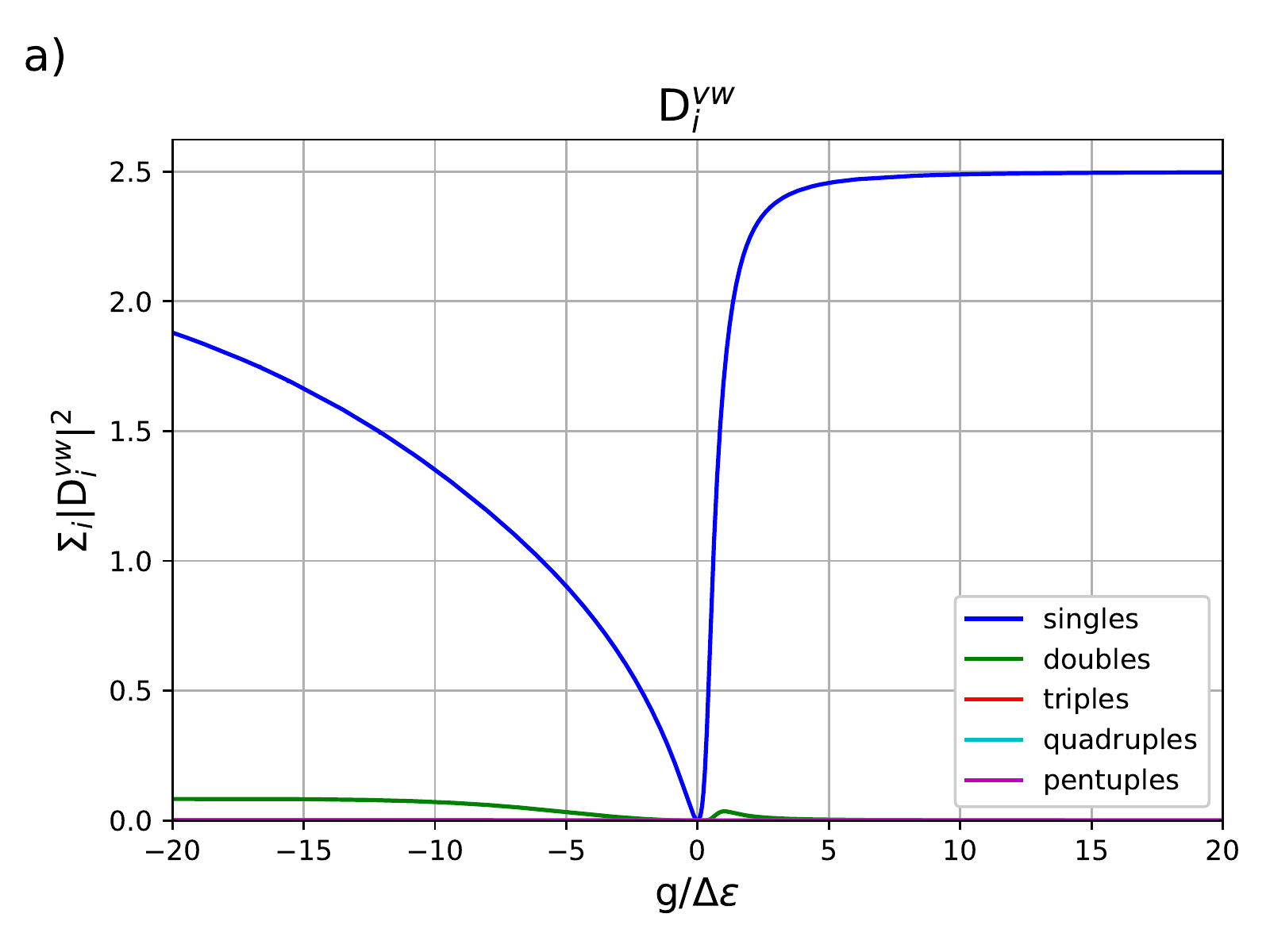}  \hfill
		\includegraphics[width=0.325\textwidth]{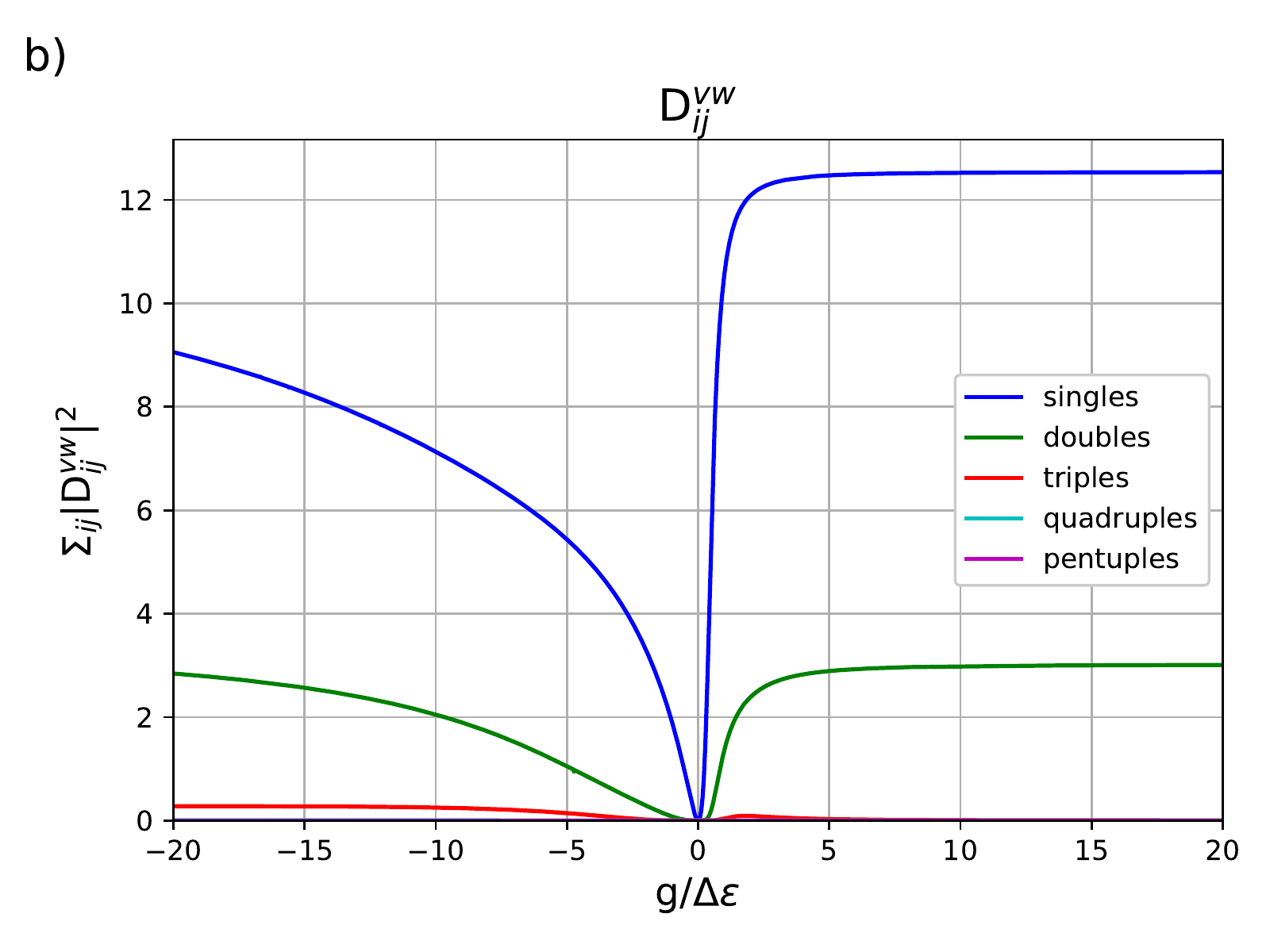} \hfill
		\includegraphics[width=0.325\textwidth]{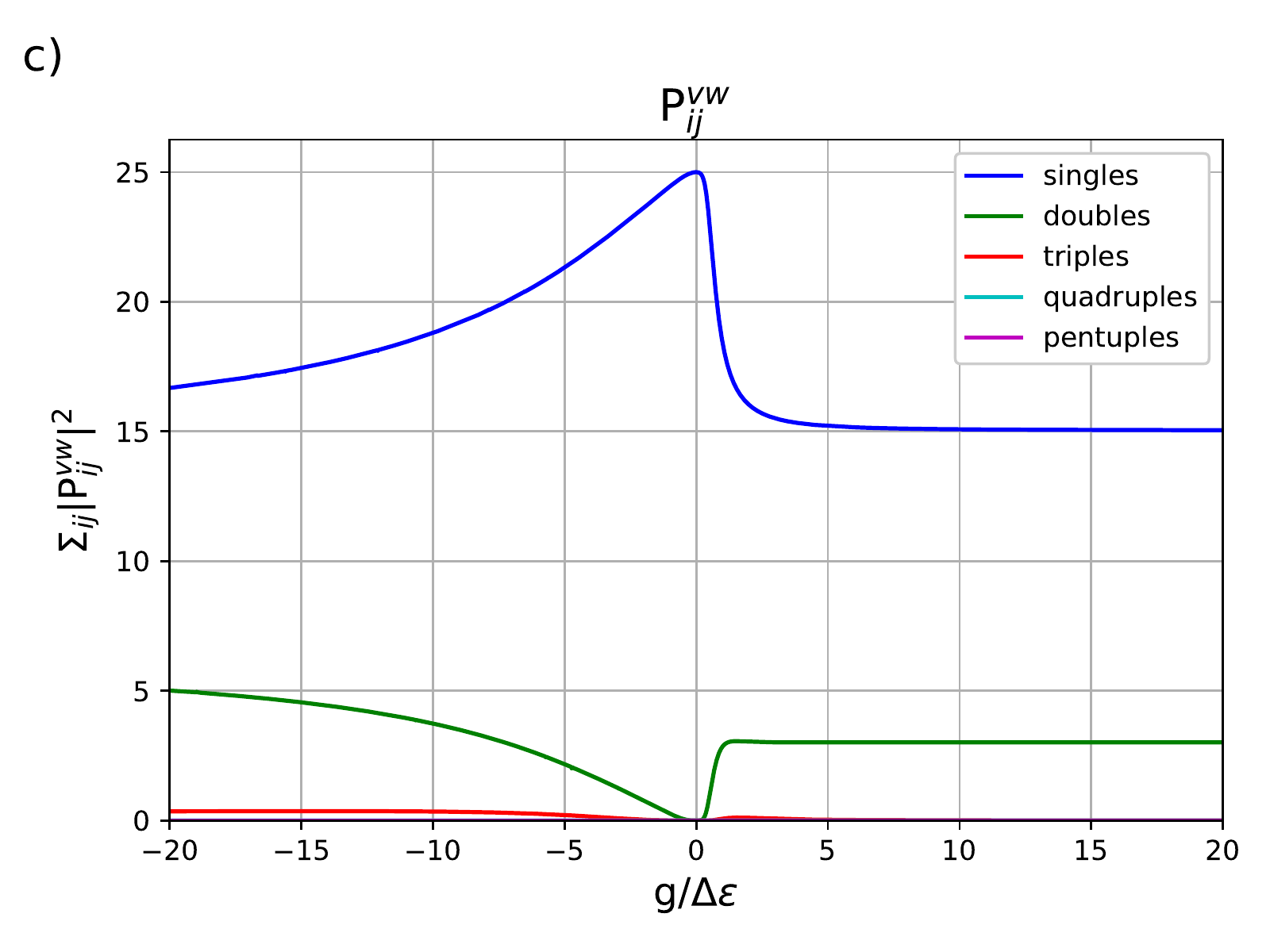}		
	\end{subfigure}
	\begin{subfigure}{\textwidth}
		\includegraphics[width=0.325\textwidth]{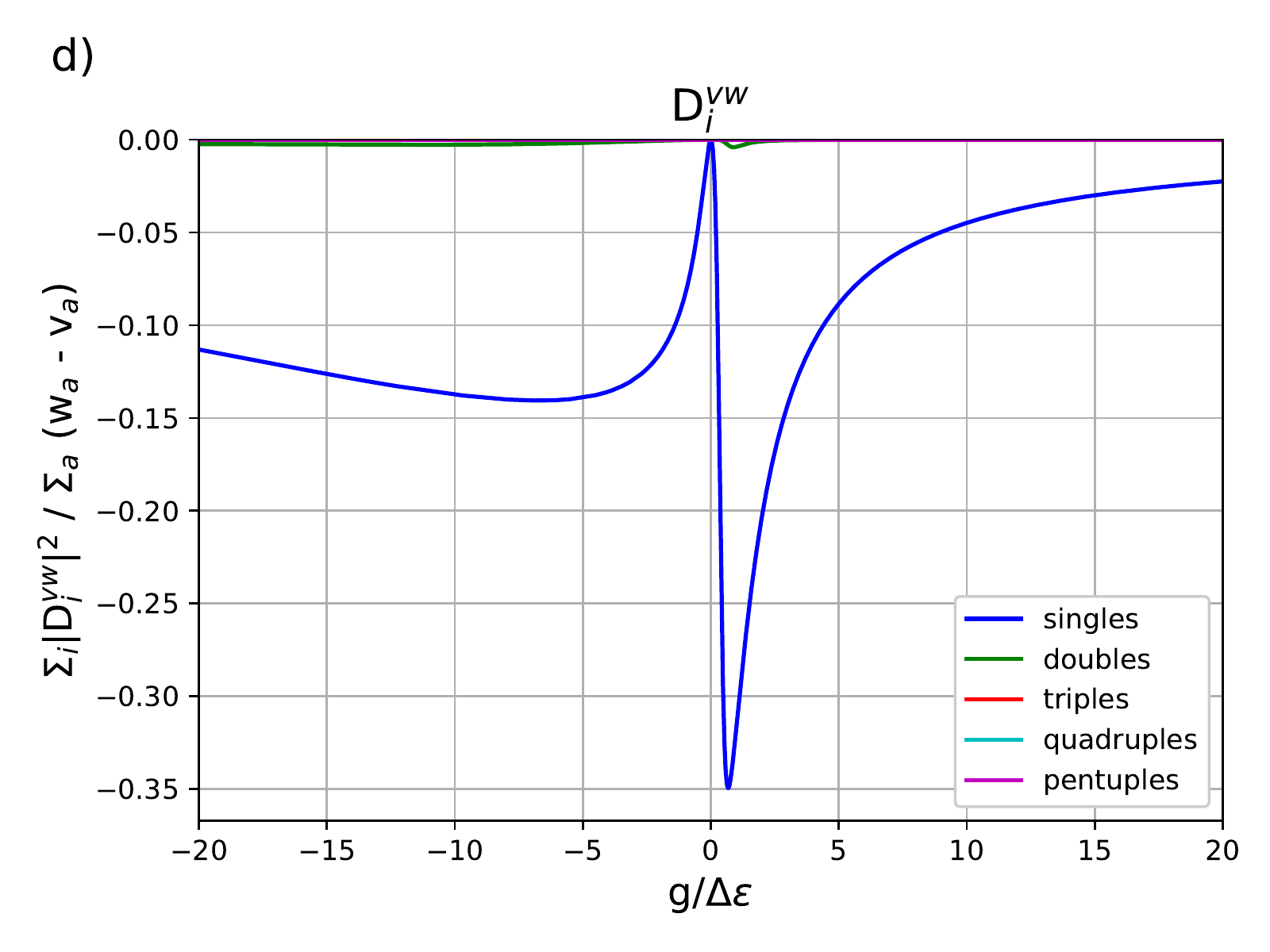}  \hfill	
		\includegraphics[width=0.325\textwidth]{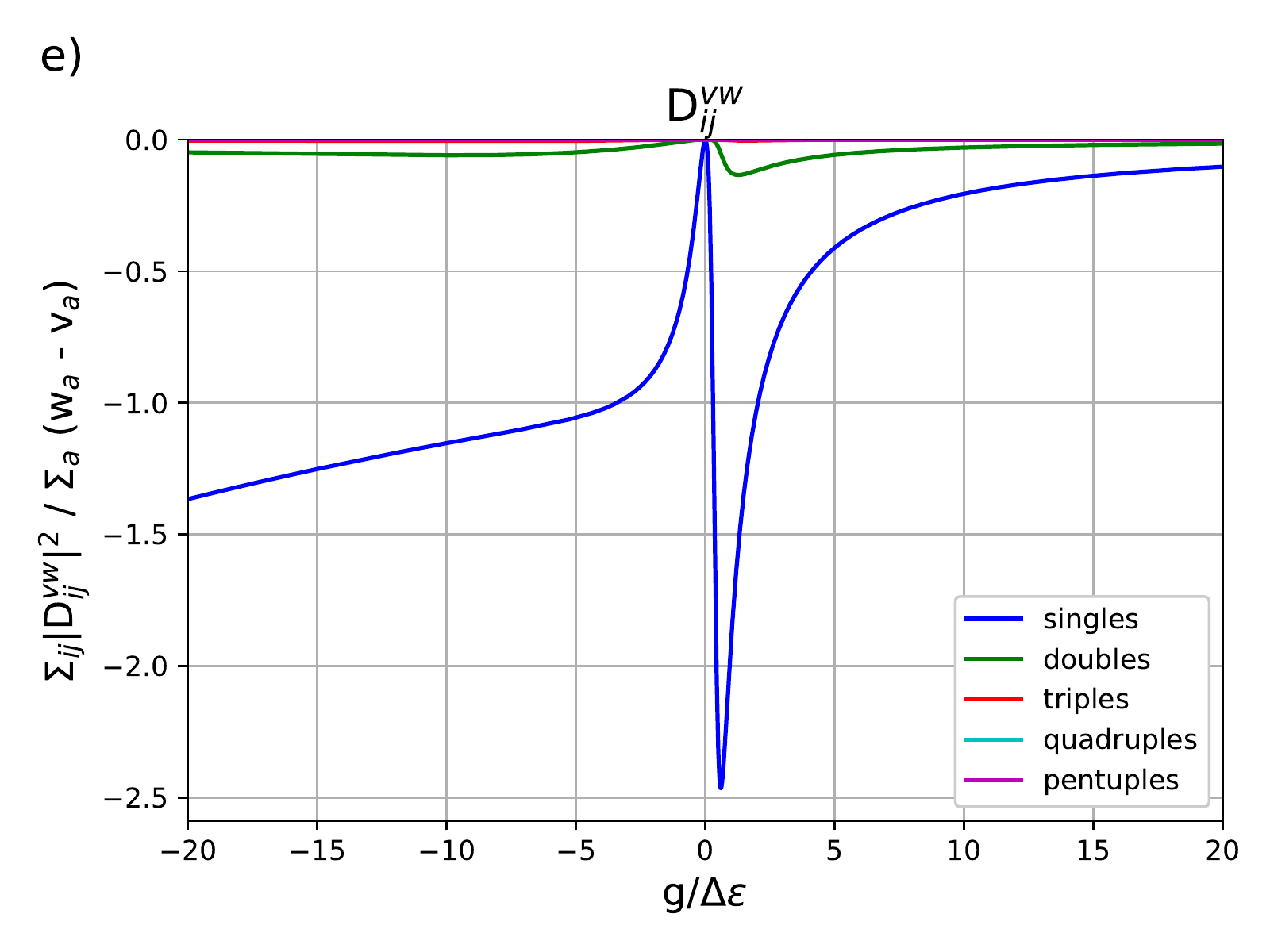} \hfill
		\includegraphics[width=0.325\textwidth]{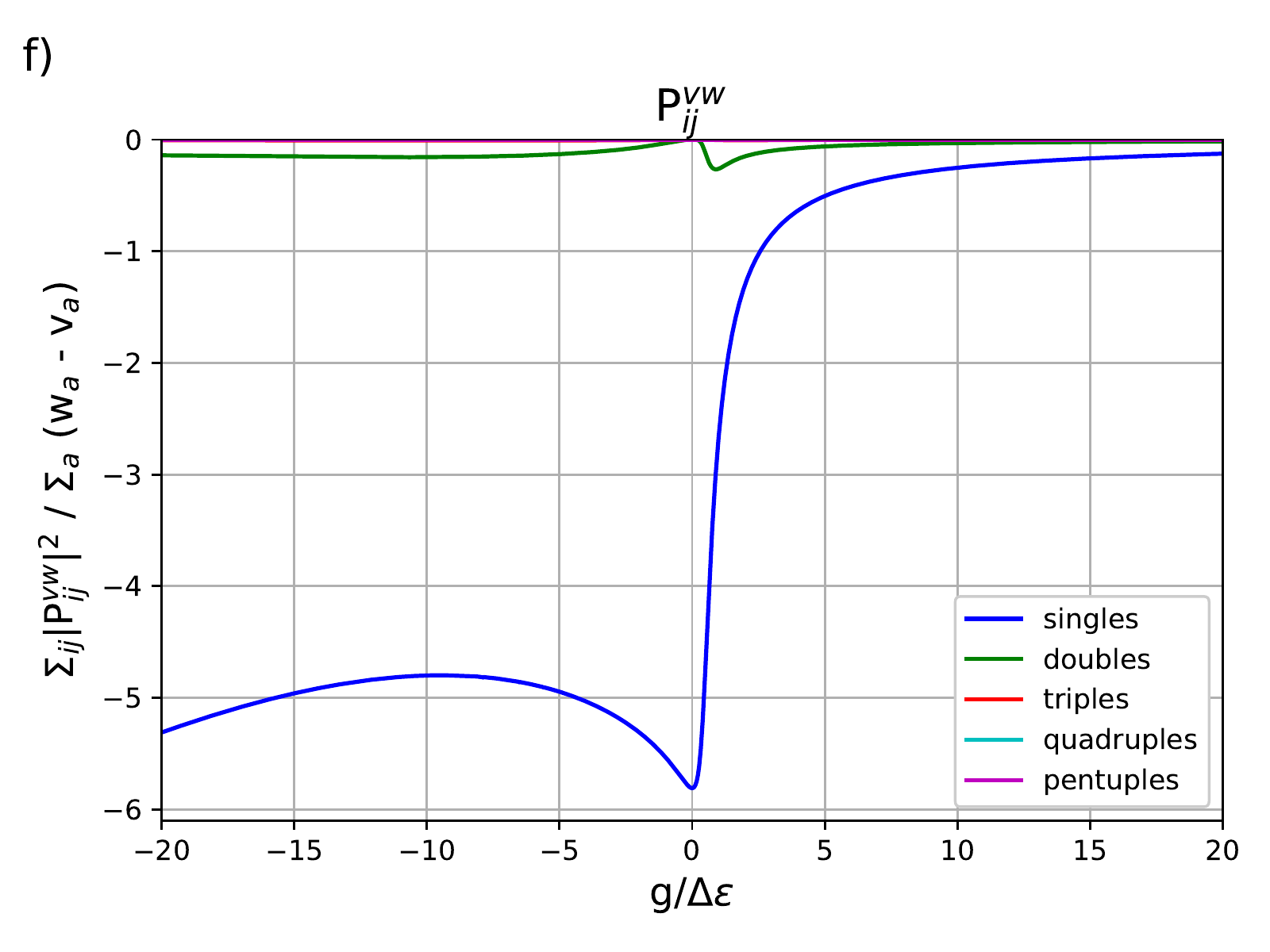}		
	\end{subfigure}
	\caption{a-c) Sums of transition probabilities for half-filled 10-site picket fence model, with energy denominators included for d-f).}	
	\label{fig:10_5_curves}
\end{figure}

Transition probabilities grouped by excitation order for the half-filled 10-site picket fence model are shown in figure \ref{fig:10_5_curves}. For $D^{vw}_i$ only the singles matter at all, and with energy denominators included everything remains finite at all couplings. For $D^{vw}_{ij}$ and $P^{vw}_{ij}$ the singles are the most important, the doubles are less important, while triples and above give almost no contribution. With energy denominators, the singles diverge linearly at strong negative couplings.

Before passing to the GB, we summarize the principal results for the PB. $D^{vw}_i$ is entirely dominated by the singles, and with an energy denominator remains finite at all couplings. $D^{vw}_{ij}$ and $P^{vw}_{ij}$ are dominated by the singles, though the doubles are relevant. With energy denominators, the singles for both $D^{vw}_{ij}$ and $P^{vw}_{ij}$ go to minus infinity linearly in the repulsive limit as the states become near degenerate while the numerators remain constant. Both are well behaved in the attractive limit.

\subsection{Gaudin Basis}
Directly plotting TDM elements or transition probabilities in the GB leads to divergences at critical points. At these points, here characterized by a specific $g$, the nature of the solutions of Richardson's equations changes from two distinct real numbers to a complex conjugate pair. \emph{At} a critical point, two rapidities are equal to one of the single-particle energies $\varepsilon$ so the TDM elements themselves diverge. However, in the transformation to the GB, the elements of the inverse of the Cauchy matrix manage these divergences so that the result is always finite. Thus, we may absorb some of the transformation using the factor
\begin{align}
q(w_a) = \frac{\prod_i (w_a - \varepsilon_i)}{\prod_{b \neq a} (w_a -w_b)}
\end{align}
to regularize the TDM elements:
\begin{align}
\bar{Z}^{vw} (w_a) &= q(w_a) Z(w_a) \\
\bar{Z}^{vw} (w_a, w_b) &= q(w_a) q(w_b) Z(w_a, w_b) \\
\bar{P}^{vw} (w_a, w_b) &= q(w_a) q(w_b) P(w_a,w_b).
\end{align}
Specifically, we plot normalized $\sum_a | \bar{Z}^{vw} (w_a) |^2$, $\sum_{ab} | \bar{Z}^{vw} (w_a,w_b) |^2$ and $\sum_{ab} | \bar{P}^{vw} (w_a,w_b) |^2$.

\begin{figure}
	\begin{subfigure}{\textwidth}
		\includegraphics[width=0.325\textwidth]{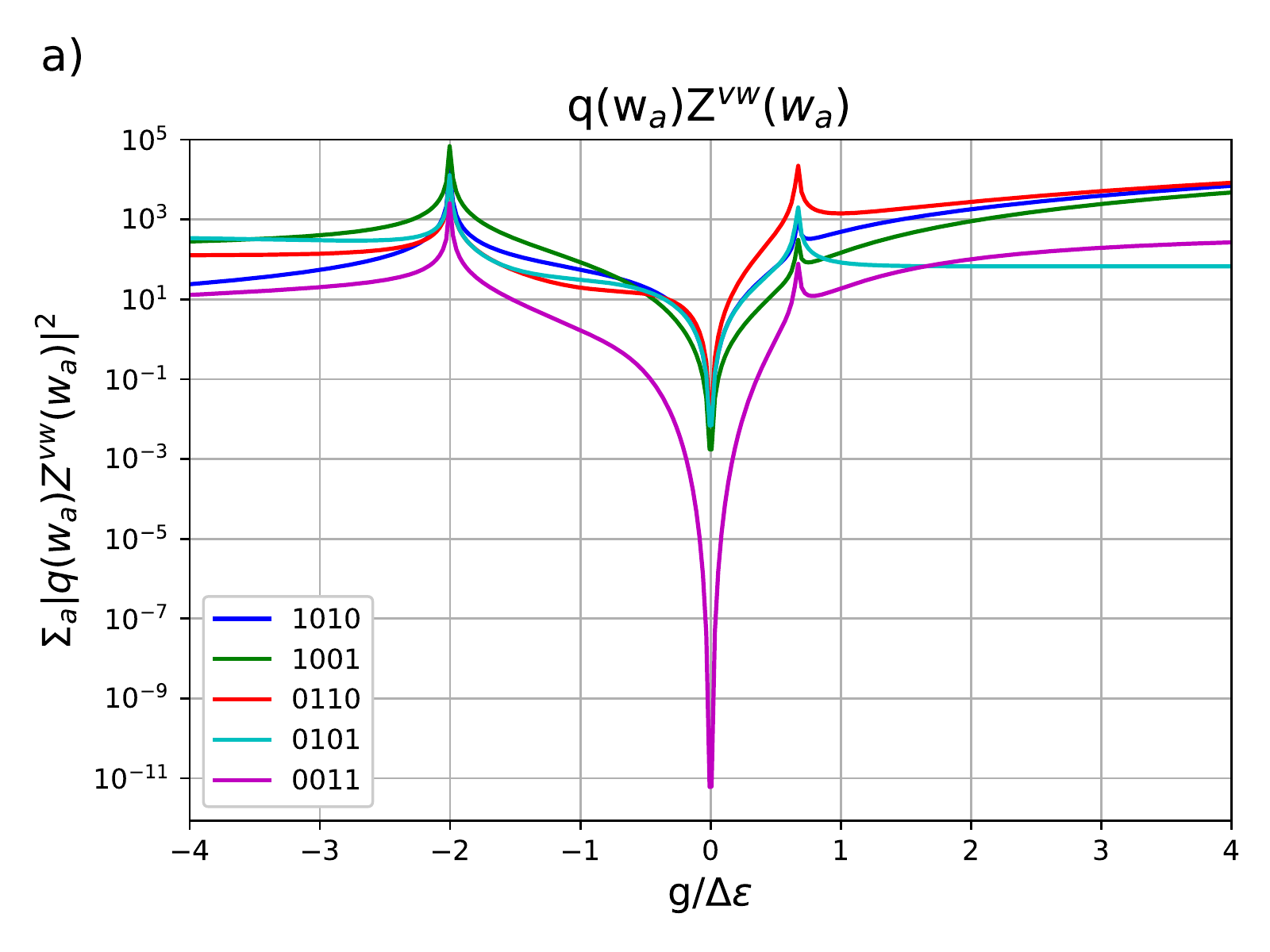}  \hfill
		\includegraphics[width=0.325\textwidth]{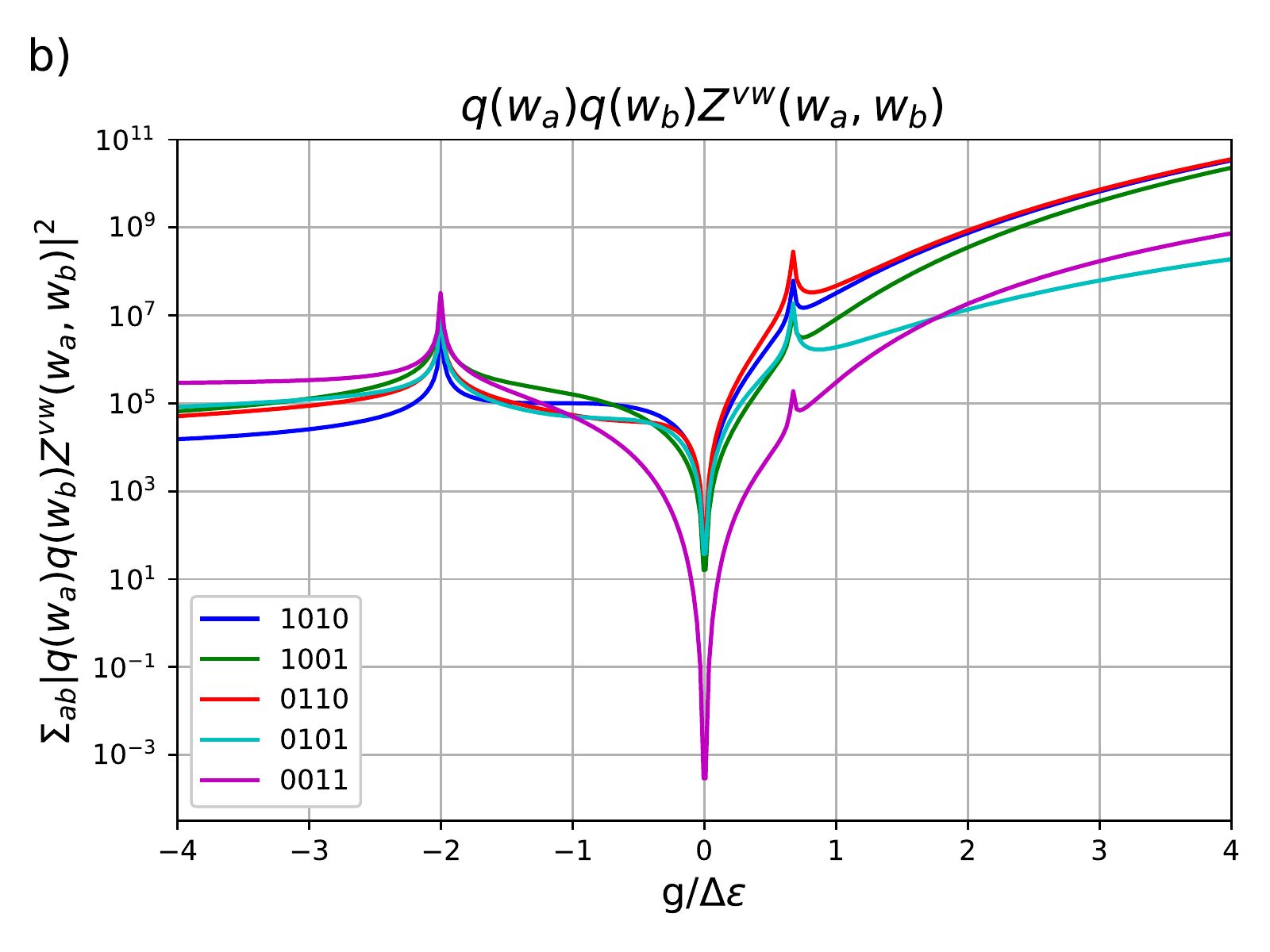} \hfill
		\includegraphics[width=0.325\textwidth]{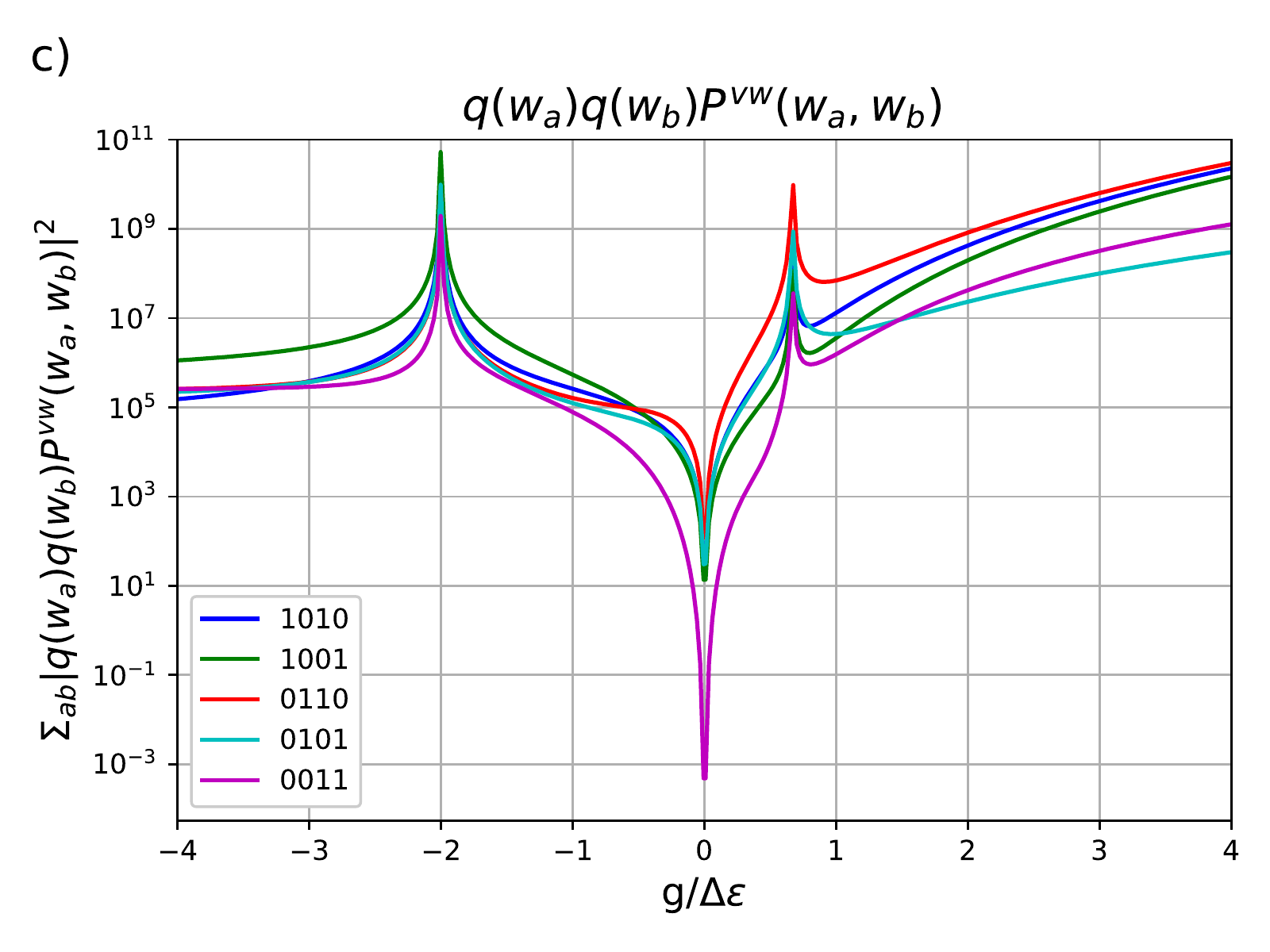}
	\end{subfigure}		
	\caption{Sums of regularized Gaudin basis transition probabilities for half-filled 4-site picket fence model.}
	\label{fig:4_2_gb}
\end{figure}

Regularized transition probabilities for the half-filled 4-site picket fence model are presented in figure \ref{fig:4_2_gb}. At large positive $g$, the attractive pairing regime, the rapidities become large, and thus the transition probabilities grow like large even polynomials. We therefore plot the results logarithmically. The cusps in the plots occur at the critical points where the nature of the rapidities change. Incorporating the energy denominator does not change the appearance of the plots. For all three types of TDM elements, the singles 1010, 1001 and 0110 are the most important in the attractive case, and the double 0011 becomes more important than the single 0101 in the strongly attractive limit. Each state is important in the strongly repulsive limit, but to different TDM elements.

\begin{figure}
	\begin{subfigure}{\textwidth}
		\includegraphics[width=0.325\textwidth]{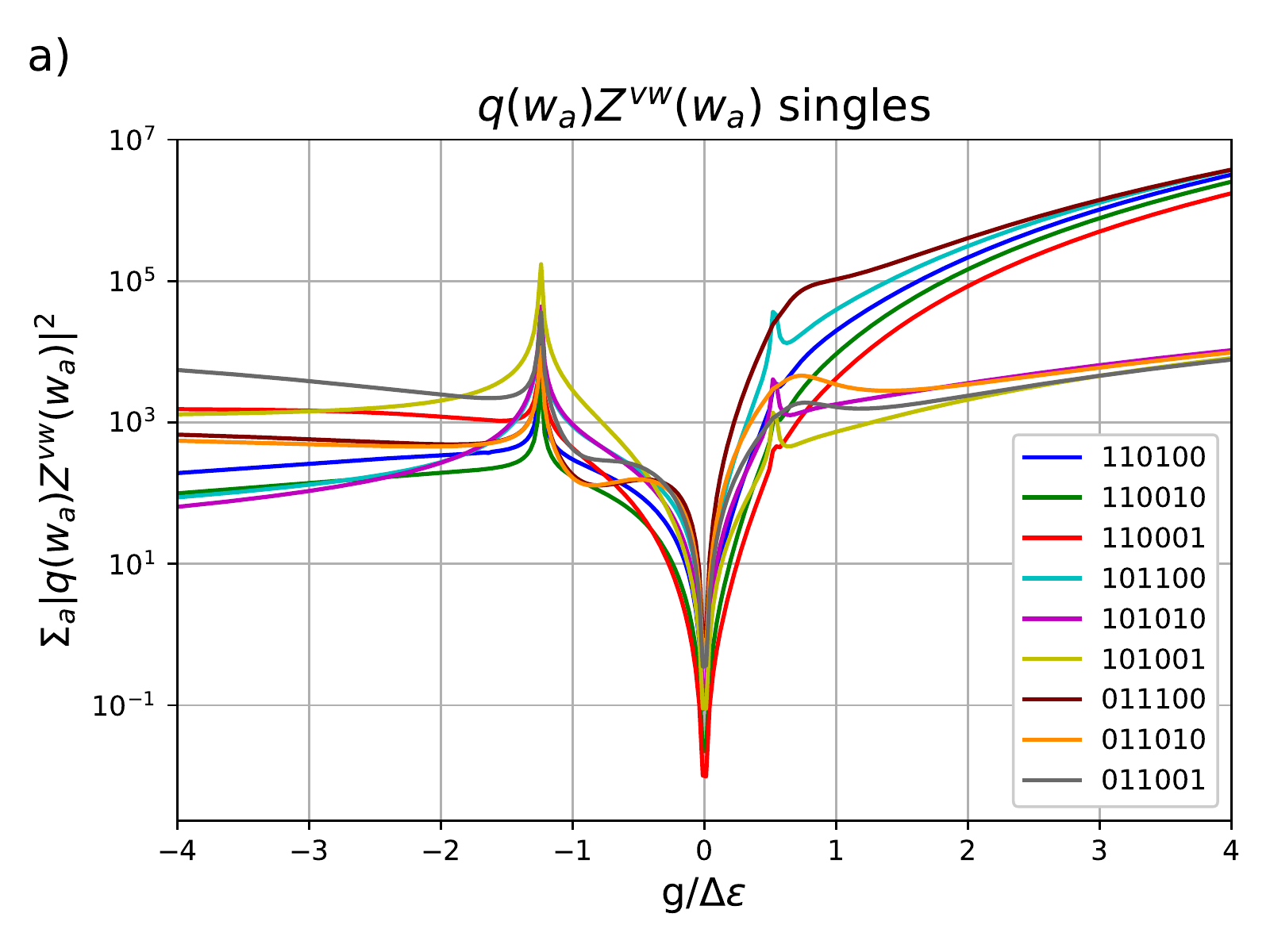}  \hfill
		\includegraphics[width=0.325\textwidth]{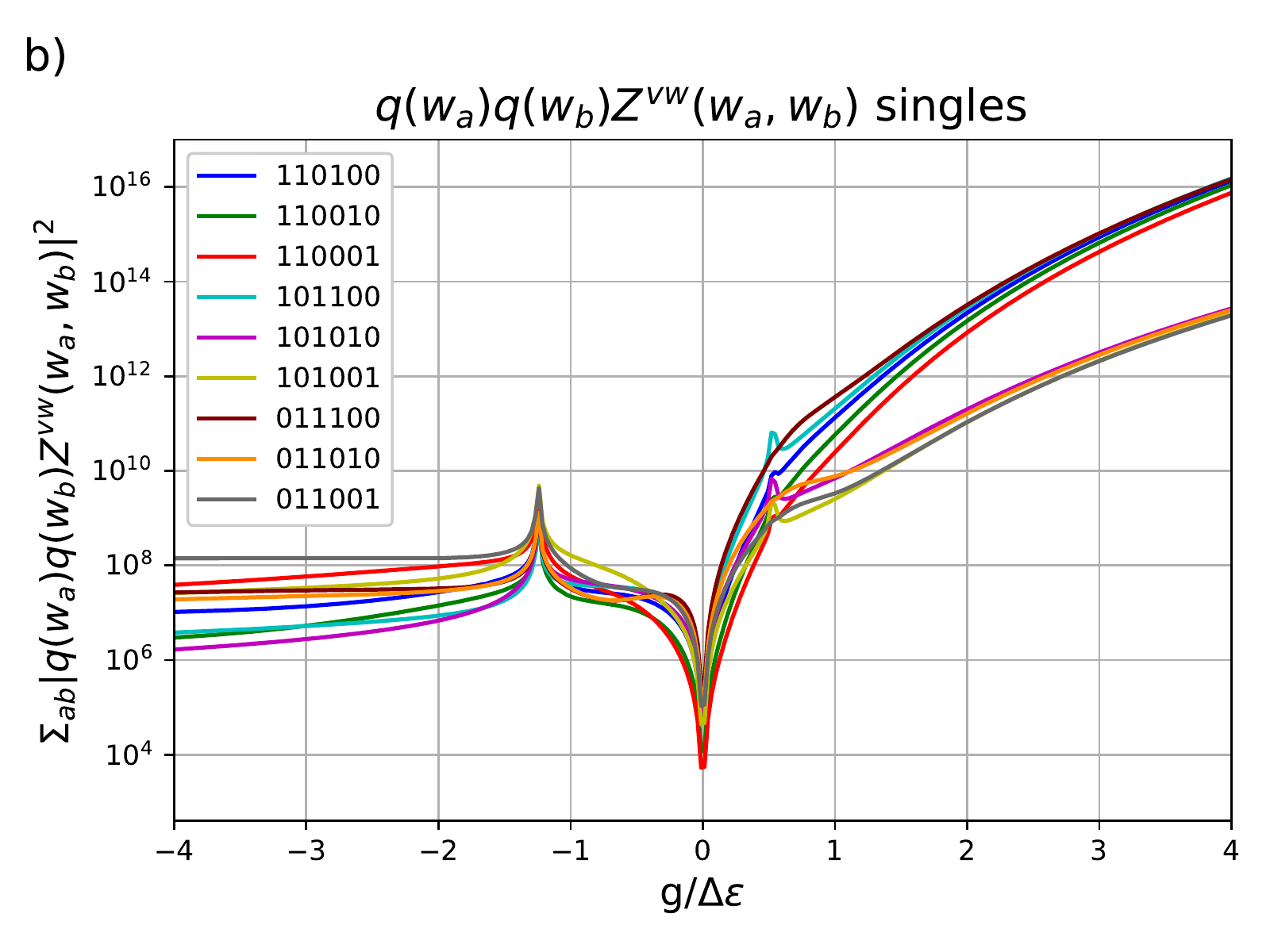} \hfill
		\includegraphics[width=0.325\textwidth]{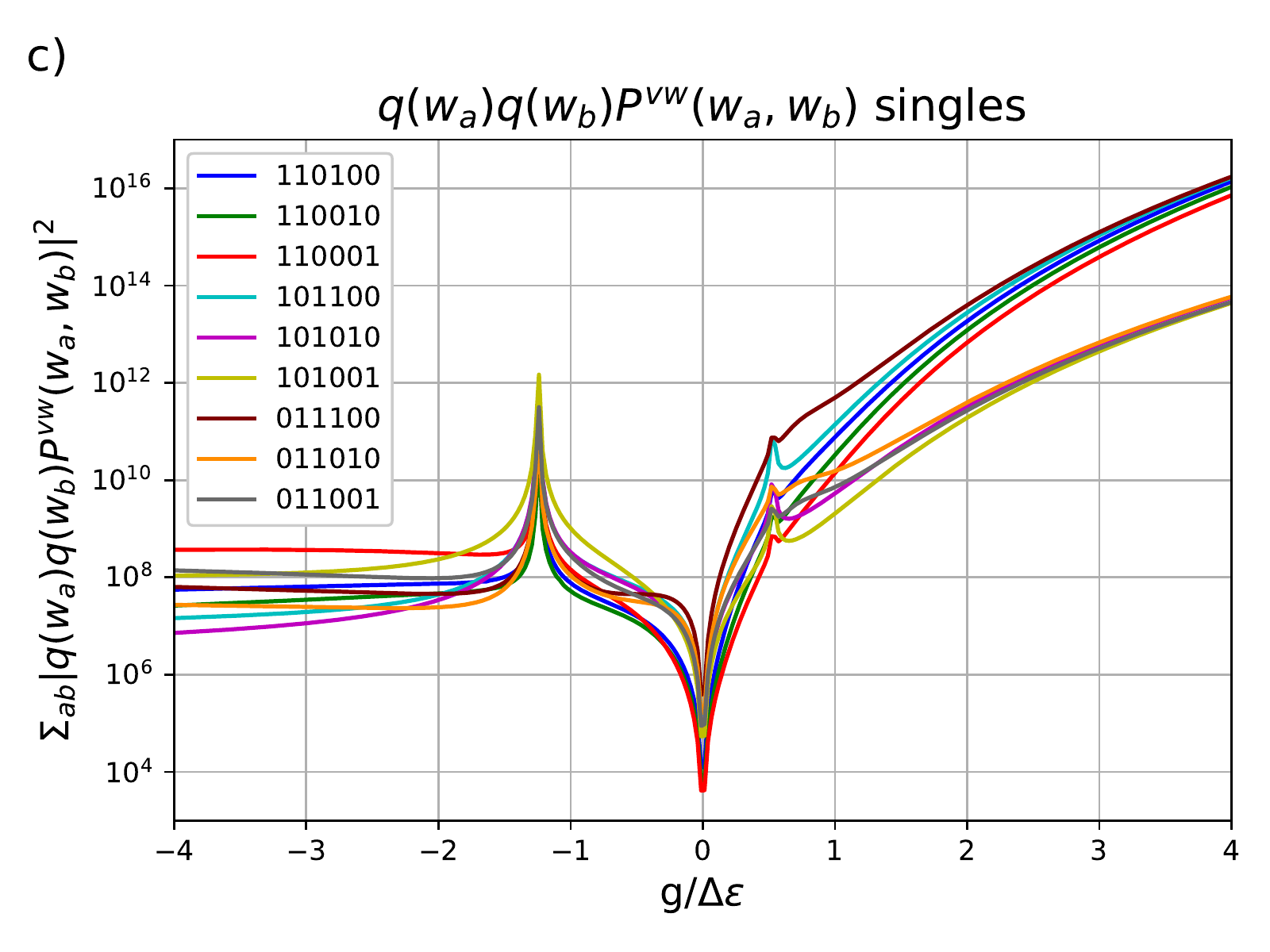}
	\end{subfigure}		
	\begin{subfigure}{\textwidth}
		\includegraphics[width=0.325\textwidth]{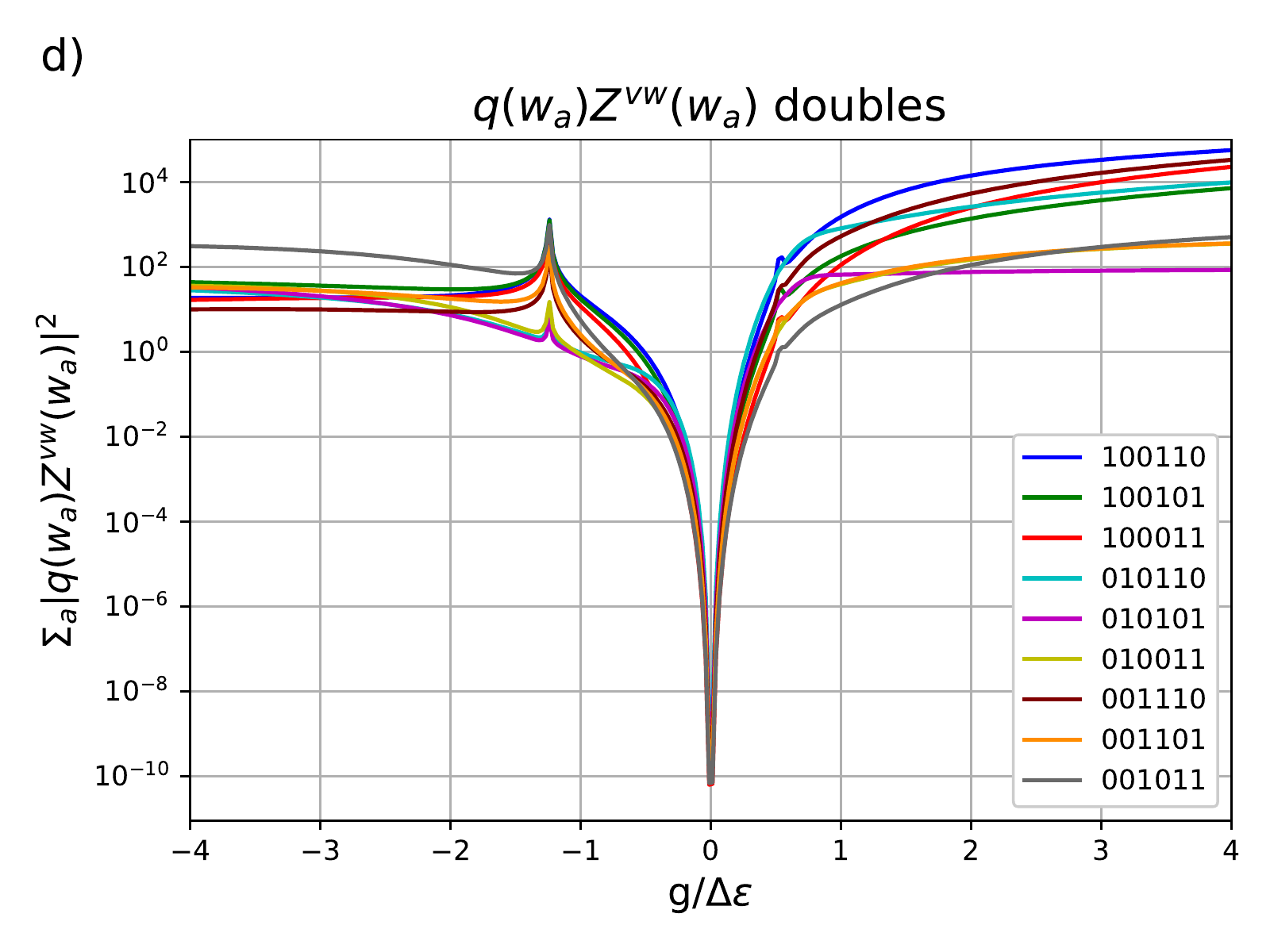}  \hfill
		\includegraphics[width=0.325\textwidth]{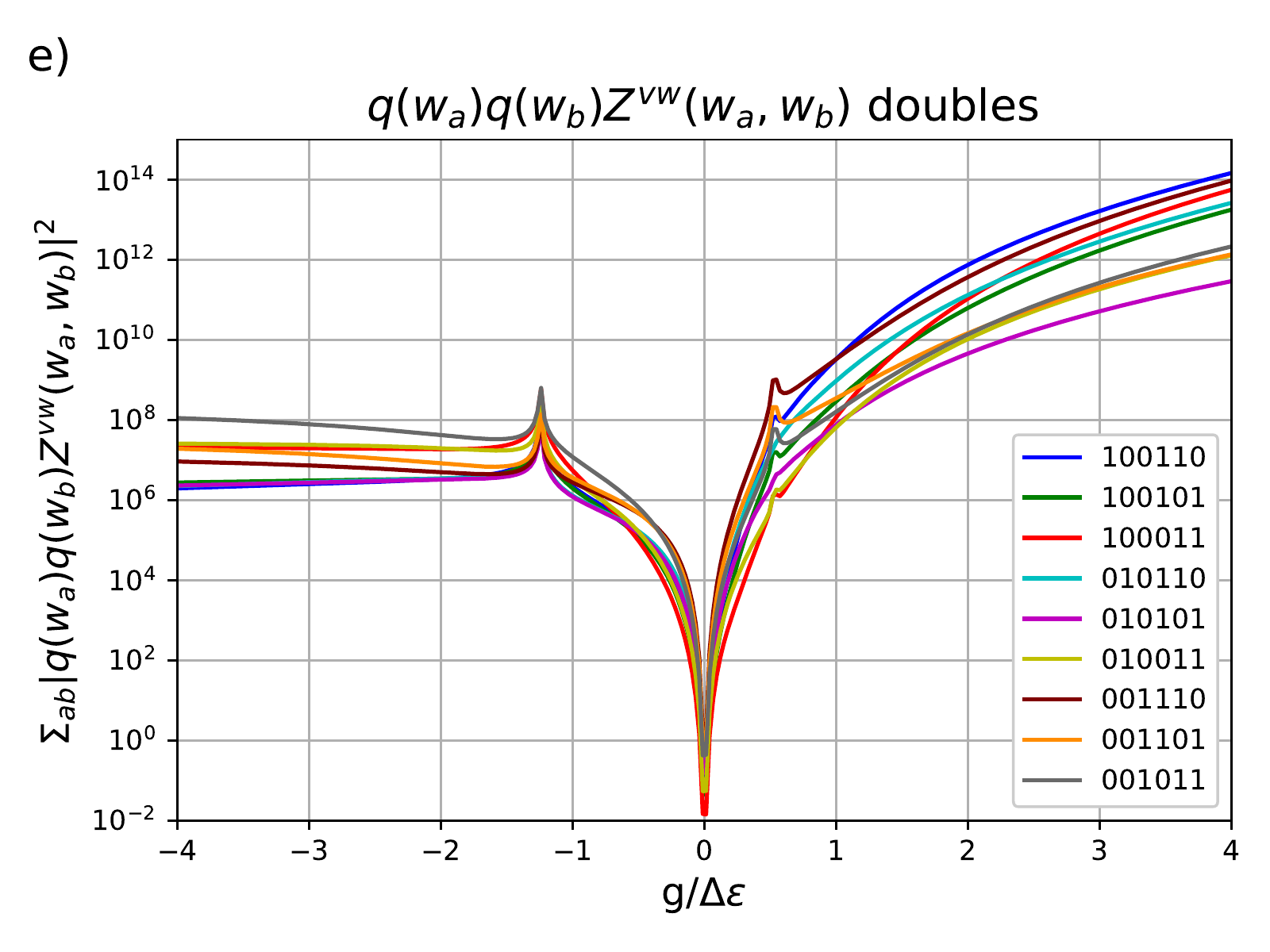} \hfill
		\includegraphics[width=0.325\textwidth]{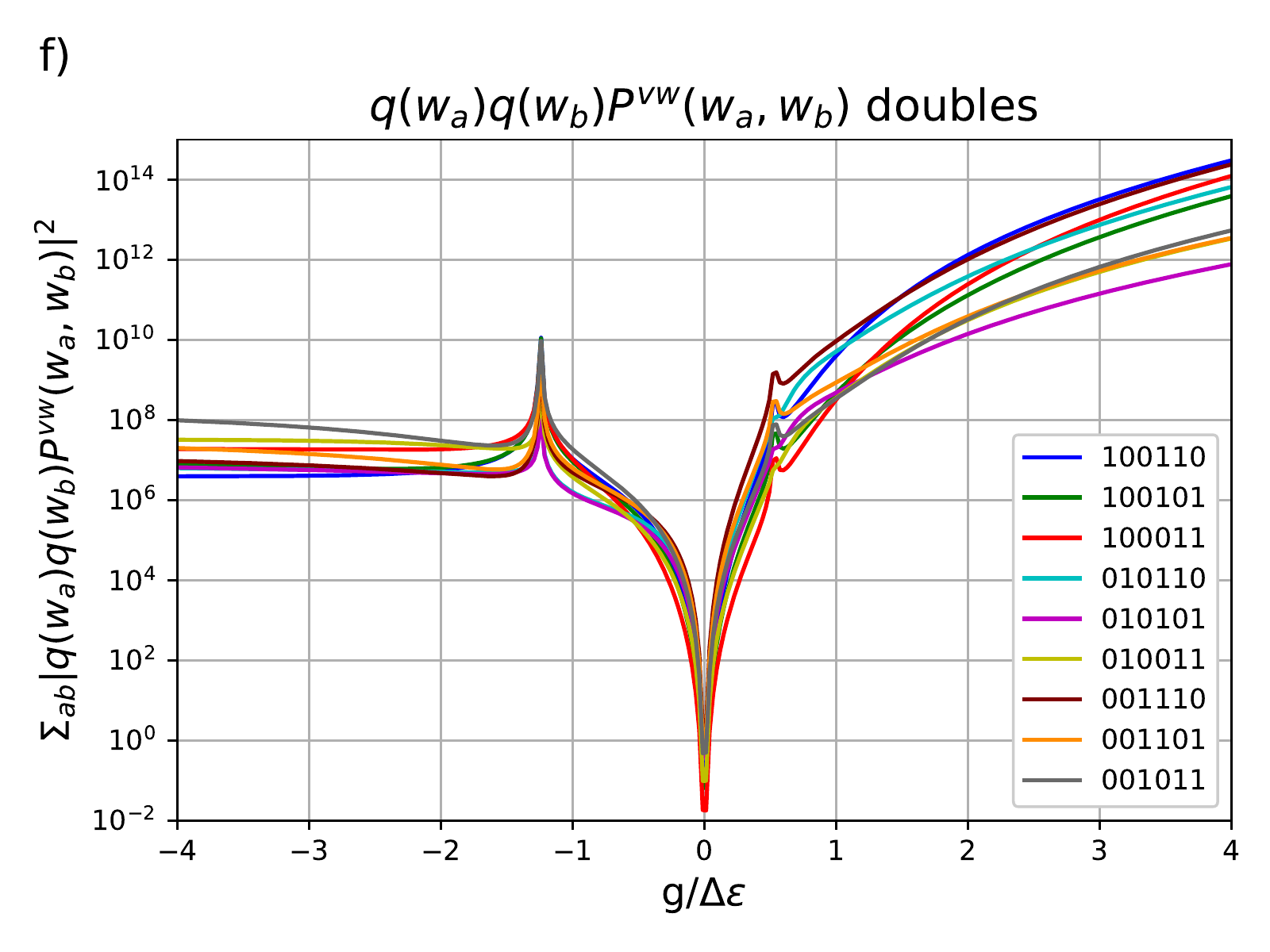}				
	\end{subfigure}
	\begin{subfigure}{\textwidth}
		\includegraphics[width=0.325\textwidth]{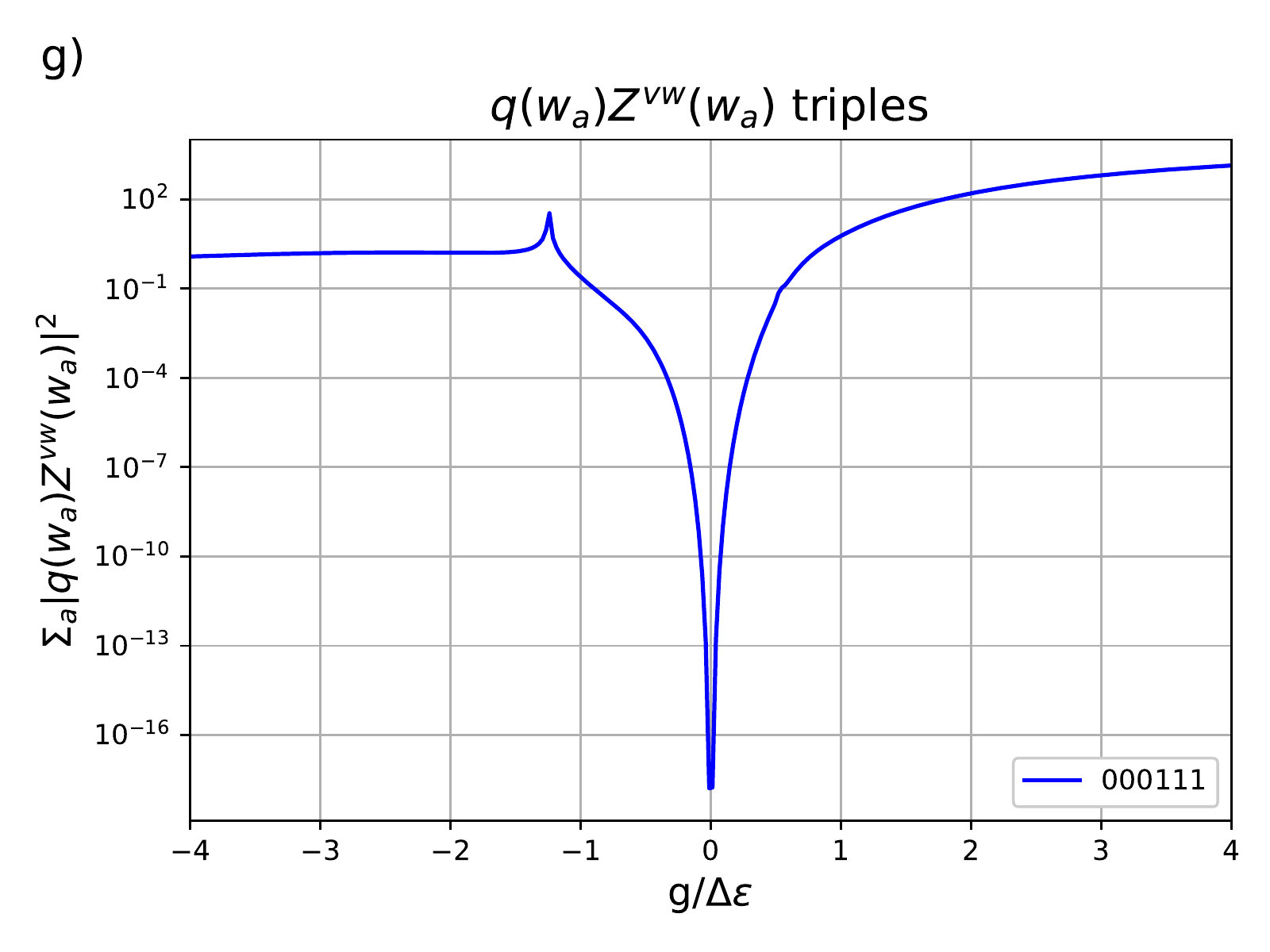}  \hfill
		\includegraphics[width=0.325\textwidth]{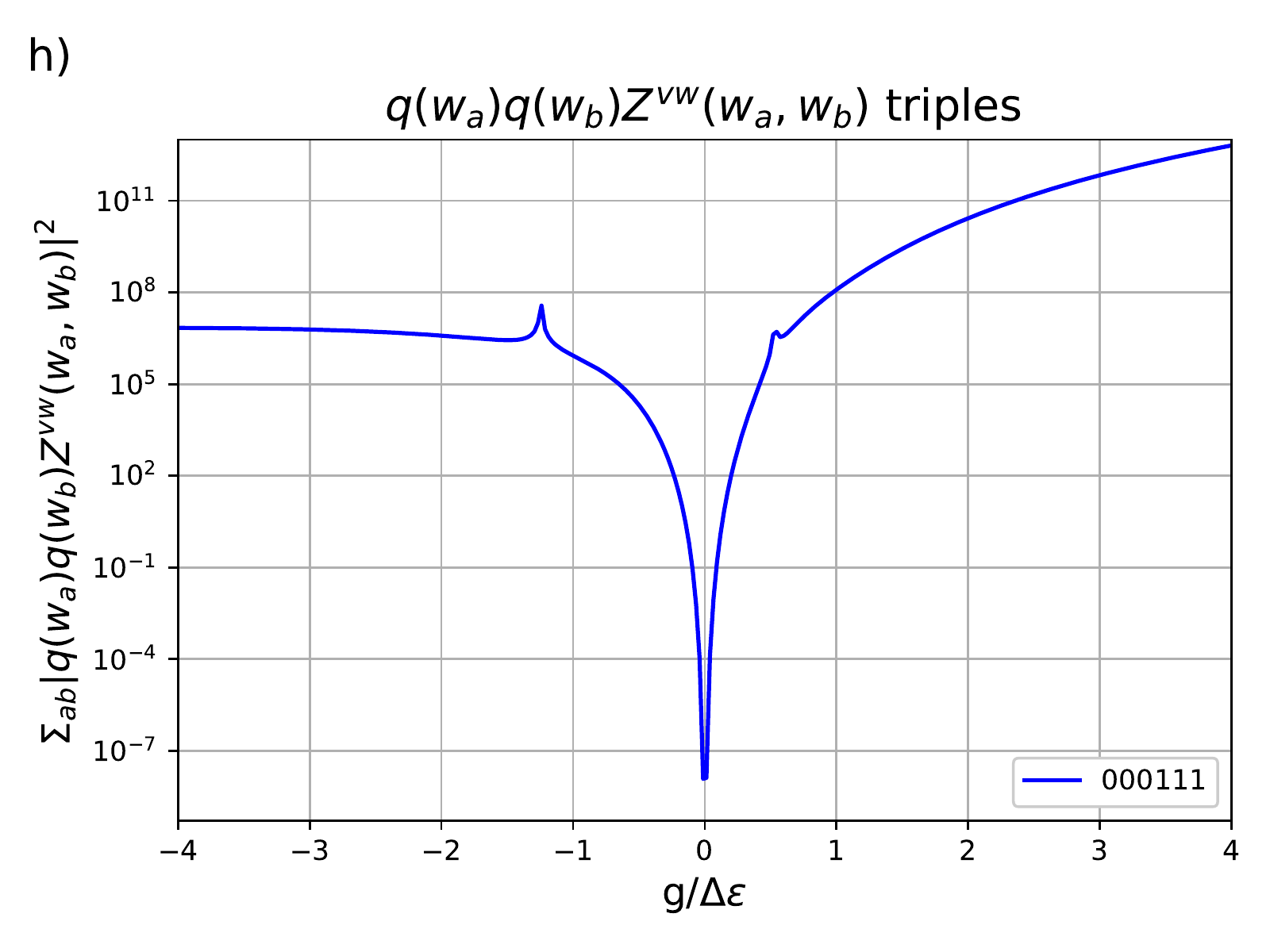} \hfill
		\includegraphics[width=0.325\textwidth]{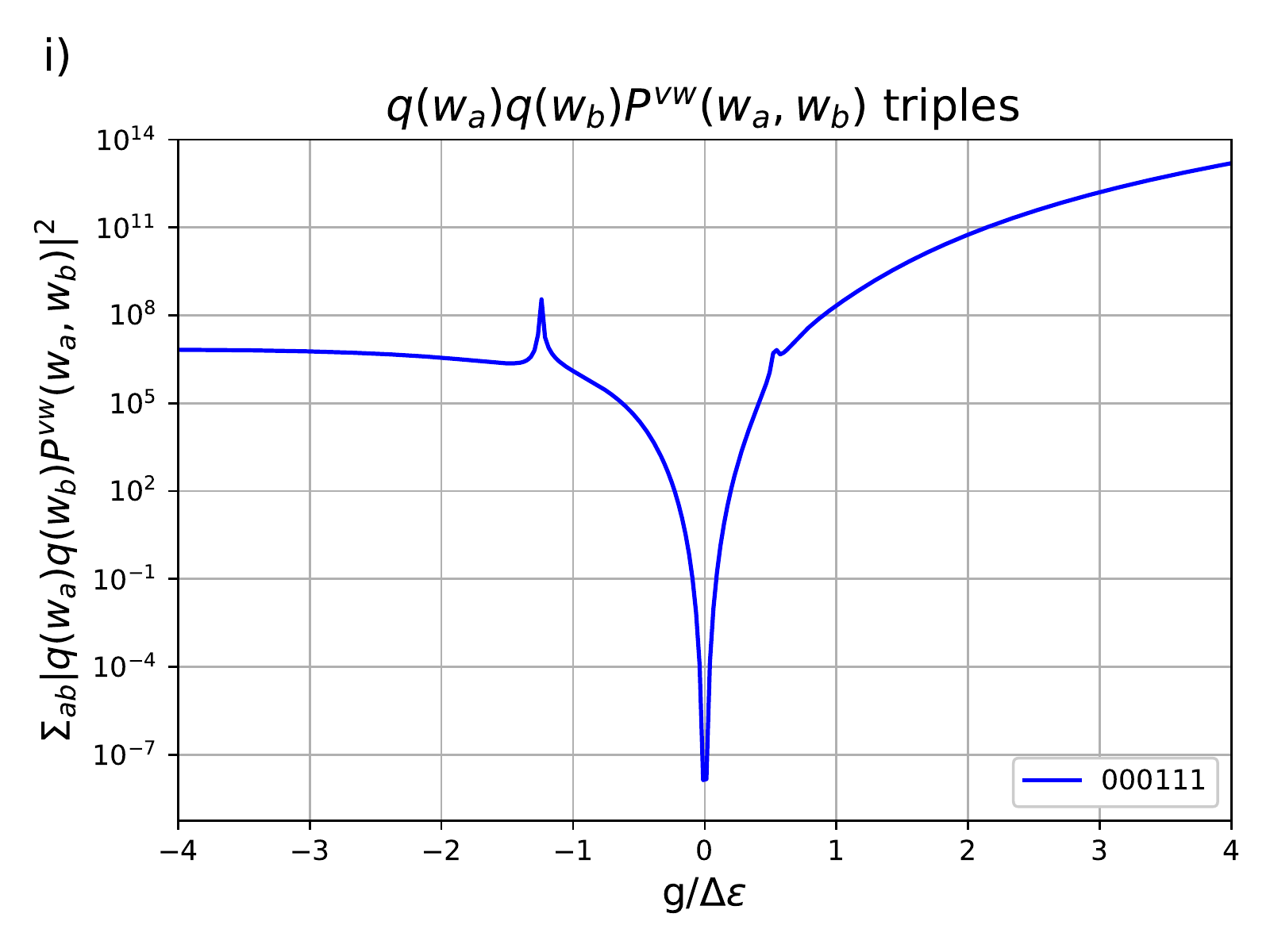}				
	\end{subfigure}
	\caption{Sums of regularized Gaudin basis transition probabilities for half-filled 6-site picket fence model for a-c) single, d-f) double and g-i) triple excitations.}
	\label{fig:6_3_gb}
\end{figure}
Regularized transition probabilities for the half-filled 6-site picket fence model are presented in figure \ref{fig:6_3_gb}.
In the repulsive regime, the three most important states are singles with a pair in the highest level. In the attractive regime, the Fermi singles are dominant, while the remaining singles are about even with the most important doubles. The triple 000111 gives no meaningful contribution.

\begin{figure}
	\begin{subfigure}{\textwidth}
		\includegraphics[width=0.325\textwidth]{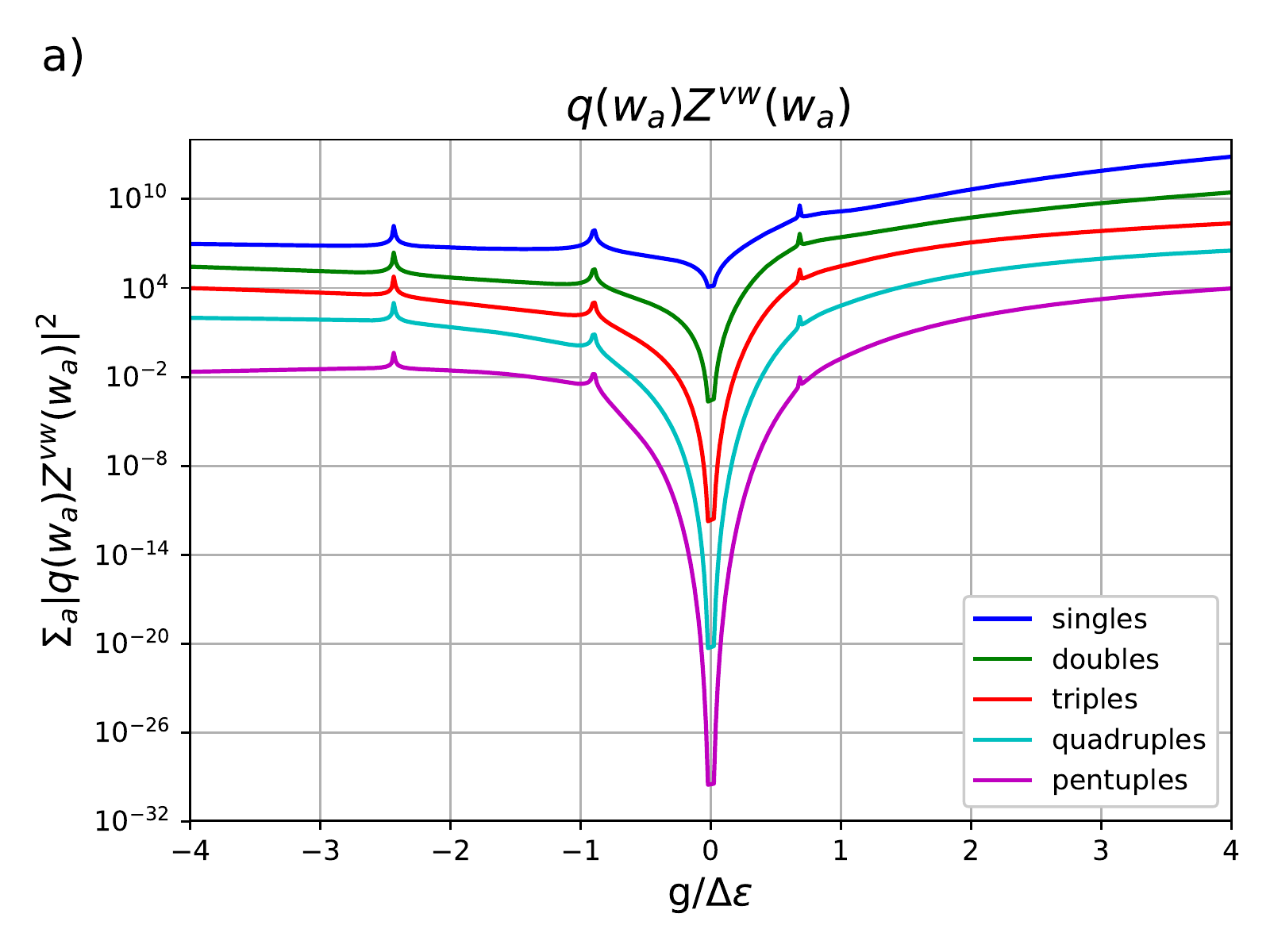}  \hfill
		\includegraphics[width=0.325\textwidth]{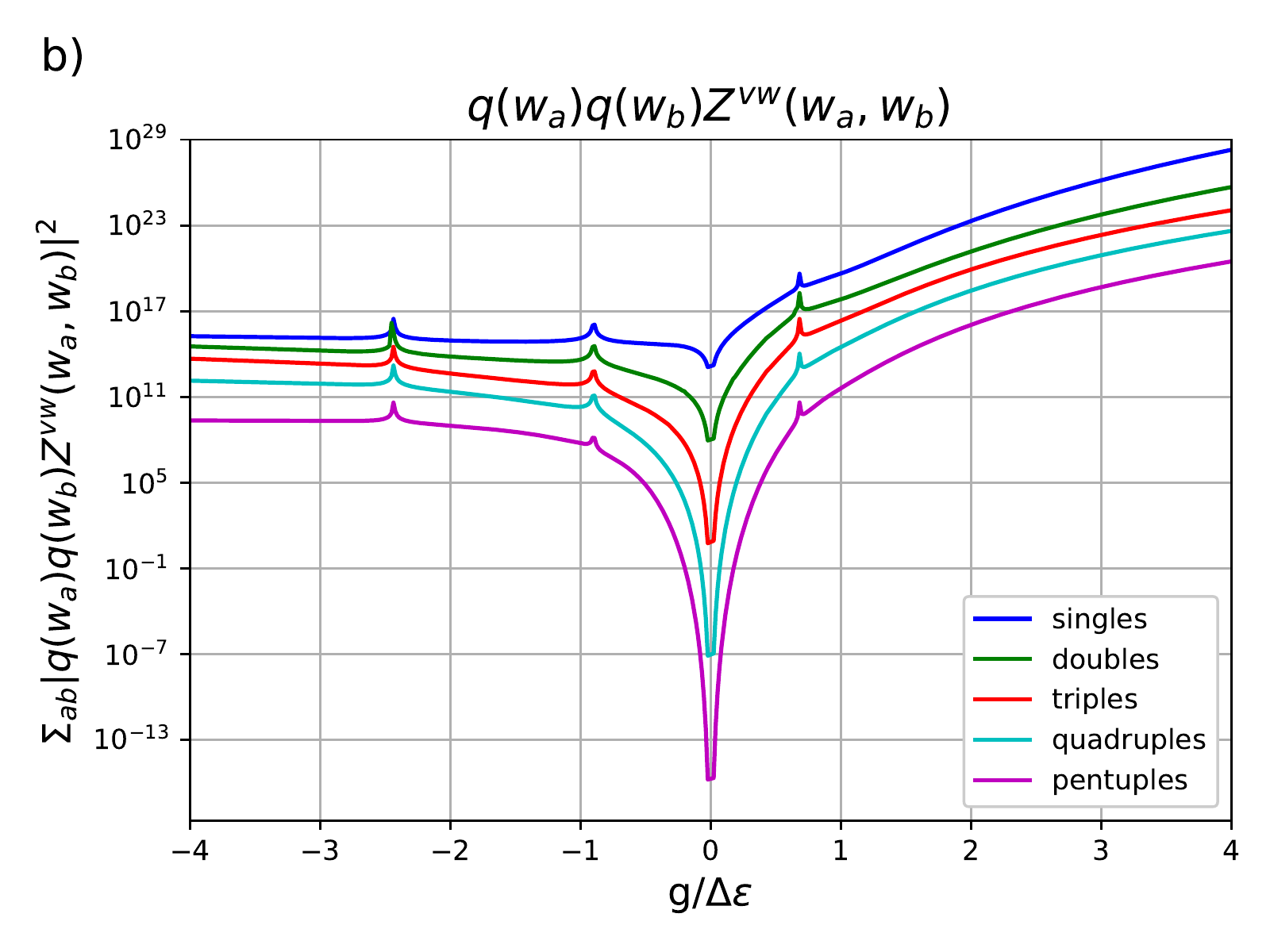} \hfill
		\includegraphics[width=0.325\textwidth]{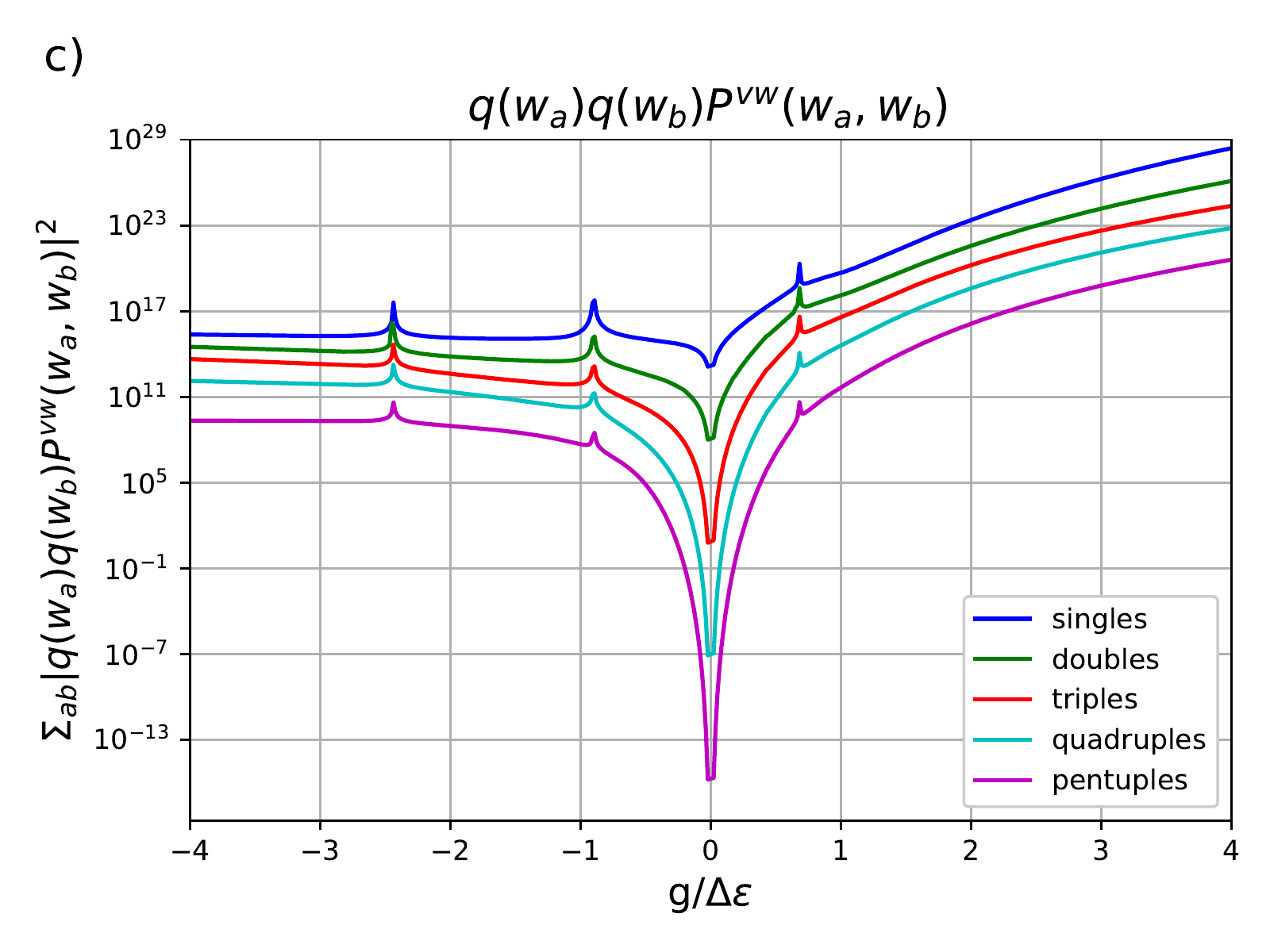}				
	\end{subfigure}
	\caption{Sums of regularized Gaudin basis transition probabilities for half-filled 10-site picket fence model.}
	\label{fig:10_5_gb}
\end{figure}
Regularized transition probabilities grouped by excitation level for the half-filled 10-site model are shown in figure \ref{fig:10_5_gb}. At all couplings the singles give the largest contribution, followed by the doubles etc.

\section{Conclusion}
The purpose of this contribution is to demonstrate that while there are not Slater-Condon rules for the TDM elements of RG states, the situation is still manageable without drastic approximations. In both the physical and Gaudin bases, contributions from the single pair excitations are by far the most important, double pair excitations should be included, but triples and beyond are much less important, especially in the physical basis. The dominant excitations in the attractive regime are the Fermi states while for the repulsive pairing regime the Rydberg states are most important. In the physical basis, $|D^{vw}_{ij}|^2$ and $|P^{vw}_{ij}|^2$ singles with energy denominators go linearly to minus infinity in the purely repulsive limit. In the Gaudin basis, all transition probabilities behave like polynomials in the attractive regime. Practically, these results suggest that for a pertubation theory it would be better to use the physical basis for attractive pairing strengths, and the Gaudin basis for repulsive pairing strengths. Generally, there is a clear decrease in importance with respect to excitation level, though the physical basis may be preferable as it avoids the rectangular transformation to the Gaudin basis. In the following contributions, we will employ these results to develop a perturbation theory starting from an RG mean-field.

\section{Acknowledgements}
We thank the Natural Sciences and Engineering Research Council of Canada for their financial support. This research was enabled in part by support from Calcul Qu\'{e}bec and Compute Canada. We also thank Alexandre Faribault for useful discussions concerning critical points of Richardson's equations.

\appendix
\section{Gaudin Algebra actions} \label{sec:GAA}
The actions of Gaudin algebra elements on an RG eigenvector were reported previously,\cite{fecteau:2020a} so we present only the final results here. For a set of $\{w\}$ that satisfy Richardson's equations, the relevant results are:

\begin{align} \label{eq:Za}
S^z (w_a) \ket{ \{w\}} = \frac{\partial S^+ (w_a)}{\partial w_a} \ket{ \{w\}_a} + \sum_{b \neq a} \frac{S^+ (w_a)}{w_a -w_b} \ket{ \{w\}_b}
\end{align}

\begin{align} \label{eq:Zaa}
S^z(w_a)S^z(w_a) \ket{\{w\}} &= \frac{1}{2} \frac{\partial^2 S^+(w_a)}{\partial w_a^2} \ket{\{w\}_a} + \sum_{c,d \neq a} \frac{S^+(w_a)S^+(w_a)}{(w_a-w_c)(w_a-w_d)} \ket{\{w\}_{c,d}} \nonumber \\
&+ \sum_{c\neq a} \frac{3}{w_a-w_c} \frac{\partial S^+(w_a)}{\partial w_a} \ket{\{w\}_c} + \sum_{c \neq a} \frac{S^+(w_a)}{(w_a-w_c)^2} \ket{\{w\}_c}
\end{align}

\begin{align} \label{eq:Zab}
S^z(w_a) S^z(w_b) \ket{\{w\}} &= \frac{\partial S^+(w_a)}{\partial w_a} \frac{\partial S^+(w_b)}{\partial w_b} \ket{\{w\}_{a,b}}  - \frac{3}{(w_a-w_b)^2} \ket{\{w\}} \nonumber \\
&+ \sum_{c\neq a,b} \left( \frac{1}{w_b-w_c}\frac{\partial S^+(w_a)}{\partial w_a}S^+(w_b) \ket{\{w\}_{a,c}} + \frac{1}{w_a-w_c}\frac{\partial S^+(w_b)}{\partial w_b}S^+(w_a) \ket{\{w\}_{b,c}} \right) \nonumber \\
&+ \sum_{c\neq a,b} \left( \frac{2}{(w_a-w_b)(w_b-w_c)}S^+(w_a)\ket{\{w\}_c} + \frac{2}{(w_b-w_a)(w_a-w_c)} S^+(w_b) \ket{\{w\}_c} \right) \nonumber \\
&+ \frac{1}{2} \sum_{c,d \neq a,b} \left( \frac{1}{(w_a - w_c)(w_d-w_b)} + \frac{1}{(w_a-w_d)(w_b-w_c)} \right) S^+(w_a)S^+(w_b) \ket{\{w\}_{c,d}} \nonumber \\
&+ \frac{1}{(w_a-w_b)^2} S^+(w_a) \ket{\{w\}_b} + \frac{1}{(w_a-w_b)^2} S^+(w_b) \ket{\{w\}_a}.
\end{align}

\begin{align} \label{eq:Paa}
S^+(w_a) S^-(w_a) \ket{\{w\}} &= -2 \left( \frac{\partial \alpha (w_a)}{\partial w_a } +  \sum_{c\neq a} \frac{1}{(w_c - w_a)^2} \right)  \ket{\{w\}} \nonumber \\
&+ 2 \sum_{c \neq a} \frac{1}{w_c - w_a}  \frac{\partial S^+(w_a)}{\partial w_a} \ket{\{w\}_{c}} 
- 2 \sum_{c\neq a} \frac{S^+(w_a)}{(w_c - w_a)^2}  \ket{\{w\}_c} \nonumber \\
&- \sum_{c,d\neq a} \frac{S^+(w_a)S^+(w_a)}{(w_a-w_c)(w_a-w_d)}  \ket{ \{w\}_{c,d} } .
\end{align}

\begin{align} \label{eq:Pab}
S^+(w_a) S^-(w_b) \ket{\{w\}} &= -2 \left( \frac{\partial \alpha (w_b)}{\partial w_b } +  \sum_{c\neq b} \frac{1}{(w_c - w_b)^2} \right) S^+(w_a) \ket{\{w\}_b} \nonumber \\
&+ 2 \sum_{c \neq b} \frac{S^+(w_a)}{w_c - w_b}  \frac{\partial S^+(w_b)}{\partial w_b} \ket{\{w\}_{b,c}} 
- 2 \sum_{c\neq b} \frac{S^+(w_a)}{(w_c - w_b)^2}  \ket{\{w\}_c} \nonumber \\
&- \sum_{c,d\neq b} \frac{S^+(w_a)S^+(w_b)}{(w_b-w_c)(w_b-w_d)}  \ket{ \{w\}_{c,d} }
\end{align}

\section{Gaudin Basis Form factors} \label{sec:GFF}
Form factors are calculated in exactly the same manner as for the RDM case: begin from Slavnov's theorem, set the free variables $u_a = w_a + h$ for some small $h$, use the geometric series to expand in powers of $h$ and take the limit $h\rightarrow 0$. For form factors of derivatives, first take the derivative of Slavnov's theorem with respect to one of the free $\{u\}$ before performing the same limiting procedure. In all cases, terms proportional to negative powers of $h$ will vanish identically, as they are determinants with repeated columns, while those with positive powers of $h$ vanish in the limit. The remaining desired terms are the result. Two expressions are presented for each form factor: the first as a limit of Slavnov's theorem, while the second is a consistency check based on the physical basis form factors, $F^i_a$ and $F^{ij}_{ab}$, which we also list here for convenience. Each form factor has been verified numerically with the two expressions.

The physical basis form factors are:
\begin{align}
F^i_a = \frac{\prod_c (v_c-\varepsilon_i)}{\prod_c (v_c-w_a)} \frac{\prod_{b\neq a} (w_a-w_b)}{\prod_{b \neq a} (\varepsilon_i - w_b)} K \det J^i_a
\end{align}

\begin{align}
F^{ij}_{ab} = \frac{(w_a-w_b)}{(\varepsilon_i-\varepsilon_j)} \frac{\prod_c (v_d-\varepsilon_i)(v_d-\varepsilon_j)}{\prod_c (v_d-w_a)(v_d-w_b)}
\frac{\prod_{d\neq a,b} (w_a-w_d)(w_b-w_d)}{\prod_{d \neq a,b} (\varepsilon_i - w_d)(\varepsilon_j - w_d)} K \det J^{ij}_{ab}
\end{align}

In the Gaudin basis, the form factors required to compute transition density matrices are:
\begin{align}
\braket{\{v\} | S^+(w_a) | \{w\}_c} &= \frac{\prod_k (v_k - w_a)  }{\prod_k (v_k - w_c)} \frac{\prod_{l \neq c} (w_c - w_l)}{\prod_{l \neq a,c} (w_a - w_l)} K \det J^a_c \\
&= \sum_i \frac{F^i_c}{w_a - \varepsilon_i}
\end{align}

\begin{align}
\braket{ \{v\} | \frac{\partial S^+(w_a)}{\partial w_a} | \{w\}_a} &= K \det J^a_a \\
&= - \sum_i \frac{F^i_a}{(w_a-\varepsilon_i)^2}
\end{align}

\begin{align}
\braket{\{v\} | S^+(w_a) S^+(w_a)  | \{w\}_{c,d}} &= \frac{\prod_k (v_k - w_a)^2 \prod_{l\neq c,d} (w_c-w_l)(w_d-w_l)}{\prod_k (v_k - w_c)(v_k - w_d) \prod_{l\neq a,c,d} (w_a-w_l)^2} (w_c-w_d) 
\frac{}{} K \det J^{\bar{a} a}_{cd} \\
&= \sum_{ij} \frac{F^{ij}_{cd}}{(w_a-\varepsilon_i)(w_a-\varepsilon_j)}
\end{align}

\begin{align}
\braket{\{v\} | S^+(w_a) S^+(w_b)  | \{w\}_{c,d}} &= \frac{\prod_k (v_k - w_a)(v_k - w_b) \prod_{l\neq c,d} (w_c-w_l)(w_d-w_l)}{\prod_k (v_k-w_c)(v_k-w_d) \prod_{l\neq a,c,d} (w_a-w_l) \prod_{m\neq b,c,d} (w_b-w_m)} 
\nonumber \\
&\frac{(w_c-w_d)}{(w_a-w_b)}
K \det J^{ab}_{cd} \\
&= \sum_{ij} \frac{F^{ij}_{cd}}{(w_a-\varepsilon_i)(w_b-\varepsilon_j)}
\end{align}

\begin{align}
\braket{ \{v\} | \frac{\partial S^+(w_a)}{\partial w_a} | \{w\}_c} &= \frac{\prod_k (v_k - w_a) \prod_{l\neq c} (w_c - w_l) }{\prod_k (v_k - w_c) \prod_{l \neq a,c} (w_a - w_l)} 
K\left( \det J^{\bar{a}}_c  + \left( k(w_a) + \frac{1}{(w_a-w_c)} \right) \det J^a_c
\right) \\
&= - \sum_i \frac{F^i_c}{(w_a-\varepsilon_i)^2}
\end{align}

\begin{align}
\braket{ \{v\} | \frac{\partial S^+(w_a)}{\partial w_a} S^+(w_b) | \{w\}_{a,c}} &= 
\frac{\prod_k (v_k - w_b) \prod_{l\neq c} (w_c-w_l)}{\prod_k (v_k - w_c) \prod_{l\neq b,c} (w_b-w_l)} K \nonumber \\
&\left(
\det J^{ab}_{ac} + \left( k(w_a) + \frac{(w_c-w_b)}{(w_a-w_c)(w_a-w_b)} \right) \det J^b_c
\right) \\
&= -\sum_{ij} \frac{F^{ij}_{ac}}{(w_a-\varepsilon_i)^2(w_b-\varepsilon_j)}
\end{align}

\begin{align}
\braket{ \{v\} | \frac{\partial S^+(w_a)}{\partial w_a} \frac{\partial S^+(w_b)}{\partial w_b} | \{w\}_{a,b}} &= K \det J^{ab}_{ab} + Kk(w_a) \det J^b_b + K k(w_b) \det J^a_a \\
&= \sum_{ij} \frac{F^{ij}_{ab}}{(w_a-\varepsilon_i)^2(w_b-\varepsilon_j)^2}
\end{align}

\begin{align}
\frac{1}{2} \braket{ \{v\} | \frac{\partial^2 S^+(w_a)}{\partial w^2_a} | \{w\}_a} & = K k(w_a) \det J^a_a + K \det J^{\bar{a}}_a \\
&= \sum_i \frac{F^i_a}{(w_a-\varepsilon_i)^3}
\end{align}

\bibliography{Babasis_2}

\bibliographystyle{unsrt}

\end{document}